# Dynamical Lattice Thermal Conductivity, Shastry's Sum Rule and Second Sound in Bulk Semiconductor Crystals

Younès Ezzahri and Karl Joulain

*Institut Pprime, CNRS-Université de Poitiers-ENSMA, Département Fluides, Thermique, Combustion, ENSIP-Bâtiment de mécanique, 2 rue Pierre Brousse, F 86022 Poitiers, Cedex, France.*

*: younes.ezzahri@univ-poitiers.fr

## ABSTRACT

The paper discusses the fundamental behavior of the dynamical lattice thermal conductivity $\kappa(\Omega)$ of bulk cubic semiconductor crystals. The calculation approach is based on solving Boltzmann-Peierls Phonon Transport Equation in the frequency domain after excitation by a dynamical temperature gradient, within the framework of the single relaxation time approximation and using modified Debye-Callaway model in which both longitudinal and transverse phonon modes are included explicitly. The system is treated as a continuum elastic isotropic linear dispersionless medium. The frequency and temperature dependences of three-phonon anharmonic Normal and Umklapp phonon scattering processes are kept the same for all semiconductors. Our model allows us to obtain a compact expression for $\kappa(\Omega)$ that captures the leading behavior and the essential features of the dynamical thermal conduction by phonons. This expression fulfills the causality requirement and leads to a convolution type relationship between the heat flux density current and the temperature gradient in the real space-time domain in agreement with Gurtin-Pipkin theory. The dynamical behavior of $\kappa(\Omega)$ is studied by changing ambient temperature as well as different intrinsic and extrinsic parameters including the effect of embedding semiconductor nanoparticles as extrinsic phonon scattering centers. Our calculations show the cut-off frequency of $\kappa(\Omega)$ to be very sensitive to the changes of some of these parameters with a notable consideration to the effect of the size and concentration of the embedded nanoparticles. Some of the parameters (Grüneisen coefficient particularly) are usually used as fitting parameters to the behavior of the steady-state thermal conductivity as a function of temperature. This fact makes their relevance on the cut-off frequency of $\kappa(\Omega)$ even more important regarding the uncertainty on their real values.

The paper investigates also the applicability of Shastry's Sum Rule (SSR) in the frame work of Boltzmann theory. It is shown that within the frame work of Callaway approximated form of the collision operator and time independent Callaway parameter, the SSR breaks down and is only valid when resistive processes dominate Normal processes, for which case, we derive an alternative expression to the classical limit of the expectation of the thermal operator introduced in Shastry's formalism.

As a by-product, we discuss also the possibility of existence and propagation of second sound in bulk cubic semiconductor crystals. We make use of Griffin's self-consistency criterion to study the dispersion relation of second sound in terms of its velocity, relaxation time and damping factor. The analysis leads to generalized *driftless* second sound properties which shed light on the fundamentally intertwining aspect between anharmonic Normal and Umklapp phonon-phonon scattering processes as well as the combined effect of these processes with other extrinsic phonon scattering processes. The behaviors of second sound velocity, relaxation time and damping factor are studied by considering the change in the same parameters as for $\kappa(\Omega)$. The study shows phonon scattering process by embedding semiconductor nanoparticles to have a significant effect on the relaxation of the coherent collective motion of the different normal phonon modes.

The first two parts of the present paper regarding the study of the fundamental behavior of the dynamical lattice thermal conductivity $\kappa(\Omega)$ of bulk cubic semiconductor crystals have been recently published in Journal of Applied Physics.[1,2] Here we aim to present our work in one piece and with a detailed description to preserve the coherence and consistency of the used assumptions and consequently obtained results. We add also a third and new part that discusses the existence and propagation of second sound in these bulk cubic semiconductor crystals.

# I. INTRODUCTION

Due to its rapidly emerging area in nanotechnologies as well as in fundamental physics studies, thermoelectric industry is still motivated in improving the efficiency of solid state energy conversion devices. This efficiency is quantified in terms of the dimensionless figure of merit $ZT = {\sigma S^2}/{\kappa} T$ where $T$ is the absolute temperature, $\sigma$, $S$ and $\kappa$ are the electrical conductivity, Seebeck coefficient and thermal conductivity of the thermoelectric material, respectively. Higher the $ZT$ is, better will be the thermoelectric energy conversion of the material.[3,4] Among different categories of materials, semiconductors (SC) are known to be the most suited and the best choice for both thermoelectric generation and refrigeration.[3,4] Current research in the field has proven the thermal conductivity to play a vital and the most fundamental role in increasing $ZT$, and almost all reported high values of $ZT$ have been interpreted, validated and confirmed as mainly due to a decrease in the thermal conductivity of the thermoelectric material regarding its geometry and composition.[5-11] In SC materials, energy (heat) is mainly carried by phonons. Understanding heat transfer and phonon behavior in these materials has become crucial and challenging in developing and engineering material structures with different dimensionalities (nanodots, nanowires, superlattices as well as nanoscale precipitates and composites) that further decrease the thermal conductivity below what can be achieved by alloying.[12-16]

The thermal conductivity $\kappa$ of bulk dielectric and SC crystal materials in which energy (heat) carriers are phonons, has been the object of many theoretical and experimental studies.[17-34] The steady-state behavior of $\kappa$ has been understood for many decades. According to the pioneer work of Debye and Peierls[17] and many experimental works later,[27-30,33,34] $\kappa$ has a universal behavior as a function of temperature: $\kappa$ depends on the size and shape of the crystal at low temperatures where the mean free path (MFP) of the phonon becomes of the order of the dimensions of the crystal. At this temperature regime, $\kappa$ mirrors the temperature behavior of the specific heat. $\kappa$ increases with temperature and reaches a maximum at about $T \approx 0.05\theta_D$, where $\theta_D$ is an average Debye temperature over all phonon polarization branches of the crystal. Above this maximum $\kappa$ is limited by scattering of phonons amongst themselves via anharmonic scattering processes, more essentially Umklapp processes (U-processes) and is characteristic of the material crystal. The effect of impurities or imperfections in the crystal to scatter phonons is particularly important near the maximum where both boundary scattering and anharmonic scattering processes are present but weak.

At short time scales, a number of unfamiliar and intriguing phenomena have been predicted and few of them have been observed. Energy transport at very short time scales where local nonequilibrium regimes appear is probably the most interesting one.[35,36] The question of energy and heat transport mechanisms at short time and length scales is the basis of numerous theoretical and experimental papers.[35-43] The study of energy and heat transport at very short time scales has even become crucial and more needed recently due to the continuously increasing of clock speeds and decreasing of feature sizes in microelectronic and optoelectronic applications.[44] Clock speeds of present microprocessors based on silicon technology are of few Gigahertz, and according to the International Technology Roadmap for Semiconductors (ITRS), devices with clock speeds of few tens of Gigahertz will be available in the next decade.[44]

In the past, only few works addressed the question of studying the time and dynamical behaviors of the lattice thermal conductivity of SC crystals. Most of the works were theoretical; the most cited ones are Guyer and Krumhansl,[45-47] Volz[48,49], Alvarez and Jou,[41,42], Hüttner[50] and Shastry.[51,52] Based on these works, the expected cut-off frequency $f_C$ of the dynamical lattice thermal conductivity has usually been theorized to be on the order of $f_C > 1/\tau$ where $\tau$ is the relaxation time of the dominant phonons during the heat transport phenomenon. Volz[49] showed that the thermal conductivity $\kappa$ of Si decreases at frequencies $f\tau > 1$. The life time at room temperature of dominant phonons in SC alloys is ~100 ps, and then the expected $f_C$ based on the former theoretical investigation is $f_C > 10GHz$. For his prediction, Volz[48,49] used two different methods, (i) molecular dynamics and (ii) the expression of $\kappa$ based on Boltzmann-Peierls equation in the grey spectrum approximation (constant relaxation time). A constant relaxation time however, seems to be a very poor approximation, especially when considering the dynamical behavior of energy transport. At the top of this short list of theoretical works, is the remarkable and interesting work of Shastry,[51,52] in which he introduced a new formalism to study the dynamical behavior of not only thermal conductivity but also other thermoelectric properties (electrical conductivity, Seebeck coefficient and Lorentz number) for different condensed matter models. Even though Shastry did not discuss the behavior of the cut-off frequency of the dynamical thermal conductivity, he introduced a new sum rule of the real part of the latter, in analogy to the well established $f$-sum rule of the real part of the electrical conductivity.[52]

Recently Koh and Cahill reported the most notable and cited experimental work so far regarding the frequency (dynamical) behavior of the thermal conductivity of SC crystal alloys.[53] Using Time domain Thermoreflectance (TDTR),[54-56] the authors measured the thermal conductivity of a number of SC crystals including alloys and single crystals as a function of the frequency of excitation of the heat source which, in the experiment was the modulation frequency of the laser pump beam.[54-56] The analysis of the dynamical behavior of the thermal conductivity at room temperature showed a cut-off frequency $f_C$ smaller than *10MHz* for SC alloys. On the other hand, the thermal conductivity of single SC crystals showed a plateau over the whole range of frequency used in the experiment *(0.6-10MHz)*.[53] This surprising result came to defy all previous theoretical investigations of the frequency or time dependence of the thermal conductivity of SC crystals.

To explain their valuable observations Koh and Cahill[53] made the statement that all phonons with MFP longer than the thermal penetration depth $\delta(\Omega)$, where $\Omega$ is the circular frequency of the excitation source, transport heat ballistically and as such won't contribute to the thermal conductivity $\kappa(\Omega)$ measured in the TDTR experiment, we will refer to this statement later in the discussion section as *Koh and Cahill statement*. Using a modified Debye-Callaway formalism as first proposed by Asen-Palmer et al[33] and Morelli et al,[34] the authors translate their assumption as a boundary scattering process that phonons would undergo at a virtual interface. This virtual interface is actually the surface of a sphere whose radius is the thermal penetration depth $\delta(\Omega)$. The authors found a good and satisfactory agreement between experimental data and the results of this phenomenological approach. We will get back to comment on this at the end of the discussion section.

The motivation behind the current work is threefold. First, we present an approach within the frame work of Boltzmann kinetic theory of phonon transport using the Callaway approximation of the collision operator in order to calculate and develop a compact formula capturing the leading dynamical behavior of the lattice thermal conductivity of bulk SC crystals $\kappa(\Omega)$, which will shed more light on the effect of different intrinsic and extrinsic parameters in influencing this dynamics. Second, we investigate the conditions under which Shastry's sum rule holds in the frame work of Boltzmann theory and we give an alternative expression to the classical limit of the expectation of the thermal operator introduced in Shastry's formalism.[51,52] Third, we discuss the possibility of existence and propagation of second sound in bulk SC crystals based on Griffin's self-consistency approach.[57]

The detail of the theoretical derivation of *κ(Ω)* is presented in the next section. In the third section, we discuss the results of this approach by analyzing the effect of varying the ambient temperature as well as different intrinsic and extrinsic parameters of the bulk SC crystal material including the effect of embedding semiconductor nanoparticles. At the end of this section we comment on the recent experimental data of Koh and Cahill,[53] then we discuss the applicability of Shastry's sum rule and the possibility of existence and propagation of second sound. We finish with summary and concluding remarks.

# II. THEORY

Our goal in this section is to develop a compact expression of the dynamical lattice thermal conductivity *κ(Ω)* of bulk SC crystals, in which heat is primarily carried by phonons, that gives an insight onto the leading behavior in their response to a dynamical temperature gradient. The latter could originate from application of a periodic heat source at the surface of the SC crystal, as was the case in the experimental work of Koh and Cahill,[53] or within its the volume.

## A. Boltzmann-Peierls Transport Equation

To develop an expression for *κ(Ω)* of a bulk SC crystal, one starts with Boltzmann-Peierls Transport Equation (BPTE). As many of the previous investigations,[17-34] the solution of this integral-differential equation is then approximated by the use of the relaxation time concept in which the phonon scattering process is expressed in terms of the single relaxation time $\tau(q,S)$ for a phonon (or more precisely a phonon wave packet[22,24,32],) of wave vector $\boldsymbol{q}$ and polarization *S*. In this case, the scattering cross sections are calculated using perturbation techniques.[17-20,22] In such an approach, the temperature and intrinsic frequency dependences of anharmonic three-phonon relaxation times are strongly dependent on the actual phonon polarization branch and on the dispersion relation of the phonon spectrum. The expressions derived for the relaxation times are only valid for specific phonons in a limited temperature range. For simplification, however, we assume the latter expressions to be valid for any temperature and we further assume an isotropic linear (Debye-like) phonon spectrum for each phonon polarization branch.

Callaway[21] approximation of the collision operator in BPTE allows a simple separation of Normal processes (N-processes) and U-processes. The pioneer purely intuitive work of Callaway[21] was investigated in detail by many authors and more robust theoretical foundations have been found.[26,31,32] For their algebraically convenient forms, the Callaway[21] and Holland[27] methods have been the most and widely used formulations for the steady-state thermal conductivity *κ(T)* that enable fitting of the experimental data for a large number of materials in which heat is carried by phonons, with only few number of adjustable parameters.

To derive *κ(Ω)*, we extend the use to a broader time dependent phenomena of the same approach used by Asen-Palmer et al[33] and later Morelli et al,[34] and we make use of the modified Debye-Callaway model to explicitly include both longitudinal and transverse phonon modes. Although this model might be not very rigorous, the treatment is to some extent justified by the reasonable agreement with experiment that has been obtained with it in the steady-state.[33,34] In this approach, the contributions of longitudinal and transverse acoustic branches are considered separately, furthermore, any conversion of normal modes between both branches (inter-transitions) is neglected; only transitions within the same acoustic branch (intra-transitions) are considered. This approach was first used by Holland[27] in his extension of Callaway model.[21]

We assume application of a temperature gradient along a direction $\vec{\imath}$, where $\vec{\imath}$ is a unit vector along a geometrical direction within the bulk SC crystal, this could be a principal crystal axis. Under the relaxation time approximation, the Callaway form of the BPTE for a phonon distribution function $n_S\left(x,\boldsymbol{q},t\right)\equiv n_{\boldsymbol{q},S}$ is given by:

$$\frac{\partial n_{\boldsymbol{q},S}}{\partial t}+\boldsymbol{V}_{\boldsymbol{q},S}.\boldsymbol{\nabla} n_{\boldsymbol{q},S}=-\frac{\partial n_{\boldsymbol{q},S}}{\partial t}\bigg|_{Coll}=\frac{n_{\boldsymbol{q},S}^{\boldsymbol{\lambda}_S}-n_{\boldsymbol{q},S}}{\tau_{q,S}^{N}}+\frac{n_{\boldsymbol{q},S}^{0}-n_{\boldsymbol{q},S}}{\tau_{q,S}^{R}} \qquad (1)$$

where $n_{\boldsymbol{q},S}^{0}=\left(e^{\frac{\hbar\omega_S(\boldsymbol{q})}{k_B T}}-1\right)^{-1}$ is the equilibrium phonon Planck distribution function to which resistive phonon scattering processes (all scattering processes that change the total phonon wave vector: Umklapp, boundary, defects, imperfections, etc.) tend to return the phonon system with a single relaxation time $\tau_R\left(q,S\right)$. $\omega_S\left(\boldsymbol{q}\right)$ is the dispersion relation of the phonon in state $(\boldsymbol{q}, S)$, $k_B$ and $T$ are the Boltzmann constant and the absolute local temperature, respectively. On the other hand, the distribution function which is stationary for N-processes (scattering processes that don't change the total phonon wave vector) is not $n_{\boldsymbol{q},S}^{0}$ but rather $n_{\boldsymbol{q},S}^{\boldsymbol{\lambda}_S}$. N-processes lead the phonon system to a displaced (*drifted*) Planck distribution function $n_{\boldsymbol{q},S}^{\boldsymbol{\lambda}_S}$ with a single relaxation time $\tau_N(q,S)$, where $\boldsymbol{\lambda}_S$ is a vector that has the dimension of a velocity times Planck constant $\hbar$ :[21,24]

$$n_{\boldsymbol{q},S}^{\boldsymbol{\lambda}_S} = \left[\exp\left(\frac{\hbar\omega_S(\boldsymbol{q}) - \boldsymbol{\lambda}_S.\boldsymbol{q}}{k_B T}\right) - 1\right]^{-1} \qquad (2)$$

$\boldsymbol{\lambda}_S/\hbar$ is called the drift velocity vector of the phonon ***(q**, S)* while $\boldsymbol{V}_{\boldsymbol{q},S} = v_{S,\iota}\boldsymbol{q}_\iota = \partial\omega_S(\boldsymbol{q})/\partial\boldsymbol{q}_\iota$ is its group velocity vector, which in general depends on the direction of $\boldsymbol{q}_\iota$. Here we assume the heat transport is in the same direction as the applied temperature gradient.

In this analysis, we assume all relaxation times describing the different intrinsic and extrinsic phonon scattering processes, to be independent of the time or frequency dependence of the applied temperature gradient.

As in Callaway analysis,[21,24] the vector $\boldsymbol{\lambda}_S$ is assumed to have a very small module. Then, to first order in $\boldsymbol{\lambda}_S$, the Taylor's series expansion of $n_{\boldsymbol{q},S}^{\boldsymbol{\lambda}_S}$ such that $\mathrm{O}(\boldsymbol{\lambda}_S^2)$ is neglected, gives:

$$\begin{aligned} n_{\boldsymbol{q},S}^{\boldsymbol{\lambda}_S} &\equiv n_{\boldsymbol{q},S}(\boldsymbol{\lambda}_S) \cong n_{\boldsymbol{q},S}(0) + \boldsymbol{\lambda}_S \cdot \left(\frac{\partial n_{\boldsymbol{q},S}(\boldsymbol{\lambda}_S)}{\partial \boldsymbol{\lambda}_S}\right)_{\boldsymbol{\lambda}_S=0} \\ &\cong n_{\boldsymbol{q},S}^0 + \frac{\boldsymbol{\lambda}_S.\boldsymbol{q}}{k_B T}\frac{\mathrm{e}^{\frac{\hbar\omega_{\boldsymbol{q},S}}{k_B T}}}{\left(\mathrm{e}^{\frac{\hbar\omega_{\boldsymbol{q},S}}{k_B T}} - 1\right)^2} = n_{\boldsymbol{q},S}^0 + \frac{(\boldsymbol{\lambda}_S.\boldsymbol{q})T}{\hbar\omega_{\boldsymbol{q},S}}\frac{dn_{\boldsymbol{q},S}^0}{dT} \end{aligned} \qquad (3)$$

In our analysis, we assume the bulk SC crystal to have a cubic symmetry and we treat it as a continuum elastic isotropic linear dispersionless medium, in which case and by symmetry consideration; $\boldsymbol{\lambda}_S$ must be a constant vector in the direction of the applied temperature gradient, so it is convenient to define still another parameter (Callaway parameter) $\beta_S$ that has the dimension of a relaxation time:[21,24]

$$\boldsymbol{\lambda}_S = -\hbar\beta_S v_{S,\iota}^2\left(\frac{\boldsymbol{\nabla}T}{T}\right) \qquad (4)$$

Since we are dealing with Debye-like phonon dispersion relation $\omega_S(\boldsymbol{q}) \equiv \omega_{\boldsymbol{q},S} = v_{S,\iota}|\boldsymbol{q}|$, so that one considers heat transport due only to acoustic phonons, we have $\boldsymbol{q} = \boldsymbol{V}_{\boldsymbol{q},S}\omega_{\boldsymbol{q},S}/v_{S,\iota}^2$ .

This implies:

$$\boldsymbol{\lambda}_S \boldsymbol{.q} = -\hbar \omega_{\boldsymbol{q},S} \beta_S \boldsymbol{V}_{\boldsymbol{q},S} \cdot \left[ \frac{\nabla T}{T} \right] \quad (5)$$

To solve Eq. (1), we continue to use two more approximations that are usually made in the treatment of the steady-state case of BPTE; (i) The distribution function $n_{\boldsymbol{q},S}$ depends on the position only through the temperature $T(x)$: $\nabla n_{\boldsymbol{q},S} = \frac{dn_{\boldsymbol{q},S}}{dT} \nabla T$, and (ii) it is assumed that deviation from equilibrium is small, i.e., $\frac{dn_{\boldsymbol{q},S}}{dT} \cong \frac{dn^0_{\boldsymbol{q},S}}{dT_0}$ where $T_0$ is the absolute local equilibrium temperature.

When the expression $\boldsymbol{\lambda}_S \boldsymbol{.q}$ is substituted into Eq. (3) then in Eq. (1), and based on the above two approximations, Eq. (1) takes the form:

$$\frac{\partial n_{\boldsymbol{q},S}}{\partial t} + \boldsymbol{V}_{\boldsymbol{q},S} \cdot \frac{dn^0_{\boldsymbol{q},S}}{dT_0} \nabla T = \frac{n^0_{\boldsymbol{q},S} - n_{\boldsymbol{q},S}}{\tau^C_{q,S}} - \frac{\beta_S}{\tau^N_{q,S}} \boldsymbol{V}_{\boldsymbol{q},S} \cdot \frac{dn^0_{\boldsymbol{q},S}}{dT_0} \nabla T$$

$$\Rightarrow \tau^C_{q,S} \frac{\partial n_{\boldsymbol{q},S}}{\partial t} + n_{\boldsymbol{q},S} = n^0_{\boldsymbol{q},S} - \tau^C_{q,S} \left[ 1 + \frac{\beta_S}{\tau^N_{q,S}} \right] \boldsymbol{V}_{\boldsymbol{q},S} \cdot \frac{dn^0_{\boldsymbol{q},S}}{dT_0} \nabla T = n^0_{\boldsymbol{q},S} - \tau^{eff}_{q,S} \boldsymbol{V}_{\boldsymbol{q},S} \cdot \frac{dn^0_{\boldsymbol{q},S}}{dT_0} \nabla T \quad (6)$$

where $\tau^C_{q,S}$ and $\tau^{eff}_{q,S}$ are respectively, the "*combined*" and the "*effective total*" relaxation times given respectively by:[21,24]

$$\frac{1}{\tau^C_{q,S}} = \frac{1}{\tau^N_{q,S}} + \frac{1}{\tau^R_{q,S}} \text{ and } \tau^{eff}_{q,S} = \tau^C_{q,S} \left[ 1 + \frac{\beta_S}{\tau^N_{q,S}} \right] \quad (7)$$

In Eq. (6), the effect of N-processes is contained in the effective total relaxation time $\tau^{eff}_{q,S}$ which is a complicated quantity, depending on $\tau^N_{q,S}, \tau^R_{q,S}$ and $\beta_S$. This complication is necessary because of the behavior of N-processes which shuffle crystal momentum back and forth between normal modes, and then contribute implicitly to the lattice thermal conductivity of a given SC crystal material.[21,24]

As for all relaxation times considered in this study, and taking into account the approximations made above, we further assume that Callaway pseudo-relaxation time $\beta_S$ is a constant independent of the time or frequency dependence of the applied temperature gradient, and thereby can be calculated similarly to the steady-state case.[21,24] This means that the dependence of the phonon gas[58] drift on time and space is contained in the expression of drift velocity $\boldsymbol{\lambda}_S / \hbar$ only through the applied dynamical temperature gradient $\nabla T(x,t)$.

For each acoustic phonon polarization branch, Callaway pseudo-relaxation time $\beta_S$ is determined by recalling that N-processes cannot change the total phonon wave vector (total crystal momentum). In the steady-state case $\left(\partial n_{\boldsymbol{q},S} / \partial t = 0\right)$, $\beta_S$ can be calculated following the same procedure of calculation as first performed by Callaway[21] and Carruthers.[24] We find (the detail of the calculation is outlined in appendix A):

$$\beta_S = \frac{\int_0^{\theta_D^S/T_0} \frac{\tau_S^C(x)}{\tau_S^N(x)} D(x)\,dx}{\int_0^{\theta_D^S/T_0} \frac{1}{\tau_S^N(x)}\left[1-\frac{\tau_S^C(x)}{\tau_S^N(x)}\right] D(x)\,dx} = \frac{\int_0^{\theta_D^S/T_0} \frac{\tau_S^C(x)}{\tau_S^N(x)} D(x)\,dx}{\int_0^{\theta_D^S/T_0} \frac{\tau_S^C(x)}{\tau_S^N(x)\tau_S^R(x)} D(x)\,dx} \qquad (8)$$

where $x = \hbar\omega / k_B T_0$, $D(x) = x^4 e^x / \left(e^x - 1\right)^2$ is Debye function and $\theta_D^S$ is Debye temperature of the acoustic polarization branch S.[59]

### B. Dynamical lattice thermal conductivity

In order to solve Eq. (6), we apply Fourier transform with respect to time to both sides. One obtains:

$$\left(1 - j\Omega\tau_{q,S}^C\right)\overline{n_{\boldsymbol{q},S}} = \overline{n_{\boldsymbol{q},S}^0} - \tau_{q,S}^{eff}\boldsymbol{V}_{\boldsymbol{q},S} \cdot \frac{dn_{\boldsymbol{q},S}^0}{dT_0}\overline{\boldsymbol{\nabla} T}$$

$$\Rightarrow \overline{n_{\boldsymbol{q},S}}(x,\Omega) = \frac{1}{1 - j\Omega\tau_{q,S}^C}\overline{n_{\boldsymbol{q},S}^0} - \frac{\tau_{q,S}^{eff}\boldsymbol{V}_{\boldsymbol{q},S} \cdot \frac{dn_{\boldsymbol{q},S}^0}{dT_0}}{1 - j\Omega\tau_{q,S}^C}\overline{\boldsymbol{\nabla} T}(x,\Omega) \qquad (9)$$

where the top bars over $n_{\boldsymbol{q},S}$, $n_{\boldsymbol{q},S}^0$ and $\boldsymbol{\nabla} T$ indicate Fourier transforms and $j$ is the complex operator $\left(j^2 = -1\right)$.

Once we know the distribution function in Fourier (frequency) domain, the next step is the calculation of the heat flux density current $\overline{\boldsymbol{J}_Q}$ in the same domain along the direction of the applied temperature gradient. $\overline{\boldsymbol{J}_Q}$ is defined as:

$$\overline{\boldsymbol{J}_Q}(x,\Omega) = \frac{1}{W}\sum_{q,S}\hbar\omega_S(\boldsymbol{q})\overline{n_{\boldsymbol{q},S}}(x,\Omega)\boldsymbol{V}_{\boldsymbol{q},S} = -\frac{1}{W}\sum_{q,S}\hbar\omega_{q,S}v_{S,\imath}^2\frac{\tau_{q,S}^{eff}\frac{dn_{\boldsymbol{q},S}^0}{dT_0}}{1 - j\Omega\tau_{q,S}^C}\overline{\boldsymbol{\nabla} T}(x,\Omega) \qquad (10)$$

where we use $W$ to denote the volume of the bulk SC crystal. We should note here that the contribution of the first term in Eq. (10) to $\overline{\boldsymbol{J}_Q}$ vanishes since the phonon equilibrium distribution $n^0_{\boldsymbol{q},S}$ can give no contribution to any energy (heat) transport.[17,22,24,32] The latter is an isotropic function in the wave vector $\boldsymbol{q}$ space while the velocity $\boldsymbol{V_q}$ is an algebraic function; the dispersion relation and the relaxation times depend on the module of the wave vector $\boldsymbol{q}$ and as such are even functions of $q$.

The density of states in the $\boldsymbol{q}$ space is very great; we can use the standard relation to replace the sum sign by an integral sign ( $\sum_{q,S} \rightarrow \frac{W}{8\pi^3} \sum_S \int d^3\boldsymbol{q}$ ):

$$\overline{\boldsymbol{J}_Q}(x,\Omega) = -\left[\frac{1}{8\pi^3}\sum_S \int \frac{\tau^{eff}_{q,S}}{1 - j\Omega\tau^C_{q,S}} v^2_{S,t} C_{Ph}(\boldsymbol{q},S)\, d^3\boldsymbol{q}\right] \overline{\nabla T}(x,\Omega) = -\kappa(\Omega)\overline{\nabla T}(x,\Omega) \qquad (11)$$

In Eq. (11), $C_{Ph}$ represents the phonon specific heat or heat capacity per normal mode:

$$C_{Ph}(\boldsymbol{q},S) = C_{Ph}(\omega_{\boldsymbol{q},S}, T_0) = \hbar\omega_{\boldsymbol{q},S} \frac{dn^0_{\boldsymbol{q},S}}{dT_0} \qquad (12)$$

By considering the dynamical temperature gradient and the heat flux density current Fourier-analyzed as in the customary way, we recognize from Eq. (11) the dynamical lattice thermal conductivity $\kappa(\Omega)$ which takes the expression:

$$\kappa(\Omega) = \frac{1}{8\pi^3}\sum_S \int \frac{\tau^{eff}_{q,S}}{1 - j\Omega\tau^C_{q,S}} v^2_{S,t} C_{Ph}(\boldsymbol{q},S)\, d^3\boldsymbol{q} \qquad (13)$$

Note that $\Omega$ represents the circular frequency which is related to the real frequency by the standard definition $\Omega = 2\pi f$ .

For simplification and further discussion (see the discussion section) we set:

$$\kappa^0_{q,S} = \frac{1}{8\pi^3} \tau^{eff}_{q,S} v^2_{S,t} C_{Ph}(\boldsymbol{q},S) \qquad (14)$$

$\kappa(\Omega)$ takes then the more compact form:

$$\kappa(\Omega) = \sum_S \int \frac{\kappa^0_{q,S}}{1 - j\Omega\tau^C_{q,S}} d^3\boldsymbol{q} = \kappa_r(\Omega) + j\kappa_i(\Omega) \qquad (15)$$

where $\kappa_r$ and $\kappa_i$ are, respectively, the real and the imaginary parts of the dynamical lattice thermal conductivity $\kappa(\Omega)$:

$$\begin{cases} \kappa_r(\Omega) = \sum_S \int \frac{\kappa^0_{q,S}}{1+\left(\Omega\tau^C_{q,S}\right)^2} d^3\boldsymbol{q} \\ \kappa_i(\Omega) = \sum_S \int \kappa^0_{q,S} \frac{\Omega\tau^C_{q,S}}{1+\left(\Omega\tau^C_{q,S}\right)^2} d^3\boldsymbol{q} \end{cases} \quad (16)$$

To simplify more the expression of $\kappa(\Omega)$, we express it, as it is customary in the modified Debye-Callaway model, as the sum over one longitudinal ($\kappa_L$) and two degenerate transverse ($\kappa_T$) phonon polarization branches:[34]

$$\kappa(\Omega) = \kappa_L(\Omega) + 2\kappa_T(\Omega) \quad (17)$$

By using the isotropy of the group velocity in the real and reciprocal spaces $v^2_{S,q_x} = v^2_{S,q_y} = v^2_{S,q_z} = \frac{1}{3}v^2_S$ and the usual change of variable $x = \hbar\omega / k_B T_0$, it is straightforward to show that the real part $\kappa_r(\Omega)$ takes the form:

$$\begin{cases} \kappa_r(\Omega) = \kappa^r_L(\Omega) + 2\kappa^r_T(\Omega) \\ \kappa^r_S(\Omega) = \kappa^r_{S1}(\Omega) + \kappa^r_{S2}(\Omega) \\ \kappa^r_{S1}(\Omega) = \frac{1}{3} C_S T_0^3 \int_0^{\theta^S_D/T_0} \frac{\tau^C_S(x)}{1+\left[\Omega\tau^C_S(x)\right]^2} D(x)\,dx \\ \kappa^r_{S2}(\Omega) = \frac{1}{3} C_S T_0^3 \beta_S \int_0^{\theta^S_D/T_0} \frac{\frac{\tau^C_S(x)}{\tau^N_S(x)}}{1+\left[\Omega\tau^C_S(x)\right]^2} D(x)\,dx \end{cases} \quad (18)$$

and similarly for the imaginary part $\kappa_i(\Omega)$:

$$\begin{cases} \kappa_i(\Omega) = \kappa^i_L(\Omega) + 2\kappa^i_T(\Omega) \\ \kappa^i_S(\Omega) = \kappa^i_{S1}(\Omega) + \kappa^i_{S2}(\Omega) \\ \kappa^i_{S1}(\Omega) = \frac{1}{3} C_S T_0^3 \int_0^{\theta^S_D/T_0} \frac{\Omega\left[\tau^C_S(x)\right]^2}{1+\left[\Omega\tau^C_S(x)\right]^2} D(x)\,dx \\ \kappa^i_{S2}(\Omega) = \frac{1}{3} C_S T_0^3 \beta_S \int_0^{\theta^S_D/T_0} \frac{\Omega\frac{\left[\tau^C_S(x)\right]^2}{\tau^N_S(x)}}{1+\left[\Omega\tau^C_S(x)\right]^2} D(x)\,dx \end{cases} \quad (19)$$

where $C_S = k_B^4 / \left(2\pi^2 \hbar^3 v_S\right)$, ($S$=$L$, $T$) and the partial conductivities $\kappa_{S1}$ and $\kappa_{S2}$ are the usual Debye-Callaway terms.[21,24] We carry the detail of the derivation of these equations in appendix B.

### C. Phonon scattering processes and their relaxation times

In SC crystals, phonons scattering processes can be divided into *intrinsic* processes arising from the anharmonicity of the interatomic forces, and *extrinsic* processes due to phonons scattering at the boundaries of the crystal and at various sorts of crystal defects and imperfections (e.g., point defects, impurities, dislocations, alloy disorder, grain boundaries, embedded nanoparticles, etc.) As first pointed out by Peierls,[17] anharmonic phonon scattering processes are of two distinct types, Normal scattering processes (N-processes) which conserve the total crystal momentum after a collision, and Umklapp scattering processes (U-processes) for which the total crystal momentum changes by a reciprocal lattice vector after a collision. On the other hand, all extrinsic scattering processes don't conserve the total crystal momentum after a collision. Because of their conservative character of the total crystal momentum, N-processes cannot by themselves lead to a finite thermal conductivity. Consequently, as pointed out by Callaway,[21] it cannot be legitimate just to add scattering rates for N-processes to those which don't conserve the crystal momentum (U-processes and all extrinsic processes). The latter processes are called resistive scattering processes because at least one of them is needed to obtain a finite thermal conductivity. The effect of N-processes is addressed with a particular attention through the use of the displaced (*drifted*) Planck distribution as we have seen in Eq. (1).

In the single relaxation time approximation, as we have presented it in the above section, each scattering process is characterized by a relaxation time which naturally is function of the phonon wave vector $q$ and polarization $S$. It depends also on the nature of the scattering mechanism through coefficients characteristic of this mechanism. We, generally, express the relaxations times as functions of the phonon intrinsic frequency instead of the wave vector.[24] Depending on the nature of the scattering mechanism, relaxation times have different expressions. In our present analysis, we will list and limit our discussion to four different scattering mechanisms, a phonon can undergo in a SC crystal, and we will give forms of their corresponding relaxation times according to the approach of Morelli et al,[34] in which every phonon scattering mechanism depends explicitly on the phonon mode. Moreover, we assume these forms to be the same for all studied SC crystals. Phonon scattering processes that are considered in our study are: *(i)* N-processes, *(ii)* U-processes, *(iii)* scattering of phonons by imperfections and *(iv)* boundary scattering.

- **Phonon-phonon Normal scattering processes**

In the early 1950, Herring[19] has established that the scattering rate of a Normal three-phonon process depends on the crystallographic class and symmetry group of the crystal as well as on phonon polarization branch and operating temperature:

$$\left[\tau_S^N(q)\right]^{-1} = q^{\eta} T^{5-\eta} \qquad (20)$$

where $\eta$ is an exponent determined by the crystal symmetry. This relation is usually expressed using the phonon intrinsic frequency. According to Morelli et al,[34] the appropriate forms for longitudinal and transverse acoustic phonons branches are given as:

$$\begin{cases} \left[\tau_S^N(\omega)\right]^{-1} = B_N^S \omega^a T^b \\ B_N^S(a,b) = \left(\dfrac{k_B}{\hbar}\right)^b \dfrac{\hbar \gamma_S^2 V^{\frac{a+b-2}{3}}}{M v_S^{a+b}} \end{cases} \qquad (21)$$

where $\gamma_S$, $M$, and $V$ are the Grüneisen parameter for the phonon acoustic polarization branch $S$, the atomic mass, and the volume per atom, respectively. Depending on the temperature range and the crystal class, different forms of the normal scattering rate have been employed in literature to fit the experimental data of the steady-state thermal conductivity.[21,23-30,33,34,53,60,61] In our case of study, we keep using the forms suggested by Morelli et al.[34] These ones are given, respectively for the longitudinal and transverse polarizations as:

$$\begin{cases} \left[\tau_L^N(\omega)\right]^{-1} = B_N^L \omega^2 T^3 \\ B_N^L(2,3) = \dfrac{k_B^3 \gamma_L^2 V}{\hbar^2 M v_L^5} \end{cases} \quad \text{and} \quad \begin{cases} \left[\tau_T^N(\omega)\right]^{-1} = B_N^T \omega T^4 \\ B_N^T(1,4) = \dfrac{k_B^4 \gamma_T^2 V}{\hbar^3 M v_T^5} \end{cases} \qquad (22)$$

- **Phonon-phonon Umklapp scattering processes**

As for N-processes, U-processes scattering rate depends also on the crystallographic class and symmetry group of the SC crystal. For a three-phonon U-process, the scattering rate is generally given as:[22,24,32]

$$\left[\tau_S^U(\omega)\right]^{-1} = B_U^S \omega^a \left(\frac{T}{\theta_D^S}\right)^b e^{-\frac{\theta_D^S}{cT}} \qquad (23)$$

where $a$, $b$, and $c$ are adjustable positive exponents, and $\theta_D^S$ is Debye temperature of the acoustic phonon polarization branch $S$. The fact that U-processes should give rise to an exponential die-out of the steady state thermal conductivity at high temperature was first proposed by Peierls in the late 1920. He suggested the form $\kappa \sim T^n e^{\frac{\theta_D}{mT}}$ with $n$ and $m$ on the order of unity.[17] On the basis of the model of Leibfried and Schlöman[34] and comparing the thermal conductivity of several pure crystals, Slack and Galginaitis suggested the following form for the three-phonon U-processes scattering rate:[28,34]

$$\begin{cases} \left[\tau_S^U(\omega)\right]^{-1} = B_U^S \omega^2 \mathrm{T} \exp\left(-\theta_D^S / 3\mathrm{T}\right) \\ B_U^S = \dfrac{\hbar \gamma_S^2}{M v_S^2 \theta_D^S} \end{cases} \qquad (24)$$

Several other expressions using other values of $a$, $b$ and $c$ for the U-processes scattering rate have appeared in the literature. [21,23-30,33,34,53,60,61] As discussed by Morelli et al,[34] the values of $a$, $b$ and $c$ are sensitive to other parameters used in the fit, especially the values of Debye temperature and Grüneisen parameter. In our analysis here, we choose to work with the form given by Eq. (24).

- **<u>Scattering of phonons by lattice imperfections</u>**

Here, we assume scattering of phonons by natural isotopes in pure single SC crystals and by alloy disorder in SC crystal alloys. In both cases, the relaxation time is calculated assuming Rayleigh scattering regime valid, and the expression of the scattering rate as derived by Klemens is given by:[20]

$$\left[\tau_S^i(\omega)\right]^{-1} = \frac{V\Gamma}{4\pi v_S^3}\omega^4 \qquad (25)$$

where $V$ is the volume per atom, and $\Gamma$ denotes the phonon scattering parameter that takes into account contributions from mass differences, atomic size differences and bond strength differences between the impurity (imperfection) and the host lattice atom. Since we are treating the SC crystal as an elastic isotropic continuum medium, no much additional errors, would be introduced by neglecting the contribution of the differences in the atomic size and bond strength and considering only mass-difference contribution. In that case $\Gamma$ will represent the mass-difference fluctuation phonon scattering parameter.

The alloy is assumed to be a random mixture of atoms with different masses and volumes arranged in a lattice. In this case of alloy disorder scattering, $\Gamma$ represents the disorder parameter and is calculated using the virtual lattice approach of Abeles.[25] According to this approximation, $\Gamma$ of a mixture of two atoms $A$ and $B$ is given by:[25]

$$\begin{cases} \Gamma = x(1-x)\left(\dfrac{\Delta M}{M}\right)^2 \\ \Delta M = M_A - M_B \\ M = xM_A + (1-x) M_B \end{cases} \qquad (26)$$

From Eq. (26), we can easily check that the alloy disorder scattering parameter becomes zero for pure material $A$ or material $B$.

- **Scattering of phonons by boundaries**

The last scattering mechanism we consider in the analysis conducted here, is the scattering of phonons due to the boundaries of the crystal. For each acoustic phonon mode polarization branch $S$, the scattering rate of this process is assumed to be independent of temperature and phonon dispersion and is simply given by:

$$\left[\tau_S^B(\omega)\right]^{-1} = \frac{v_S}{L_C} \qquad (27)$$

where $L_C$ is a characteristic length of the crystal in the direction of the phonon transport. For all SC crystals considered in our study, we take $L_C$ to be constant, $L_C=5mm$. The value of $L_C$ is assumed to be long enough for the SC crystals to be treated as bulk materials.[34]

The inverse of the total resistive relaxation time $\tau_S^R$ accounting for all phonon scattering processes that destroy the total crystal momentum is given according to Mathiessen's rule:

$$\left[\tau_S^R(\omega)\right]^{-1} = \left[\tau_S^U(\omega)\right]^{-1} + \left[\tau_S^i(\omega)\right]^{-1} + \left[\tau_S^B(\omega)\right]^{-1} \qquad (28)$$

# III. RESULTS AND DISCUSSION

## 1. Behavior of the dynamical lattice thermal conductivity

The physical picture we are interested to in the current analysis is related to the behavior of the phonon gas in a region of the bulk SC crystal subject to a dynamical temperature gradient or a time or frequency dependent temperature disturbance resulting from the application of an external source.

In the theory section, we made the assumption that the Callaway pseudo-relaxation time $\beta_S$, describing the effect of N-processes, does not depend on time and that this approximation should preserve the essential features of the dynamical thermal conduction by phonons especially at temperatures above the maximum in the steady-state thermal conductivity ($T \approx 0.05\theta_D$). This assumption is plausible if one takes into consideration the smallness of $\lambda_S$, but might be questionable at low temperatures and can eventually be relaxed to explore the effect of possible time dependence of $\beta_S$. We should note however that, in their investigation of the conditions of manifestation of the second sound in solid dielectrics, Guyer and Krumhansl[45] gave a thorough discussion based on solving BPTE in the time domain where $\beta_S$ was taken an explicitly time dependent function. The authors found that the dispersion relation of the second sound in solids obtained in both cases, with and without time dependent $\beta_S$, continues to exist with similar damping terms. This constitutes a robust argument to neglect the time dependence of $\beta_S$ in our analysis. It is worthwhile to mention that in the simplest case of the grey spectrum approximation (GSA) when all phonon modes belonging to different polarization branches have the same relaxation times for each scattering process independent of the wave vector $q$, Eq. (8) gives $\beta_S = \tau_S^R$. This shows the very fundamental intertwining between anharmonic N-processes and resistive processes; the implicit effect of N-processes in the onset of a noninfinite thermal conductivity is taken account of through the resisting causing collisions namely the relaxation time of the resistive processes which effect is explicit.

The expression of the dynamical lattice thermal conductivity $\kappa(\Omega)$ as given by Eq. (15), shows that $\kappa(\Omega)$ is an analytical function on the upper frequency complex plane. As a matter of fact, starting from the expressions of the real and imaginary parts $\kappa_r(\Omega)$ and $\kappa_i(\Omega)$ as given by Eq. (16), it is straightforward to show that theses expressions are Hilbert transforms of each other (see appendix C for detail of the proof), the Kramers-Kronig relations are then verified and as such the causality requirement is fulfilled where the dynamical temperature gradient is the driving potential force (*cause*) and the heat flux density current is the thermodynamically corresponding conjugate (*effect*).

When using the GSA, the effective total relaxation time $\tau_{q,S}^{eff}$ [Eq. (7)] reduces to the resistive wave vector independent relaxation time $\tau_{q,S}^{eff} = \tau_S^R$ and the expression of $\kappa(\Omega)$ becomes:

$$\kappa(\Omega) = \frac{\kappa_0}{1 - j\Omega\tau_C} \qquad (29)$$

where $\kappa_0$ is the steady-state thermal conductivity and $\tau_C$ is an effective wave vector independent combined relaxation time. Eq. (29) is Cattaneo's expression of $\kappa(\Omega)$ which one can derive starting from BPTE in the GSA and solving directly in the frequency domain the moment equation giving the heat flux density current.

A very remarkable and interesting result that follows from Eq. (11) is obtained by going back to the time domain and performing an inverse Fourier transform; the latter transforms the natural product into a convolution product. The heat flux density current can be written in a convolution form in the real space-time domain as:

$$\boldsymbol{J}_Q(x,t) = -\int_{-\infty}^{+\infty} K(t-t')\nabla T(x,t')\,dt' = -K \otimes \nabla T(x,t) \qquad (30)$$

where" $\otimes$ "represents the convolution product. Eq. (30) says simply that the response at time $t$ ($\boldsymbol{J}_Q$) is related to the previously applied driving potential force ($\nabla T$) as is required in all natural processes. The actual form of $\boldsymbol{J}_Q$ in Eq. (30) is similar to the form derived by Gurtin and Pipkin in their theory of heat conduction in solids in the linear regime,[62] where $K(t)$ is the heat flux relaxation function or *heat flux kernel* that takes the form:

$$K(t) = \int_0^{\infty} \kappa(\Omega) e^{-j\Omega t} d\Omega = \sum_S \int \frac{1}{\tau_{q,S}^C} e^{-\frac{t}{\tau_{q,S}^C}} \kappa_{q,S}^0 d^3\boldsymbol{q} \qquad (31)$$

In order to discuss the behavior of the dynamical lattice thermal conductivity $\kappa(\Omega)$ of bulk SC crystals, as a function of ambient temperature as well as different intrinsic and extrinsic parameters, we consider *5* different SC materials in our analysis; (i) natural Si, (ii) natural Ge, (iii) $Si_{0.7}Ge_{0.3}$ alloy, (iv) $In_{0.53}Ga_{0.47}As$ alloy and (v) $In_{0.49}Ga_{0.51}P$ alloy. The choice of these materials is based on their relevance and importance in microelectronic and optoelectronic industry especially in high frequency devices.[63,64] Besides $Si_{0.7}Ge_{0.3}$ alloy is known to be one of the best SC materials suited for thermoelectric energy conversion and more importantly in the thermoelectric generation process especially at high temperatures.[3,4,65] Tables I and II recapitulate, respectively, the different geometrical and physical properties of the SC crystals used in our calculations. We assume all physical properties of the SC crystal materials to be independent of temperature. As we could not find documented values of the Debye temperatures of $In_{0.53}Ga_{0.47}As$ and $In_{0.49}Ga_{0.51}P$ alloys in literature, we calculated them for both longitudinal and transverse acoustic phonon polarization branches assuming the Debye cut-off frequency for each phonon mode to be given by $\omega_D^S = v_S q_D^S = {v_S \pi}/{a}$ where *a* is the SC crystal lattice constant (see Table II).

When they are considered as fixed values, the Grüneisen parameters for longitudinal and transverse phonon acoustic polarization branches are taken to be the same for all single crystals and alloys; $\gamma_L=1$ and $\gamma_T=0.7$.[53]

**Table I :** Geometrical properties of the 5 bulk semiconductor crystal materials used in the simulation of the steady-state and dynamical behaviors of the lattice thermal conductivity as a function of temperature and frequency of modulation of the introduced heat source.

| Material | Lattice constant *a* (A) | Atomic mass $M_a$ (kg) $\times 10^{-26}$ | Volume per atom V ($m^3$) $\times 10^{-29}$ | Density ($kg/m^3$) |
|---|---|---|---|---|
| Si | 5.431 | 4.66 | 2 | 2329 |
| Ge | 5.658 | 12 | 2.27 | 5332 |
| $Si_{0.7}Ge_{0.3}$ | 5.493 | 6.9 | 2.07 | 3332* |
| $In_{0.53}Ga_{0.47}As$ | 5.868 | 13.8 | 2.52 | 5500 |
| $In_{0.49}Ga_{0.51}P$ | 5.653 | 10 | 2.24 | 4470 |

*: Calculated based on the properties of individual elements at *T=300K* considering the (100) direction.[66]

**Table II :** Physical properties of the 5 bulk semiconductor crystal materials used in the simulation of the steady-state and dynamical behaviors of the lattice thermal conductivity as a function of temperature and frequency of modulation of the introduced heat source.

| Material | Longitudinal sound velocity $v_L$ (m/s) | Transverse sound velocity $v_T$ (m/s) | Longitudinal Debye temperature $\theta_D^L$ (K) | Transverse Debye temperature $\theta_D^T$ (K) | Longitudinal Grüneisen parameter $\gamma_L$ | Transverse Grüneisen parameter $\gamma_T$ | Rayleigh mass-difference fluctuation phonon scattering parameter $\Gamma$ |
|---|---|---|---|---|---|---|---|
| Si | 8430[a] | 5840[a] | 586 [a] | 240 [a] | 1[b] | 0.7[b] | $2\times10^{-4}$[a] |
| Ge | 4920[a] | 3540[a] | 333 [a] | 150 [a] | 1 [b] | 0.7 [b] | $6.08\times10^{-4}$[a] |
| $Si_{0.7}Ge_{0.3}$ | 6812[c] | 4769[c] | 510[d] | 213[d] | 1 [b] | 0.7 [b] | 0.2403[f] |
| $In_{0.53}Ga_{0.47}As$ | 4267[c] | 2984[c] | 175[e] | 122[e] | 1 [b] | 0.7 [b] | 0.0357[b] |
| $In_{0.49}Ga_{0.51}P$ | 5208[c] | 3609[c] | 221[e] | 153[e] | 1 [b] | 0.7 [b] | 0.0675[b] |

[a]: Ref 34.
[b]: Ref 53.
[c]: Calculated based on the properties of individual elements at $T=300K$ considering the (100) direction.[66]
[d]: Calculated using the weighted average approach from Ref 16.
[e]: Calculated from the sound velocities assuming the Debye cut-off wave vector $=\pi/a$ where $a$ is the lattice constant: $\theta_D^S = \hbar v_S q_D^S / k_B$ with $q_D^S = \pi/a \Rightarrow \theta_D^S = h v_S / 2 k_B a$ .
[f]: Calculated using the virtual medium approach from Ref 25.

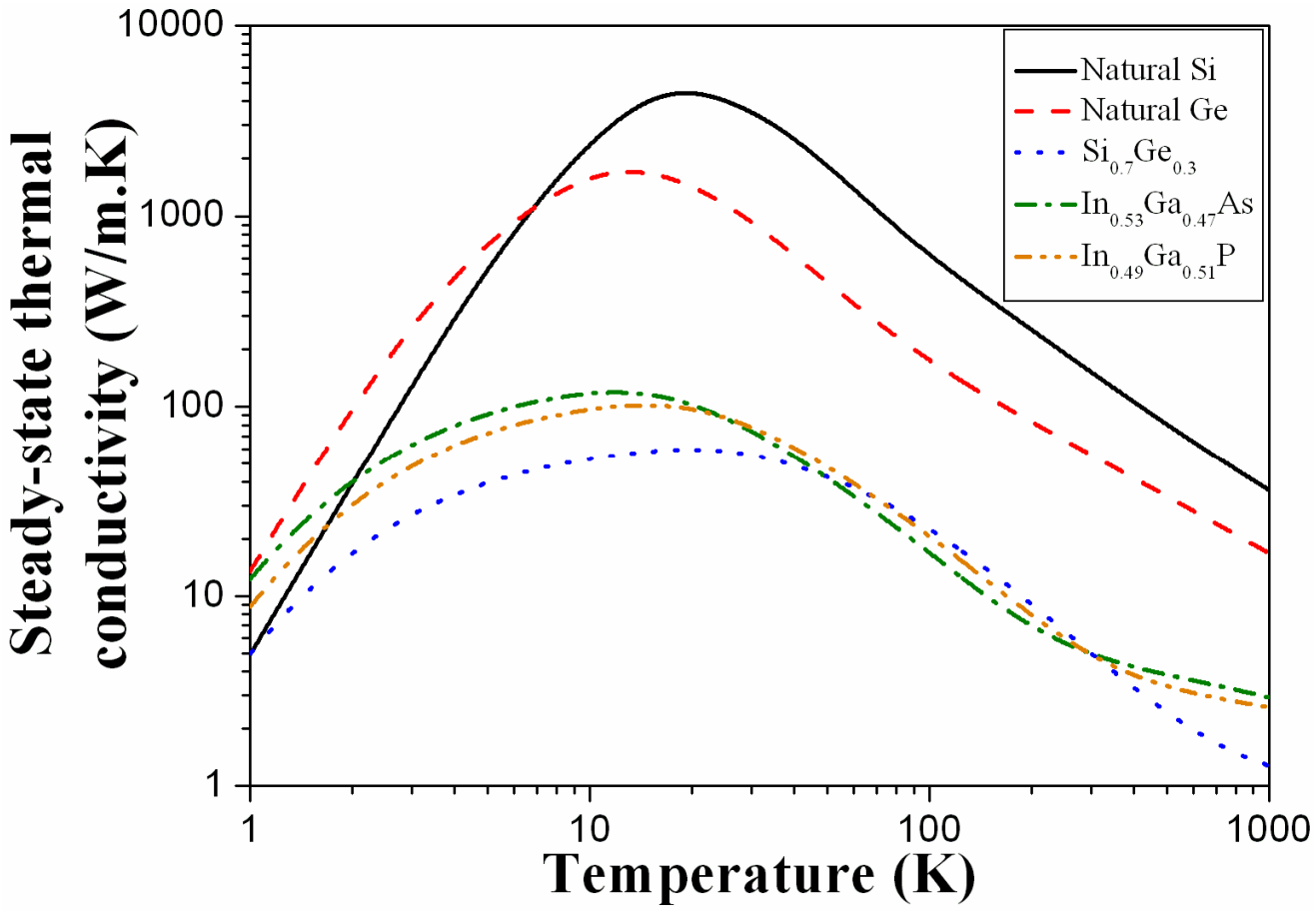

**Figure 1 :** Computed behavior of the steady-state lattice thermal conductivity $\kappa(0)$ of the 5 different bulk SC crystals as a function of temperature.

Figure 1 shows the calculated behavior of the steady-state lattice thermal conductivity $\kappa(0)$ of the 5 different bulk SC crystals as a function of temperature. A typical bell-shape behavior is reproduced, in which $\kappa(0)$ follows an almost $T^3$ power law behavior at low temperatures, reaches a maximum then starts to fall off at high temperatures due mainly to anharmonic phonon-phonon scattering processes.[21-34] The peak value of $\kappa(0)$ of each SC material is found to be achieved for a temperature of about *($T \approx 0.06\theta_D$)* in agreement with the aforementioned estimation in the introduction, where $\theta_D$ is an average Debye temperature over longitudinal and transverse phonon acoustic polarization branches $\theta_D = \left(\theta_D^L + 2\theta_D^T\right)/3$. $Si_{0.7}Ge_{0.3}$ alloy shows the lowest peak value of $\kappa(0)$. The calculated $T$-behavior of $\kappa(0)$ is in a very good agreement with reported experimental data for all 5 bulk SC crystals; Si and Ge,[28,29,33,34] $Si_{0.7}Ge_{0.3}$, $In_{0.53}Ga_{0.47}As$ and $In_{0.49}Ga_{0.51}P$.[7,53]

In Figs 2(a) and 2(b), we report respectively, the calculated behaviors of the frequency evolution of the real part and amplitude of the dynamical lattice thermal conductivity $\kappa(\Omega)$ for the *5* different bulk SC materials at room temperature, over a frequency interval [*0.1Hz-160THz*]. The insets in Figs 2(a) and 2(b) illustrate the behaviors of the imaginary part and phase of $\kappa(\Omega)$, respectively. As expected, the amplitude of $\kappa(\Omega)$ shows a plateau in the low frequency regime and then starts to fall off rapidly as the frequency gets higher; a typical first order low-pass filter thermal behavior. The phonon gas can't follow the thermal perturbation when the frequency of the latter becomes very high; the SC crystal becomes a thermal insulator. We can see also that the beginning of the falling off occurs at different threshold frequencies depending on the SC crystal; the SC alloy crystals seem to be characterized by lower threshold frequencies than the single SC crystals. The inset in Fig 2(a) shows the frequency behavior of the imaginary part $\kappa_i(\Omega)$. The latter manifests a Lorentzian-like shape behavior describing a resonance phenomenon of the phonon gas in the SC crystal at a resonance frequency $f_R = 1/2\pi\tau_m$, where $\tau_m$ is a certain weighted average relaxation time over all phonon scattering mechanisms and polarizations. SC alloy crystals are characterized by lower resonance amplitudes than single SC crystals. On the other hand, the inset in Fig 2(b) shows the phase of $\kappa(\Omega)$ increasing as a function of frequency to saturate at a value of $\pi/2$ in the high frequency regime. This is also a typical behavior of the phase that describes the delay between the cause (dynamical temperature gradient) and the effect (heat flux density current) in a linear system.

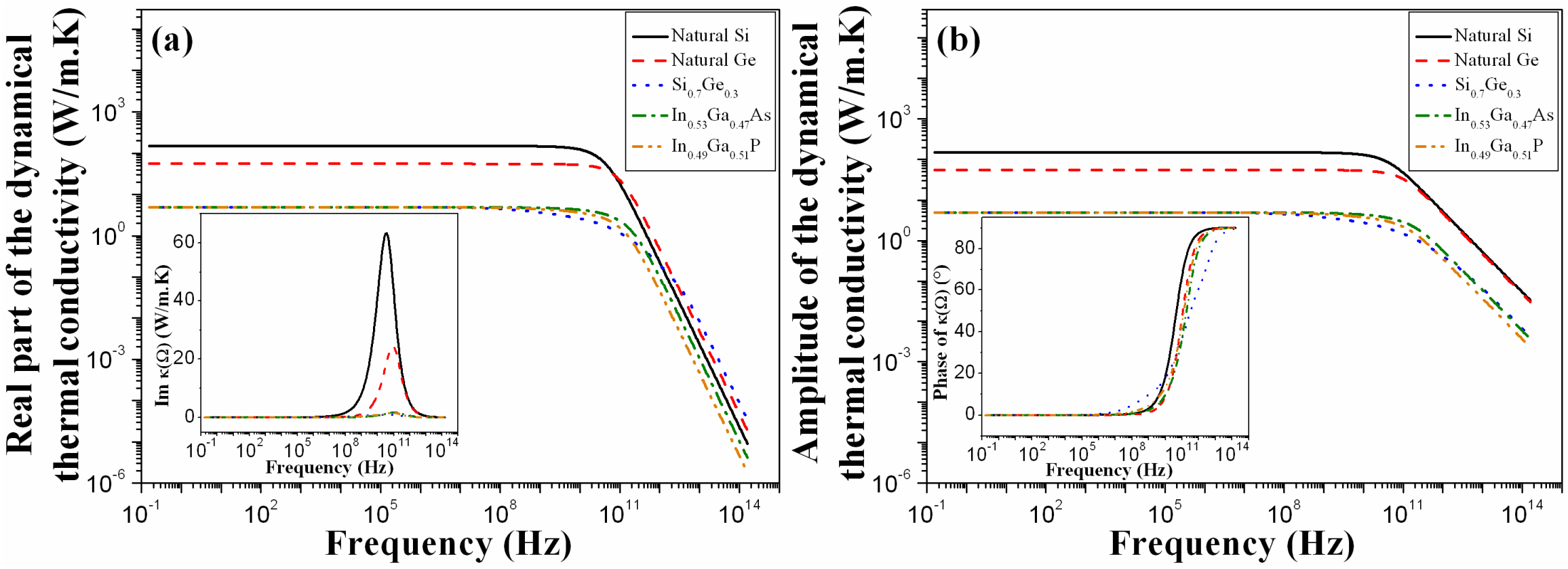

**Figure 2 :** Computed behavior of the dynamical lattice thermal conductivity *κ(Ω)* of the 5 different bulk SC crystals at room temperature as a function of frequency. (a) real part (imaginary part in the inset) and (b) amplitude (phase in the inset).

Because of the important role, $Si_{0.7}Ge_{0.3}$ SC alloy plays in thermoelectricity as well as in microelectronics, we consider this SC crystal as a test bulk material to study the effect of changing temperature as well as different intrinsic and extrinsic parameters on the behavior of *κ(Ω)*. The first parameter to consider is temperature. We report, respectively, in Figs 3(a) and 3(b) the calculated dynamical behaviors of the real part and amplitude of *κ(Ω)* of $Si_{0.7}Ge_{0.3}$ SC bulk alloy at different temperatures. While the low frequency regime behavior mirrors the steady-state behavior, we can see that at each temperature, the values of the thermal conductivity in the high frequency regime are reduced drastically in comparison with the low frequency regime values, so that the SC alloy becomes almost a perfect thermal insulating material. For instance, at *T=300K*, the amplitude of *κ(Ω)* decreases by almost 3 orders of magnitude when it is compared to the reference bulk value of *~5W/m.K*. As the temperature increases, the reduction rate decreases and the deviation threshold frequency from the plateau shape increases, which leads to an increase of the cut-off frequency $f_C$ of *κ(Ω)* as a function temperature. As it is customary in microelectronics, $f_C$ is defined as the frequency at which $Amp\left[\kappa(f_C)\right] = Max\left\{Amp\left[\kappa(f)\right]\right\}/\sqrt{2} = \kappa(0)/\sqrt{2}$ and can formally be expressed as $f_C = 1/2\pi\tau_m^C\left(\omega_D^L, \omega_D^T\right)$ where $\tau_m^C$ is a weighted average combined relaxation time of $\tau_L^C$ and $\tau_T^C$ evaluated at the cut-off Debye frequencies $\omega_D^L$ and $\omega_D^T$, respectively. Even though it is not really systematic, but we can always find a relation between $\tau_m$ and $\tau_m^C$ that relates the position of the resonance peak in the imaginary part $\kappa_i(\Omega)$ to the cut-off frequency $f_C$ of *κ(Ω).*

Fig 3(c) reports the behavior of $f_C$ as a function of temperature for the *5* SC bulk crystals considered in our study, and Fig 3(d) illustrates the behavior of the combined relaxation time $\tau_S^C$ [Eq. (7)] in the case of $Si_{0.7}Ge_{0.3}$ SC alloy, for both longitudinal and transverse (inset) acoustic phonon polarization branches, as a function of the intrinsic phonon frequency ω (dispersion relation) and at different temperatures. Over the temperature range considered in our analysis [*1-1000K*], $f_C(T)$ shows similar trends for all *5* SC bulk crystals and it increases as the temperature is increased. It seems that there is a threshold temperature in the $f_C(T)$ behavior at which the increasing rate of $f_C$ suddenly gets faster. A simple look to Fig 1 suggest that this threshold point in the $f_C(T)$ behavior corresponds to the temperature value at which the steady-state thermal conductivity $\kappa(0)$ reaches a maximum. This interesting feature might be attributed to the interplay between all phonon scattering processes that take place at this particular temperature for each SC crystal. We can see also that for all *5* SC crystals, $f_C$ varies from *100kHz* up to few *THz*. $f_C$ of $Si_{0.7}Ge_{0.3}$ SC alloy shows an interesting trend; it has the highest increasing rate as a function of temperature among all *5* SC crystals for $T \leq 20K$, then this rate becomes the lowest for $20 \leq T \leq 770K$. For this SC alloy, $f_C$ is as low as *12MHz* at *T=100K* and is still less than *2GHz* at room temperature (*T=300K*). Even at a temperature as high as *T=600K*, $f_C$ is still less than 100GHz, this latter value will be soon within the reach of high frequency microelectronic devices according to the ITRS.[44] The very low value of $f_C$ in the low *T* regime indicates that the dominant mean relaxation time of phonon scattering in this regime is on the order of microseconds. These results shed light on how crucial understanding the dynamical behavior of the thermal conductivity has become, in order to better control energy and heat transport in low and high operating temperature microelectronic and optoelectronic devices.

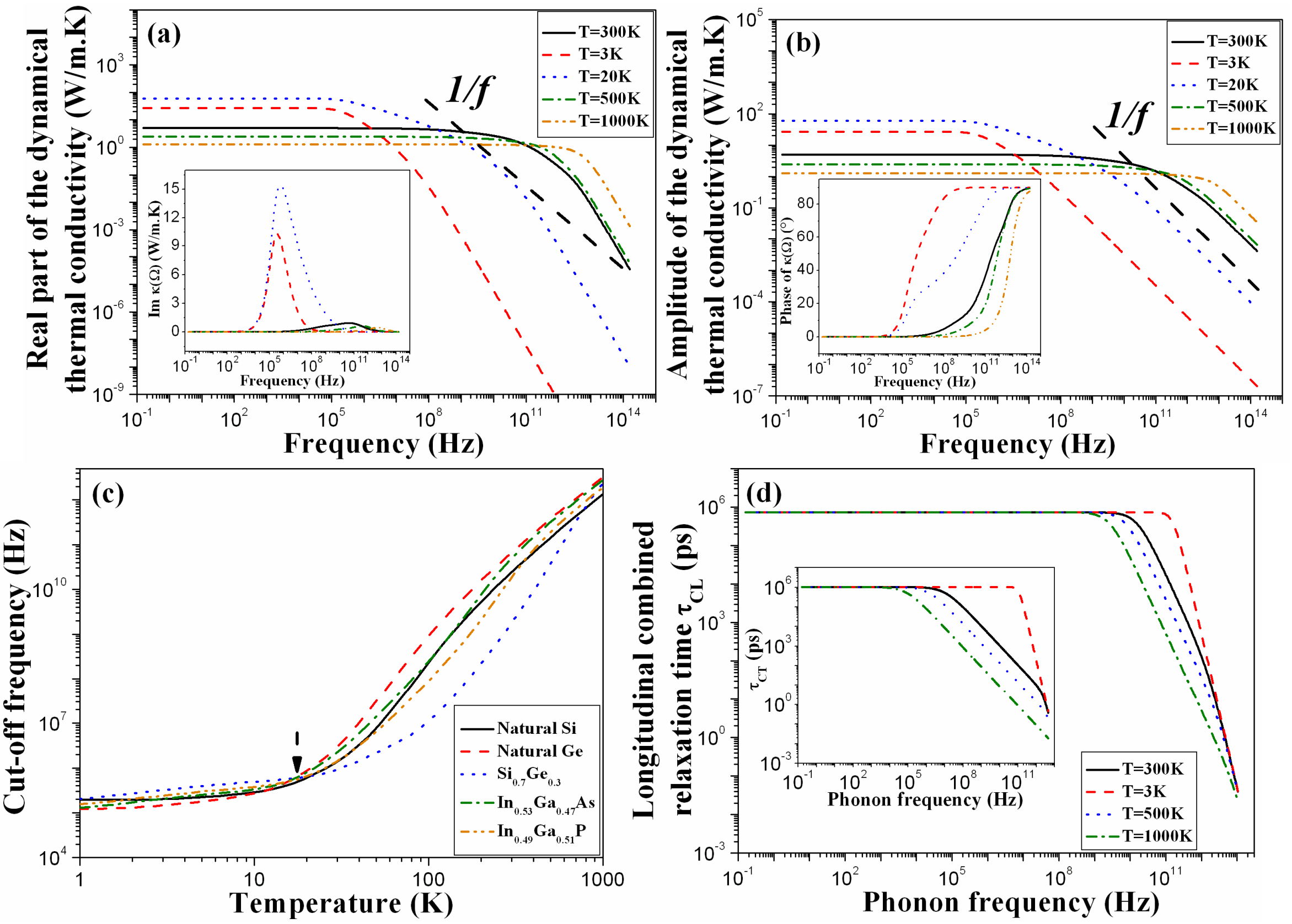

**Figure 3 :** (a) Computed behavior of the real part $\kappa_r(\Omega)$ of the dynamical lattice thermal conductivity $\kappa(\Omega)$ of $Si_{0.7}Ge_{0.3}$ SC alloy as a function of frequency for different temperatures *T*, the inset shows the imaginary part $\kappa_i(\Omega)$. (b) Computed behavior of the amplitude of $\kappa(\Omega)$ of $Si_{0.7}Ge_{0.3}$ SC alloy as a function of frequency for different *T*, the inset shows the phase. (c) Computed behavior of the cut-off frequency $f_C$ of $\kappa(\Omega)$ for the 5 different bulk SC crystals as a function of *T*. (d) Computed behavior of the combined relaxation time $\tau_S^C$ in the case of $Si_{0.7}Ge_{0.3}$ SC alloy, for both longitudinal and transverse (inset) acoustic phonon polarization branches, as a function of the intrinsic phonon frequency ω and at different *T*.

In contrary to the high frequency behavior of the real part, the high frequency behavior of the amplitude can be fitted with a very satisfying $f^{-1}$ power law at each temperature. This confirms the previous analysis of Volz[49] who found the same asymptotic behavior in his study of $\kappa(\Omega)$ of Silicon using Molecular Dynamics method based on spectral Green-Kubo approach.[49] The $f^{-1}$ power law is expected to be valid in the behavior of the amplitude of $\kappa(\Omega)$ for $f \geq f_C$. As a matter of fact, in the high frequency regime, the transport of phonon is predominantly ballistic, this leads to a relaxation time independent response, for which the temperature dependence is mostly governed by the specific heat temperature dependence, i.e., low temperature quantum effect, as discussed earlier by Volz,[49] and acoustic velocity captures the dynamics.

The low operating temperature regime where long wavelength phonons dominate is known to be a place where very interesting and a variety of fundamental phonon transport phenomena occur, particularly ballistic phonon transport and second sound propagation in which the interplay between anharmonic phonon-phonon scattering N-processes and U-processes plays a key role.[45] Low cut-off frequency of $\kappa(\Omega)$ is another phenomenon to be added to the list, and might be fundamentally connected to both aforementioned phenomena.

The frequency behavior of the imaginary part and the phase of $\kappa(\Omega)$ are reported in the insets of Figs 3(a) and 3(b), respectively. By decreasing temperature, both the position of the resonance peak and its amplitude change; in a way these changes mirror the behavior of the stead-state thermal conductivity as a function of temperature. On the other hand, the phase seems to increase and reach the saturation value of $\pi/2$ faster as the temperature decreases.

Among the extrinsic parameters that we consider the effect of their variations on $\kappa(\Omega)$, we have both longitudinal and transverse Grüneisen parameters $\gamma_L$ and $\gamma_T$ as well as the mass-difference fluctuation parameter $\Gamma$. We report, respectively, in Figs 4(a) and 4(b) the calculated dynamical behaviors at room temperature of the real part and the amplitude of $\kappa(\Omega)$ of $Si_{0.7}Ge_{0.3}$ SC alloy for different values of $\gamma_L$ while in Figs 4(c) and 4(d), we report, respectively, the dynamical behaviors of the same functions for different values of $\gamma_T$. The insets in Figs 4(a) and 4(c) illustrate the behavior of the imaginary part of $\kappa(\Omega)$. On the other hand, the behavior of the phase of $\kappa(\Omega)$ is illustrated in the insets of Figs 4(b) and 4(d). Rigorously speaking, the mode Grüneisen parameters $\gamma_L$ and $\gamma_T$ depend on the phonon intrinsic frequency (dispersion relation). Depending on the SC crystal crystallographic class and symmetry group, $\gamma_L$ and $\gamma_T$ can be calculated using *ab-initio* lattice dynamical models.[34]

In our analysis, we consider the average values of $\gamma_L$ and $\gamma_T$ to vary from 0.5 to 1.5 and from 0.2 to 1.2, respectively. We remind here, that in previous figures, we assumed fixed values of $\gamma_L=1$ and $\gamma_T=0.7$ in our calculations. By varying the Grüneisen parameter, the strengths of both anharmonic phonon-phonon U-processes and N-processes scattering rates change. According to Eqs. (23) and (25), both scattering rates strengths have a quadratic dependence on the value of the Grüneisen parameter.

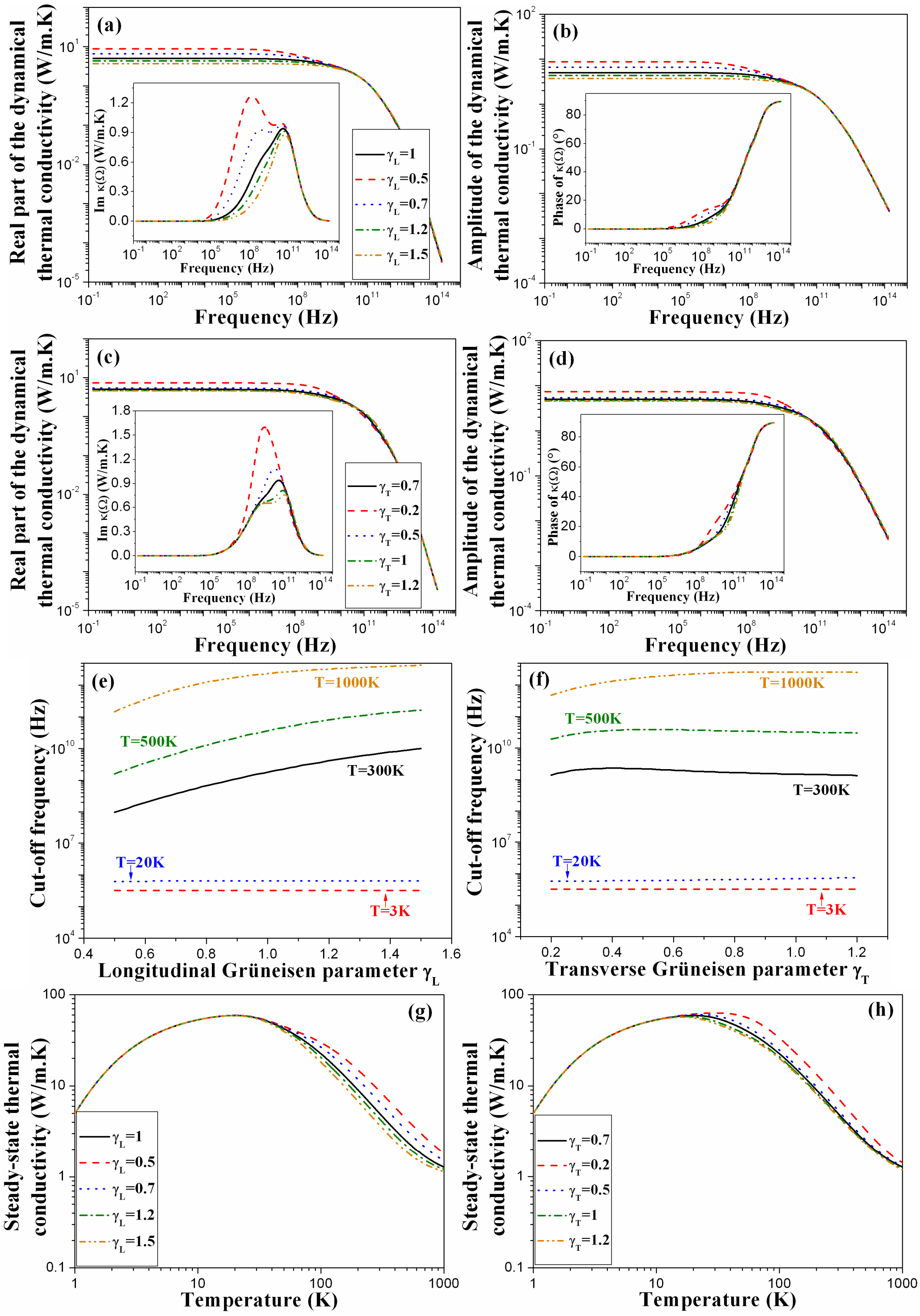

**Figure 4 :** Computed behavior of the real part $\kappa_r(\Omega)$ (a, c) and amplitude (b, d) of $\kappa(\Omega)$ of $Si_{0.7}Ge_{0.3}$ SC alloy at room temperature as a function of frequency for different values of the (longitudinal, transverse) Grüneisen parameter ($\gamma_L$, $\gamma_T$). The insets in (a) and (c) show the imaginary part $\kappa_i(\Omega)$, while the insets in (b) and (d) show the phase. Computed behavior of $f_C$ of $\kappa(\Omega)$ of $Si_{0.7}Ge_{0.3}$ SC alloy as a function of $\gamma_L$ (e) and $\gamma_T$ (f) for different $T$. Computed behavior of $\kappa(0)$ of $Si_{0.7}Ge_{0.3}$ SC alloy as a function of $T$ for different values of $\gamma_L$ (g) and $\gamma_T$ (h).

As can be seen in Figs 4(a-d), changing $\gamma_L$ (resp. $\gamma_T$) seem to have an effect mainly in the low frequency regime and particularly near the resonance frequency in the behavior of the imaginary part, while no significant effect occur in the high frequency regime where all curves almost collapse. By decreasing $\gamma_L$ (resp. $\gamma_T$), the strength of the scattering rate of both N-processes and U-processes decreases which lead to an increase in the amplitude of $\kappa(\Omega)$ in the low frequency regime, this effect mirrors again the effect of changing $\gamma_L$ (resp. $\gamma_T$) on the steady-state thermal conductivity where the effect is manifested for temperatures above the maximum and completely disappear for temperatures below [Fig 4(g) for $\gamma_L$ and Fig 4(h) for $\gamma_T$]. Probably the most interesting and intriguing effect of varying $\gamma_L$ (resp. $\gamma_T$) is illustrated in the dynamical behavior of the imaginary part $\kappa_i(\Omega)$, and may be less importantly in the behavior of the phase. For both functions, the effect occurs on the same frequency interval. A double resonance peak shape seems to appear in $\kappa_i(\Omega)$ as we decrease $\gamma_L$, while the same effect happens by increasing $\gamma_T$. On the other hand, the amplitude of the resonance peak increases by decreasing either $\gamma_L$ or $\gamma_T$.

In Figs 4(e) and 4(f), we report the behavior of the cut-off frequency $f_C$ of $\kappa(\Omega)$ as a function of $\gamma_L$ and $\gamma_T$, respectively, for different temperatures $T$. At low $T$, $f_C$ is almost constant independent of $\gamma_L$ (resp. $\gamma_T$), and as long as we increase $T$, different behaviors of $f_C$ start to occur. In the high $T$ regime, $f_C$ keeps increasing as $\gamma_L$ increases, while for increasing $\gamma_T$, $f_C$ seems to reach quickly a saturation value that increases with $T$.

The effects of varying $\gamma_L$ and $\gamma_T$ on $\kappa(\Omega)$ and particularly on $\kappa_i(\Omega)$ capture the essence of the fundamental interplay between anharmonic phonon-phonon scattering N-processes and U-processes.

In Figs 5(a), and 5(b) we report, respectively, the calculated dynamical behavior at room temperature of the real part $\kappa_r(\Omega)$ and the steady-state behavior of $\kappa(0)$ of $Si_{0.7}Ge_{0.3}$ SC alloy for different values of the mass-difference fluctuation parameter $\Gamma$. $\Gamma$ is assumed to change from its initial alloy disorder value of $\Gamma \approx 0.24$ to $\Gamma = 0.36$. Increasing $\Gamma$ can be obtained by isotopically enriching the SC alloy and/or incorporating additional impurities. The inset of Fig 5(a) illustrates the dynamical behavior of the imaginary part $\kappa_i(\Omega)$. Varying $\Gamma$ does not seem to have a significant effect on the behavior of the three functions; neither does it on the behavior of the cut-off frequency $f_C$, the behavior of which as a function of $\Gamma$ for different temperatures is reported in Fig 5(c). At each temperature, $f_C$ is almost a constant independent of $\Gamma$ and increases as one increases $T$ [Fig 3(c)].

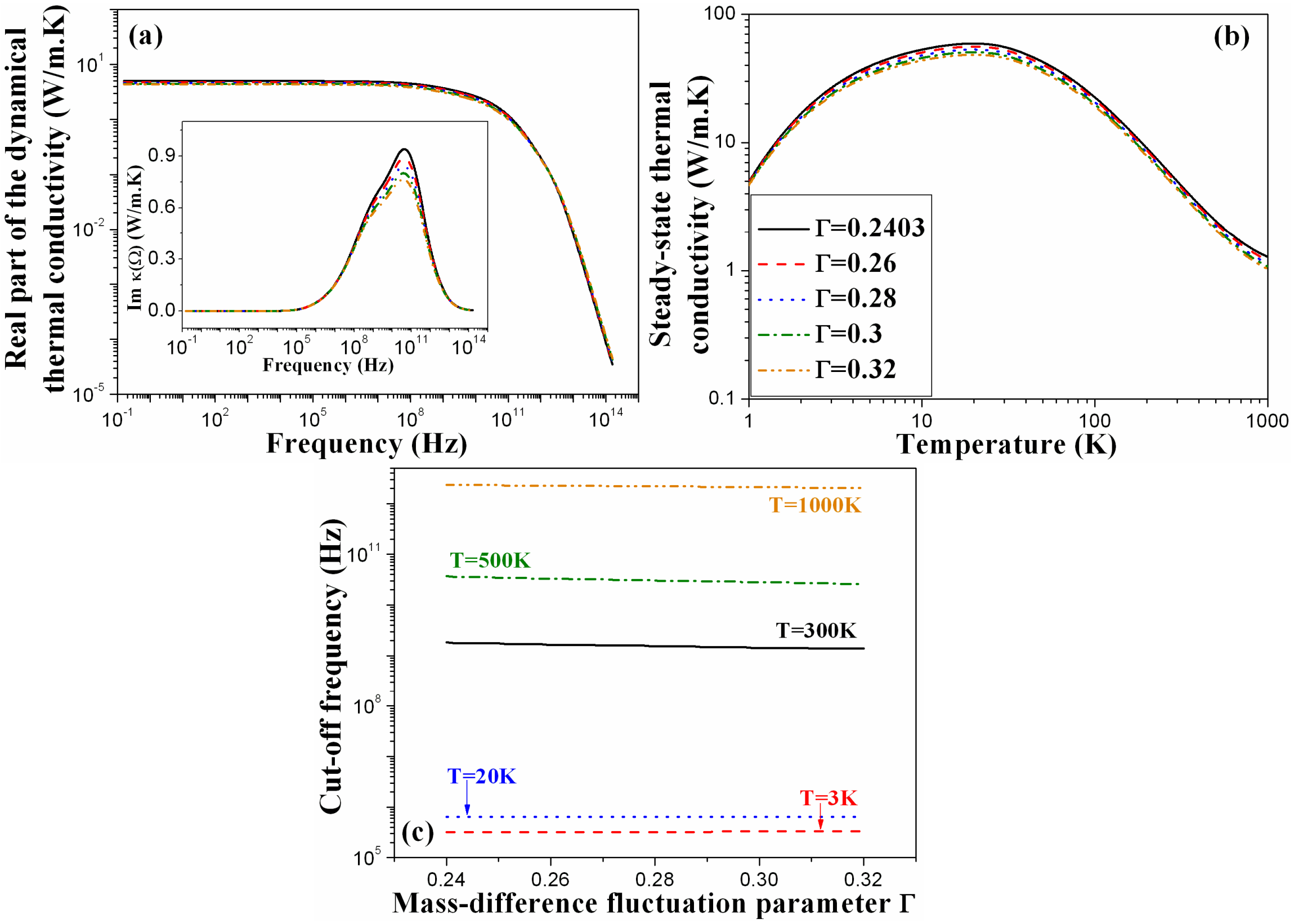

**Figure 5 :** (a) Computed behavior of the real part $\kappa_r(\Omega)$ of $Si_{0.7}Ge_{0.3}$ SC alloy at room temperature as a function of frequency for different values of the mass-difference fluctuation parameter $\Gamma$, the inset show the imaginary part $\kappa_i(\Omega)$. (b) Computed behavior of $\kappa(0)$ of $Si_{0.7}Ge_{0.3}$ SC alloy as a function of temperature $T$ for the same different values of $\Gamma$. (c) Computed behavior of $f_C$ of $\kappa(\Omega)$ of $Si_{0.7}Ge_{0.3}$ SC alloy as a function of $\Gamma$ for different $T$.

## 2. Embedding nanoparticles in a SC alloy matrix

As we mentioned in the introduction, almost all recent reports on increasing *ZT* of SC thermoelectric materials,[3-11] have been proven to be due essentially to a considerable reduction in the thermal conductivity $\kappa$ of these materials. One of the most successful methods to achieve such a reduction in $\kappa$ has been "*nanostructuration*" of the thermoelectric SC crystal material.[12-16] Even though, it has been difficult to beat the "*alloy limit*" in crystals without creating defects, dislocations and voids.

Recently, a clever approach called "*Nanoparticle-in-alloy*" has been suggested to beat the alloy limit of $\kappa$ in SC crystals. This approach was experimentally demonstrated in the case of ErAs nanoparticles embedded in InGaAs matrix with a *50%* reduction in $\kappa$ and a corresponding *100%* increase in *ZT*.[7] Mingo et al[16,67,68] investigated this approach with different other material combinations and found very promising results that might have huge and potential implications in SiGe-based technology both in thermoelectricity and microelectronics. The implantation of this approach is not however trivial. Nanoparticles, which could be either metallic, semimetallic or semiconductor, have to be lattice-matched with the SC alloy crystal matrix in such a way that they can be grown inside the matrix without defects and dislocations. They also need to scatter more efficiently phonons with a negligible effect on electron mobility.[16,67] In SC alloy crystals, alloy disorder scatter phonons due either to mass difference and/or size difference and/or bond strength difference. In the Rayleigh scattering regime, the scattering cross section varies as $\sigma \sim \lambda^{-4}$ [Eq. (25)] where $\lambda$ is the phonon wavelength. This means that short wavelength phonons are scattered much more efficiently than mid and long wavelength phonons; hence heat conduction in an alloy is mainly dominated by these latter phonons. By incorporating nanoparticles in a SC alloy crystal, mid and long wavelength phonons are also scattered which reduces the thermal conductivity below the alloy limit.[7]

In a recent paper, A. Kundo et al[68] analyzed and pointed out the effect of light and heavy embedded nanoparticles on the thermal conductivity $\kappa$ of a SC alloy matrix using an atomistic *ab-initio* Green's function approach. The authors considered $Si_{0.5}Ge_{0.5}$ SC alloy with embedded Si or Ge nanoparticles and found that the effect on $\kappa$ of incorporating the nanoparticles is different depending on whether the nanoparticles are relatively heavier or lighter than the hosting matrix. The calculation predicts that heavier nanoparticles (Ge) should be more efficient than lighter ones (Si) in reducing $\kappa$ of $Si_{0.5}Ge_{0.5}$. The difference is due to the fact that heavy nanoparticles scatter mid and long wavelength phonons (phonons that make the largest contribution to $\kappa$) more strongly than do the light ones.

It is worth to mention that, besides acting as phonon scattering centers, embedded nanoparticles may also act as dopants donating carriers, which will reduce the needed amount of dopant impurities in the SC matrix and thus enhance the electron mobility.[69] Additionally, energy dependence scattering rates of electrons are sensitive to the long range potential screening-tails of the nanoparticles which may result in an increased Seebeck coefficient.[70]

In our analysis of the effect of "*Nanoparticle-in-alloy*" approach on the behavior of the dynamical lattice thermal conductivity $\kappa^{np}(\Omega)$ of a SC alloy matrix, we assume a $Si_{0.7}Ge_{0.3}$ SC alloy crystal matrix in which we incorporate spherical Ge nanoparticles.[16,67] In contrast with the previous study by Mingo et al[16,67] in which phonon scattering N-processes were disregarded, we consider in our analysis all phonon scattering processes to be active. In this case, in addition to N-processes, we have four resistive phonon scattering processes to take into account; the three mentioned above in the theory section (Umklapp scattering, alloy-disorder scattering and boundary scattering), to which we add scattering of phonons by the embedded nanoparticles.

In this case, the total resistive scattering rate is given according to Mathienssen's rule as:

$$\left[\tau_S^R(\omega)\right]^{-1} = \left[\tau_S^U(\omega)\right]^{-1} + \left[\tau_S^a(\omega)\right]^{-1} + \left[\tau_S^B(\omega)\right]^{-1} + \left[\tau_S^{np}(\omega)\right]^{-1} \qquad (32)$$

Umklapp scattering rate $\tau_S^U$, alloy disorder scattering rate $\tau_S^a$ and boundary scattering rate $\tau_S^B$ are as given by Eqs. (24), (25) and (27), respectively, where $\Gamma$ in Eq. (25) is calculated according to the virtual lattice approach[25] using Eq. (26) in which only mass-difference contribution is accounted for.

Nanoparticle scattering rate is calculated by assuming a spherical shape and using a Mathienssen type interpolation between the long and short wavelength scattering regimes:[16,67,71]

$$\left[\tau_S^{np}(\omega)\right]^{-1} = \frac{v_S\left[\sigma_{sS}^{-1} + \sigma_{lS}^{-1}\right]^{-1}}{V_{np}} f_{np} \qquad (33)$$

where $V_{np} = 4/3\,\pi R^3$ is the nanoparticle volume of radius $R$, $\sigma_{sS}$ and $\sigma_{lS}$ are the cross sections of the short and long wavelength scattering regimes, respectively, and $f_{np}$ is the nanoparticle volume fraction. Eq. (33) holds in the linear regime that is valid for small values of $f_{np}$; the case we assume here in our treatment of $\kappa^{np}(0)$ and $\kappa^{np}(\Omega)$. The scattering cross section limits are given by:[16,22,67,71]

$$\begin{cases} \sigma_{sS} = 2\pi R^2 \\ \sigma_{lS}(\omega) = \dfrac{4\pi}{9}\left(\dfrac{\rho-\rho_{np}}{\rho}\right)^2 R^6 \left(\dfrac{\omega}{v_S}\right)^4 \end{cases} \qquad (34)$$

where $\rho$ and $\rho_{np}$ are the densities of the host matrix material and the embedded nanoparticle, respectively. As it is well known, the short wavelength limit scattering cross section is twice the geometrical cross section,[71] while the long wavelength limit is given according to Rayleigh's expression.[22] Eq. (34) takes into account both phonon scattering contributions due to mass and size differences of the nanoparticles relative to the host SC matrix. Since we are treating the SC crystal as an elastic isotropic continuum medium, no much additional error would be introduced by neglecting the contribution due to bond strength differences.[16,67]

Through this subsection, we assume Ge nanparticles of density ($\rho_{Ge}$≈$5.323g/cm^3$) to be embedded in $Si_{0.7}Ge_{0.3}$ SC alloy matrix of density ($\rho_{Si0.7Ge0.3}$≈$3.332g/cm^3$). We assume that including the nanoparticles will not affect the intrinsic physical properties of the SC host alloy. We will first analyze the case of a fixed nanoparticle volume fraction of $f_{np}=2\%$. We assume this value of $f_{np}$ won't affect negatively the electron mobility.[16,67] In fact, as has been discussed by Mingo et al,[16, 67] the effect of nanoparticles on the electron mobility of the SC alloy matrix will be negligible if the contribution to the electron mean free path (MFP) due to this scattering process is much larger than the intrinsic MFP due to the pure inelastic alloy scattering. According to Mingo et al,[16,67] the MFP of electrons due to nanoparticle scattering is given by $\ell_{np} \sim R/3f_{np}$.[16,67] On the other hand, the MFP due to the pure inelastic alloy scattering is energy independent and can be estimated from the measurements of the electron mobility. Using the relation between the electrical mobility and the relaxation time $\mu(T) \sim e\tau(k_BT)/m^* = e\ell_{alloy}/\sqrt{3k_Bm^*T}$,[22] the aforementioned condition to neglect the effect of nanoparticle scattering on the electron mobility becomes then:

$$\ell_{np} > \ell_{alloy} \Rightarrow f_{np} < R/3\ell_{alloy} \sim \frac{eR}{3\mu(T)\sqrt{3k_Bm^*T}} \qquad (35)$$

where $e$ is the electron elementary absolute charge and $m^*$ is the electron (hole) effective mass at the bottom of the conduction band (top of the valence band) depending on whether the SC alloy is $N$ or $P$ type. According to Slack and Hussain,[65] the total electron mobility in $Si_{(1-x)}Ge_x$ SC alloy can be estimated using:

$$\begin{cases} \mu(x,T_0)_{Tot}^{-1} = \mu_V^{-1} + \mu_{alloy}^{-1} \\ \mu_{alloy}(x,T_0) = \mu_0 \Big/ \left[4x(1-x)\sqrt{T_0/300}\right] \text{ cm}^2\text{/Vs} \\ \mu_V(x,T_0=300K) = \left[86x + (1-x)40.7\right] \text{ cm}^2\text{/Vs} \end{cases} \qquad (36)$$

where $\mu_{alloy}$ is the alloy mobility term and $\mu_V$ is the virtual crystal mobility term, $T_0$ is the local equilibrium absolute temperature and $\mu_0$~$150$ for a $N$ type and ~$140$ for a $P$ type SC material.[65]

At room temperature, this will give electron mobility for the $Si_{0.7}Ge_{0.3}$ SC material of about *~41cm²/Vs* for both *N* and *P* types. The effective mass $m^*$ can be a complicated function of the doping level, temperature as well as the content of Ge in the $Si_{(1-x)}Ge_x$ SC crystal alloy. For the present calculations, we take the effective mass of conductivity for both electrons and holes in $Si_{0.7}Ge_{0.3}$ SC alloy to be the same $m^* \sim m_0$ where $m_0$ is the free electron mass.[65] We obtain an alloy limited MFP of $\ell_{alloy} \approx 2.7nm$. By using this value, the condition on the nanoparticle volume fraction [Eq. (35)] becomes $f_{np} < R/8nm$. This result means that for $f_{np}=2\%$, nanoparticles of radius $R$ larger than *0.2nm* will not have any significant effect on the electronic MFP. This is very comforting since we are considering $R$ to vary over an interval *[1-100nm]*. We should mention here, that since the nanoparticle contribution to the electron scattering rate does not depend on temperature, the above discussion remains valid at high temperature.[16] Nevertheless, electron mobility in the low $T$ regime might be affected.

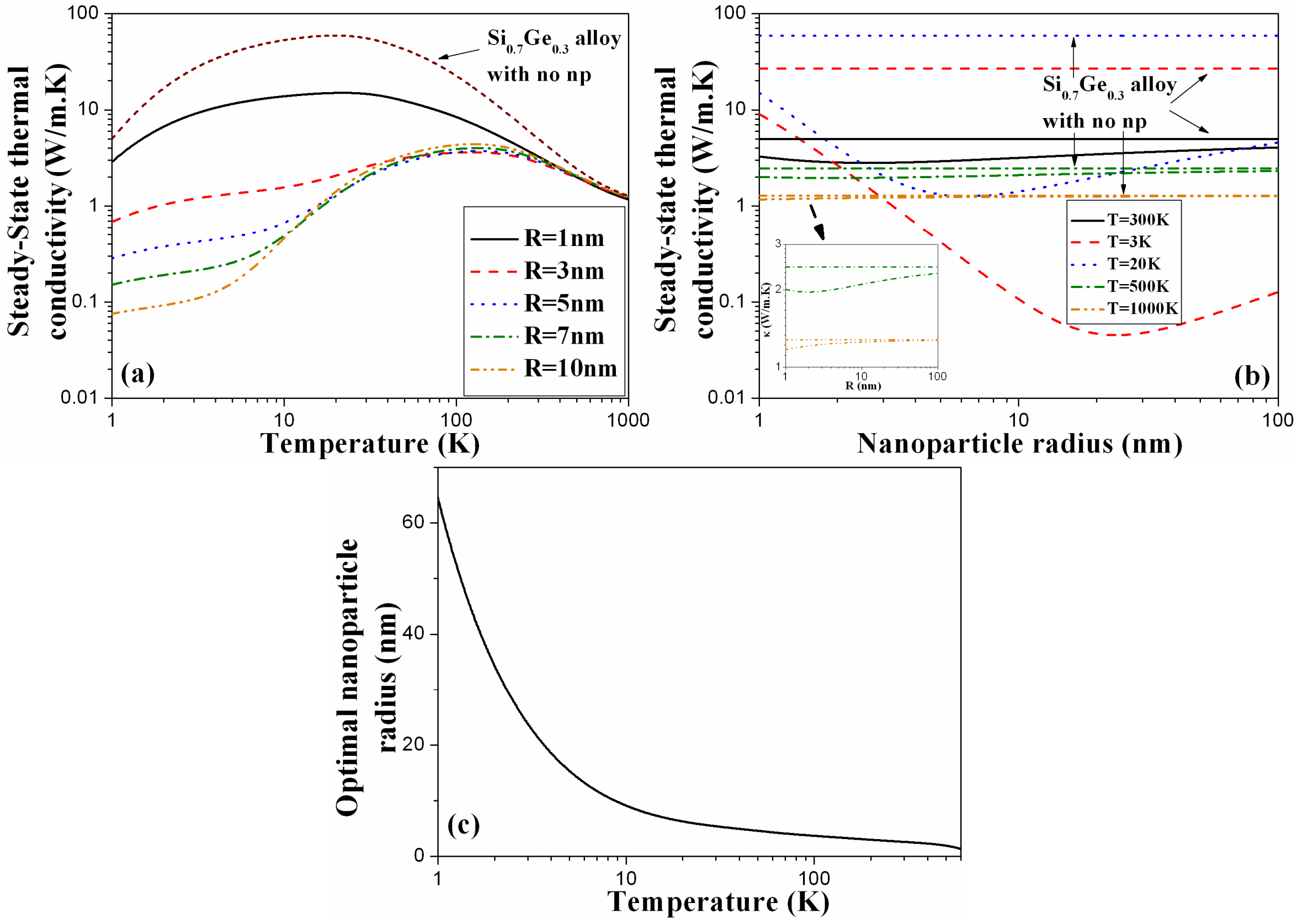

**Figure 6 :** Computed behaviors of $\kappa^{np}(0)$ of $Si_{0.7}Ge_{0.3}$ SC alloy with embedded Ge nanoparticles as a function of temperature $T$ for different values of the nanoparticles radius $R$ (a) and as a function of $R$ for different $T$ (b). Also are reported the values of $\kappa(0)$ of $Si_{0.7}Ge_{0.3}$ SC alloy without Ge nanoparticles. (c) Computed behavior of the optimal Ge nanoparticles radius $R_{min}$ that minimizes $\kappa^{np}(0)$ of $Si_{0.7}Ge_{0.3}$ SC alloy as a function of $T$.

In Figs 6(a) and 6(b), we report, respectively, the calculated steady-state behavior of the thermal conductivity $\kappa^{np}(0)$ of $Si_{0.7}Ge_{0.3}$ SC alloy matrix as a function of ambient temperature $T$ for different values of Ge nanoparticles radius $R$ and as a function of $R$ for different $T$. Also is added to both figures, the calculated behavior of $\kappa(0)$ of $Si_{0.7}Ge_{0.3}$ alloy with no nanoparticles. As we can see in Fig 6(a), $\kappa^{np}(0)$ decreases by embedding nanoparticles and the position of the peak value tends to shift to the right by increasing $R$. The effect is more noticeable and even drastic in the low temperature regime where the reduction of $\kappa^{np}(0)$ can be of more than an order of magnitude and where $\kappa^{np}(0)$ continues to decrease as $R$ increases. Furthermore, we can see that in this temperature regime, the behavior of $\kappa^{np}(0)$ tends to depart from the specific heat per unit volume $T^3$ power law to follow another $T^r$ ($r$ is some small number more likely to be rational) power law as the nanoparticle radius $R$ increases. On the other hand, all curves seem to superimpose in the high temperature regime. In Fig 6(b), we reproduce the same interesting feature discussed by Mingo et al,[16,67,68] namely the existence of an optimal nanoparticle radius $R_{min}$ that minimizes the steady-state thermal conductivity $\kappa^{np}(0)$. The existence of $R_{min}$ is a result of the interplay between long and short wavelength scattering regimes [Eqs. (33) and (34)]. As a matter of fact, it is easy to verify from the latter equations that the nanoparticles scattering rate $\left[\tau_S^{np}\left(\omega,R\right)\right]^{-1}$ for a phonon mode at polarization $S$, when it is treated as a function of the nanoparticle radius $R$, has a maximum value $\left[\tau_S^{np}\left(\omega,R_0\right)\right]^{-1}$ for a certain radius $R_0$; a maximum $\left[\tau_S^{np}\left(\omega,R_0\right)\right]^{-1}$ leads to a minimum $\kappa^{np}(0)$. It is worth to mention here that due to averaging over all phonon modes and polarizations, $R_{min}$ is different from $R_0$. Fig 6(b) shows also that the effect of embedding nanoparticles diminishes as $T$ increases. In order to gain more insight onto the behavior of $R_{min}$ as a function of $T$, we report in Fig 6(c) the calculated behavior of $R_{min}$ as a function of $T$; $R_{min}$ decreases monotonically as $T$ increases. While $R_{min}$ is about $20nm$ at $T=3K$, it reduces to $2nm$ at $T=500K$. Nanoparticles scattering rate is temperature independent as given by Eqs. (33) and (34); the $T$-behavior of $R_{min}$ can thus be understood by taking into consideration the fact that in the low $T$ regime, the wavelength and MFP of phonons increase so that larger nanoparticles would be more effective in scattering them than smaller ones, which explains why $R_{min}$ increases as $T$ decreases. On the other hand, in the high $T$ regime, scattering of phonons is due predominantly to anharmonic N-processes and U-processes so that the effect of phonon scattering by embedded nanoparticles becomes of the same order of magnitude as phonon scattering by impurities and defects in the Rayleigh regime.

The minimum of $\kappa^{np}(0)$ is very wide which is very advantageous from the technological point of view; this means that a precise control of the nanoparticles size is not essential to achieve lower thermal conductivities and consequently higher *ZTs*.

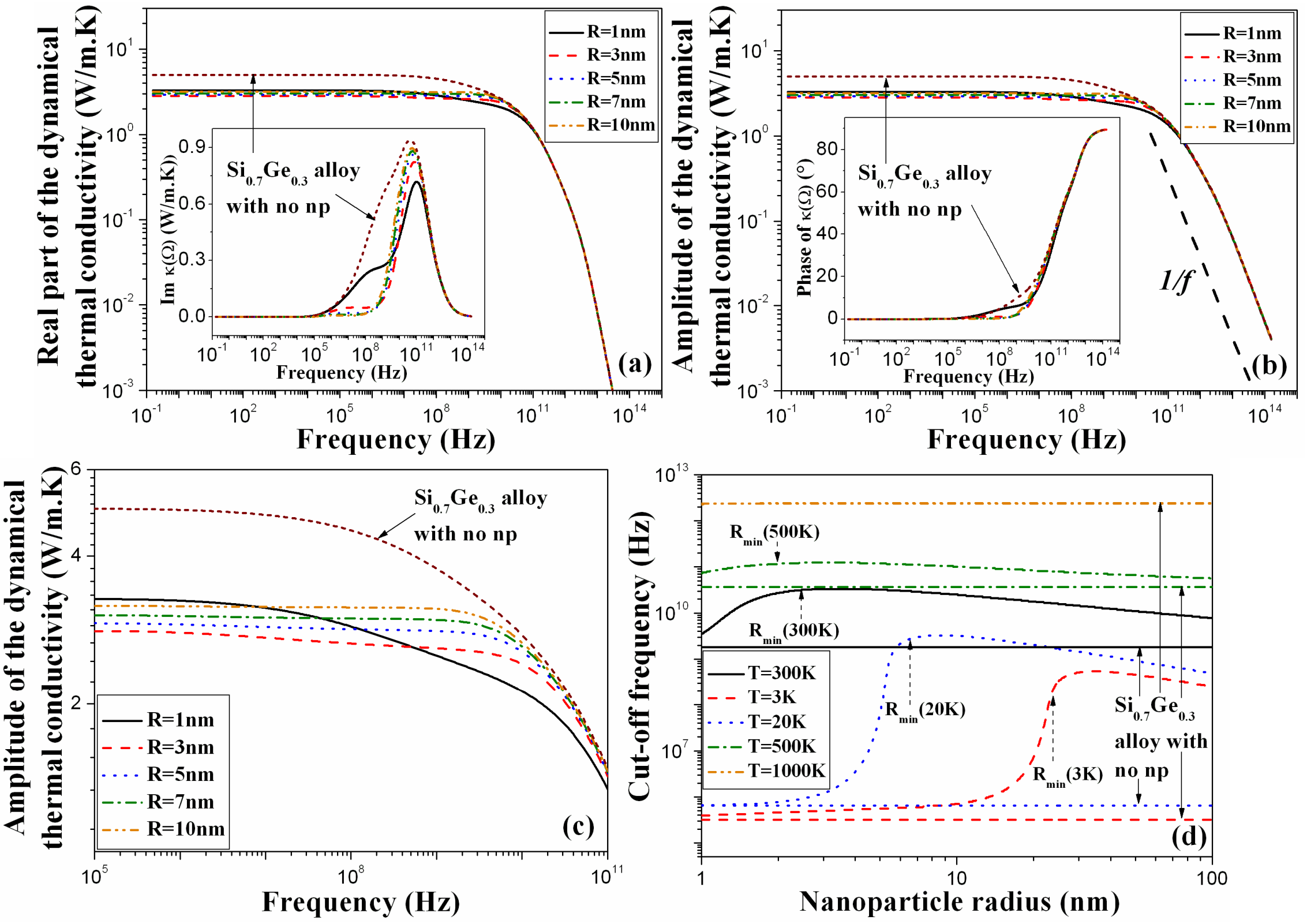

**Figure 7 :** Computed behaviors of the real part (a) and amplitude (b) of $\kappa^{np}(\Omega)$ of $Si_{0.7}Ge_{0.3}$ SC alloy with embedded Ge nanoparticles at room temperature as a function of frequency for different values of the nanoparticles radius *R*. The inset in (a) shows the imaginary part of $\kappa^{np}(\Omega)$ while the inset in (b) shows the phase. (c) A zoom of (b) over [$10^5$-$10^{11}$*Hz*] frequency interval. (d) Computed behavior of $f_C$ of $\kappa^{np}(\Omega)$ of $Si_{0.7}Ge_{0.3}$ SC alloy with embedded Ge nanoparticles as a function of *R* for different *T*. Also are reported the values of $f_C$ of $\kappa(\Omega)$ of $Si_{0.7}Ge_{0.3}$ SC alloy without Ge nanoparticles. The dashed arrows indicates the position of the optimal nanoparticles radius $R_{min}$ that minimizes $\kappa^{np}(0)$ at each *T*.

Now, let us shed some light on the dynamical behavior of $\kappa^{np}(\Omega)$. We report, respectively, in Figs 7(a) and 7(b) the calculated room temperature dynamical behaviors of the real part and the amplitude of $\kappa^{np}(\Omega)$ of $Si_{0.7}Ge_{0.3}$ SC alloy matrix with embedded Ge nanoparticles for different values of the radius *R* of the latter. As in the steady-state case, we added to both figures the calculated dynamical behaviors of $\kappa(\Omega)$ of $Si_{0.7}Ge_{0.3}$ with no nanoparticles. There is a slight difference in the dynamical behaviors between $\kappa^{np}(\Omega)$ and $\kappa(\Omega)$. This difference is mainly noticeable at the end of the plateau where the amplitude (real part) of $\kappa^{np}(\Omega)$ seems to manifest a slight second-order-like low pass filter thermal behavior, while the amplitude (real part) of $\kappa(\Omega)$ manifests a first-order low pass filter thermal behavior [Fig 7 (c)]. Here again, we see that in the low frequency regime, the behavior as a function of the nanoparticles radius $R$ mirrors the steady-state behavior of $\kappa^{np}(0)$.

The effect of changing $R$ appears on a certain intermediate range of frequency, and the high frequency $f^{-1}$ power law in the behavior of the amplitude of $\kappa(\Omega)$ is seemingly preserved where we can see that all curves relative to different $R$ values superimpose in the high frequency regime. The calculated dynamical behaviors of the imaginary part and phase of $\kappa^{np}(\Omega)$ are reported in the insets of Figs 7(a) and 7(b), respectively. Here also, the dynamical behaviors of these two functions, captures clearly the effect of changing $R$ on $\kappa^{np}(\Omega)$. For each value of $R$, the imaginary part of $\kappa^{np}(\Omega)$ manifests a secondary resonance peak at the left of the primary one, the amplitudes and positions of both peaks vary by increasing $R$; the amplitudes vary in an opposite way; by increasing $R$, the amplitude of the secondary peak decreases, while the amplitude of primary one increases. On the other hand, the positions of both peaks seem to shift left by increasing $R$.

We illustrate in Fig 7(d), the calculated behavior of the cut-off frequency $f_C$ of $\kappa^{np}(\Omega)$ of $Si_{0.7}Ge_{0.3}$ SC alloy matrix as a function of embedded Ge nanoparticles radius $R$ for different ambient temperatures $T$. We report also the corresponding values of $f_C$ of $\kappa(\Omega)$ of $Si_{0.7}Ge_{0.3}$ with no nanoparticles. $f_C$ shows a very interesting and curious trend that looks like a tilted step function behavior especially for low temperatures where the effect of nanoparticles scattering is more dominant. This behavior smoothes out as $T$ is increased. In the low temperature regime, the behavior of $f_C$ as function of $R$ can be cast into three different regimes characterized by two threshold points; for small values of $R$, $f_C$ increases slightly as $R$ is increased. Suddenly when $R$ reaches a certain value (first threshold point), the trend changes and $f_C$ starts to increase faster with a higher increasing rate almost in a tilted step-like jump to reach a maximum at another larger value of $R$ (second threshold point). After the maximum, $f_C$ starts falling off as $R$ is increased with a low reduction rate. Between the two threshold points, a small change in $R$, on the order of nanometer fluctuations, leads to a huge variation in $f_C$ of a few orders of magnitude. As $T$ is increased, the positions of the threshold points, shift left and the steepness of $f_C$ increasing rate between these two points tends to be less pronounced; for temperatures higher than $300K$, the first threshold point occurs for values of $R<1nm$. The two threshold points in the behavior of $f_C$ as a function of $R$ seem to correspond to values of $R$ surrounding the optimal $R_{min}$ value that leads to the minimum value of $\kappa^{np}(0)$ as indicated by the arrows. This behavior is again a manifestation of the interplay between long and short wavelength scattering regimes [Eqs. (33) and (34)] as mentioned previously. Investigation of Fig 7(d) shows that for all temperatures and over the range of $R$ values [*1-100nm*], $f_C$ of $\kappa^{np}(\Omega)$ with embedded nanoparticles is higher than the corresponding $f_C$ of $\kappa(\Omega)$ without nanoparticles; this effect is expected since by incorporating nanoparticles, one increases the total phonon scattering rate. The average difference decreases as $T$ gets higher, where we can see that for $T=1000K$, $f_C$ of $\kappa^{np}(\Omega)$ becomes insensitive to $R$ and indistinguishable from the corresponding $f_C$ of $\kappa(\Omega)$. This behavior corroborates the fact that in the high $T$ regime, scattering of phonons is mainly dominated by anharmonic phonon-phonon scattering processes.[17,22,24,32]

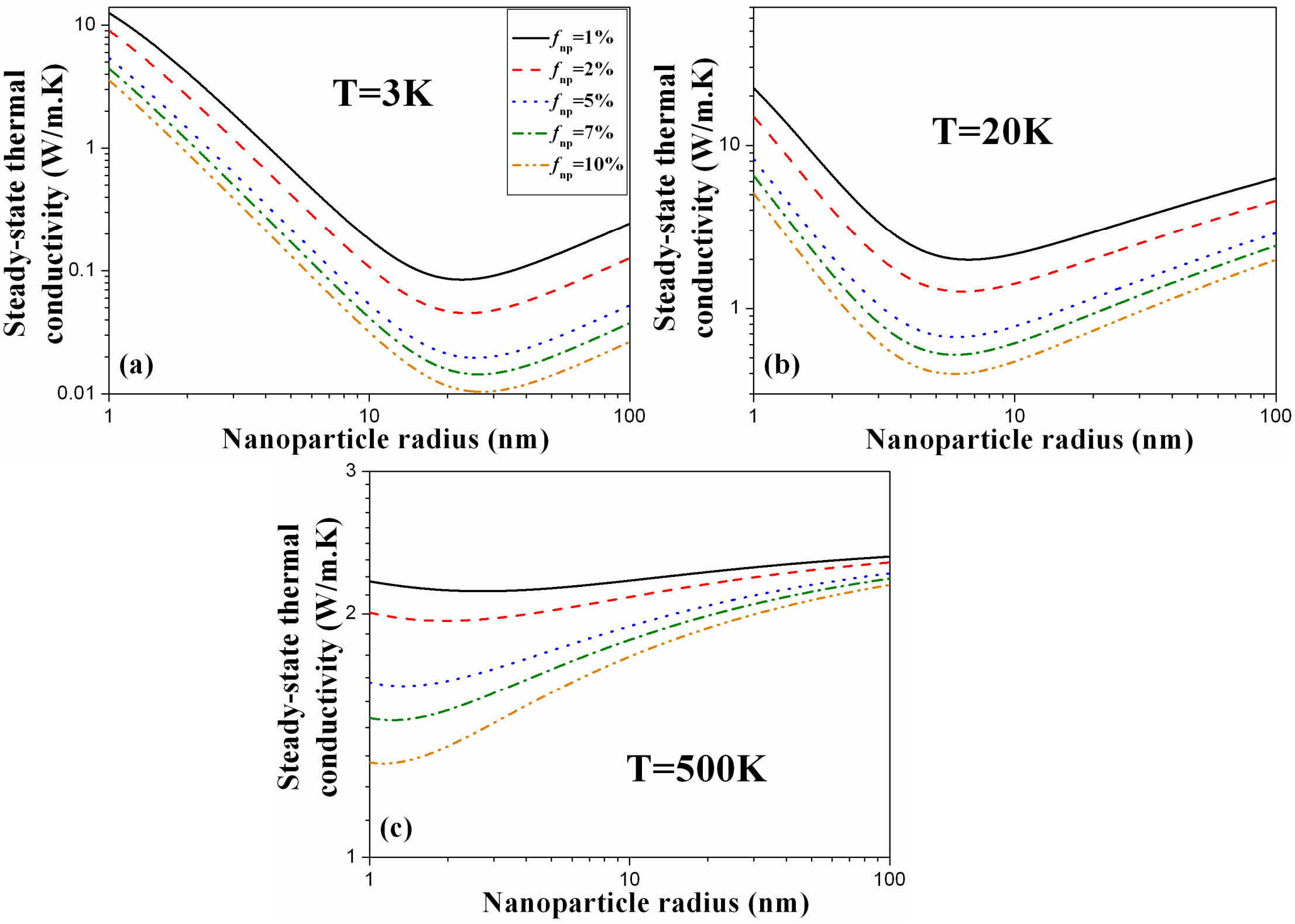

**Figure 8 :** Computed behaviors of $\kappa^{np}(0)$ of $Si_{0.7}Ge_{0.3}$ SC alloy with embedded Ge nanoparticles as a function of the nanoparticles radius $R$ for different values of the nanoparticles volume fraction $f_{np}$ at: (a) $T=3K$, (b) $T=20K$ and (c) $T=500K$.

Let's now see what happens when one changes the nanoparticles volume fraction $f_{np}$ at different ambient temperatures $T$. As we mentioned earlier, in the high $f_{np}$ regime, the nanoparticle scattering rate is no longer linear to $f_{np}$ due essentially to multiple scattering effects. Besides, a high nanoparticle concentration may lead to undesirable additional scattering of electrons in the SC matrix, hence negatively affecting the electron mobility.[16]

In our case, we assume $f_{np}$ to vary from *1%* to *10%* and we continue to make the hypothesis of a linear relation linking the nanoparticle scattering rate and $f_{np}$ [Eq. (33)]. For a maximum value of $f_{np}=10\%$, the effect of nanoparticles on the electronic MFP would be negligible if their radius $R$ is larger than *0.8nm*. This condition is still fulfilled regarding the interval of variation of $R$. Embedding Ge nanoparticles with $f_{np}$ varying from *1%* to *10%* will continue to affect mainly phonon transport.

We report in Figs 8 (a), (b) and (c) the calculated steady-state behaviors of the thermal conductivity $\kappa^{np}(0)$ of $Si_{0.7}Ge_{0.3}$ SC alloy matrix as a function of the embedded Ge nanoparticles radius $R$ for different values of the nanoparticles volume fraction $f_{np}$ at $T=3K$, *20K* and *500K*, respectively. As these figures show, at each ambient temperature, $\kappa^{np}(0)$ decreases as $f_{np}$ is increased; an expected effect regarding the proportionality between the nanoparticle scattering rate and $f_{np}$ [Eq. (33)]. For each value of $R$, the decreasing rate is higher at low $T$ and tends to decrease as $T$ increases. Besides, we can see that the position of the optimal nanoparticle radius $R_{min}$ that leads to the minimum $\kappa^{np}(0)$ shifts either right or left depending on the value of $T$. While in the low $T$ regime, $R_{min}$ increases by increasing $f_{np}$ (right shift), this tendency reverses as $T$ increases where we can see that for both $T=20K$ and $T=500K$, $R_{min}$ decreases by increasing $f_{np}$ (left shift).

These results are confirmed in Figs 9 (a) and 9 (c) that report, respectively the calculated behavior of $R_{min}$ as a function of $T$ for different values of $f_{np}$ and as a function of $f_{np}$ for different values of $T$. In these figures, around $T_i=10K$ appears to be the inversion temperature region, below which $R_{min}$ increases as a function of $f_{np}$ and above which $R_{min}$ decreases as a function of $f_{np}$. These figures show also that in both $T$ regimes, the rate of increasing or decreasing of $R_{min}$ as a function of $f_{np}$, increases by further decreasing or increasing $T$.

If it is straightforward to explain the behavior of $R_{min}$ as a function of $T$ for a fixed $f_{np}$ due to the interplay between extrinsic and intrinsic phonon scattering processes depending on the $T$ regime, on the other hand understanding of the interesting behavior of $R_{min}$ as function of $f_{np}$ at different $T$ is seemingly not trivial regarding the assumptions made, particularly the linearity between the nanoparticles scattering rate and $f_{np}$. The latter assumption forbids any multiple scattering effects.

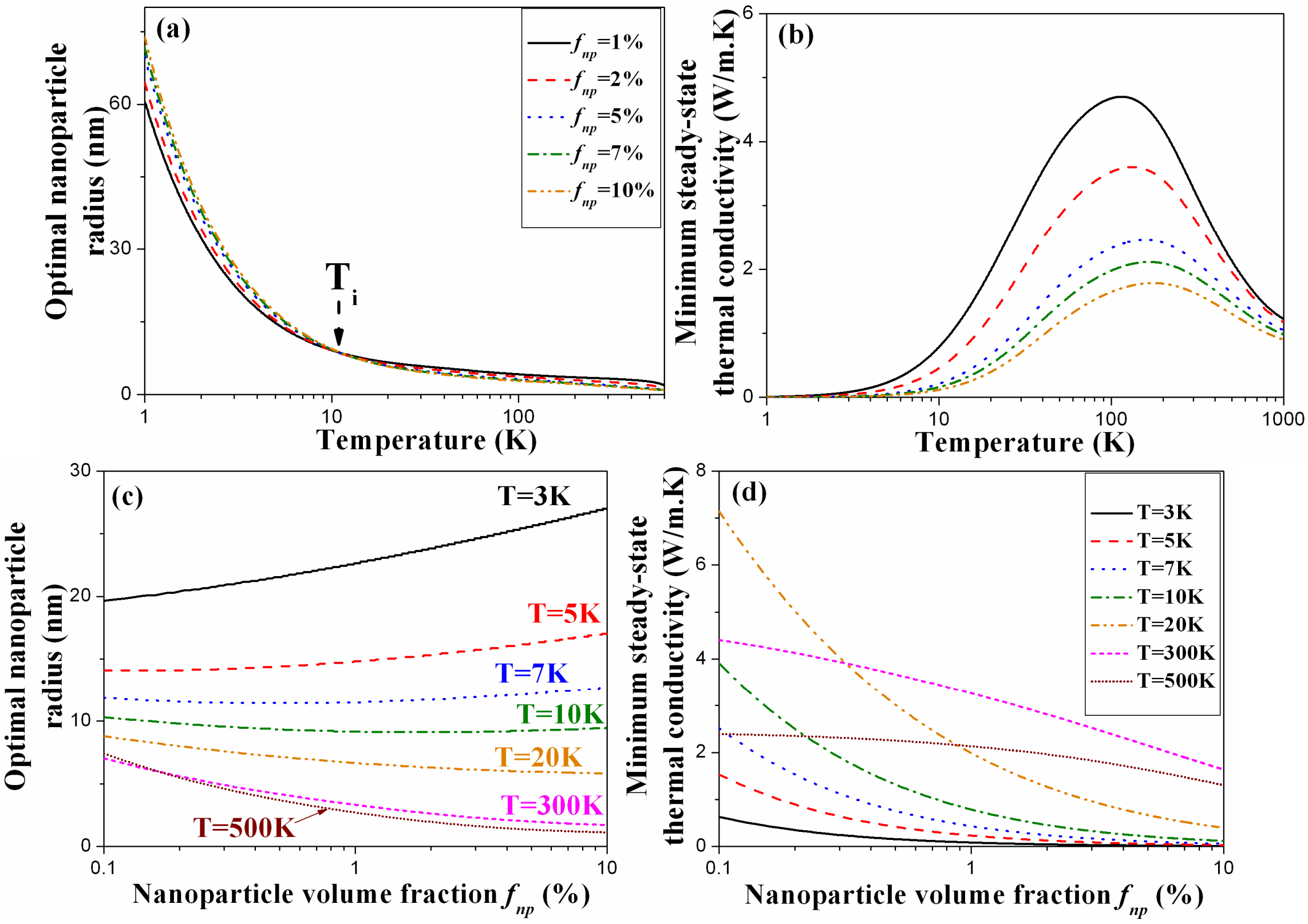

**Figure 9 :** Computed behaviors of the optimal Ge nanoparticles radius $R_{min}$ [(a) and (c)] and the corresponding minimum of $\kappa^{np}(0)$ [(b) and (d)] of $Si_{0.7}Ge_{0.3}$ SC alloy as a function of $T$ and $f_{np}$ for different values of $f_{np}$ and $T$. The arrow in Fig 9(a) points the inversion temperature region at which the behavior of $R_{min}$ as a function of $f_{np}$ switches. This is confirmed in Fig 9(c) (see text for description).

In Figs 9 (b) and 9 (d), we report, the calculated minimum of $\kappa^{np}(0)$ as a function of $T$ for different values of $f_{np}$ and as a function of $f_{np}$ for different values of $T$, respectively. $Min[\kappa^{np}(0)]$ as a function of $T$ manifests a Gaussian-like shape with a peak amplitude (position) that decreases (increases) by increasing $f_{np}$. It is seen also, that the variation around the peak gets wider as $f_{np}$ increases. On the other hand, Fig 9 (d) shows that $Min[\kappa^{np}(0)]$ decreases by increasing $f_{np}$ in a monotonic way that tends to saturate however by increasing the ambient temperature $T$ when embedding nanoparticles becomes less effective.

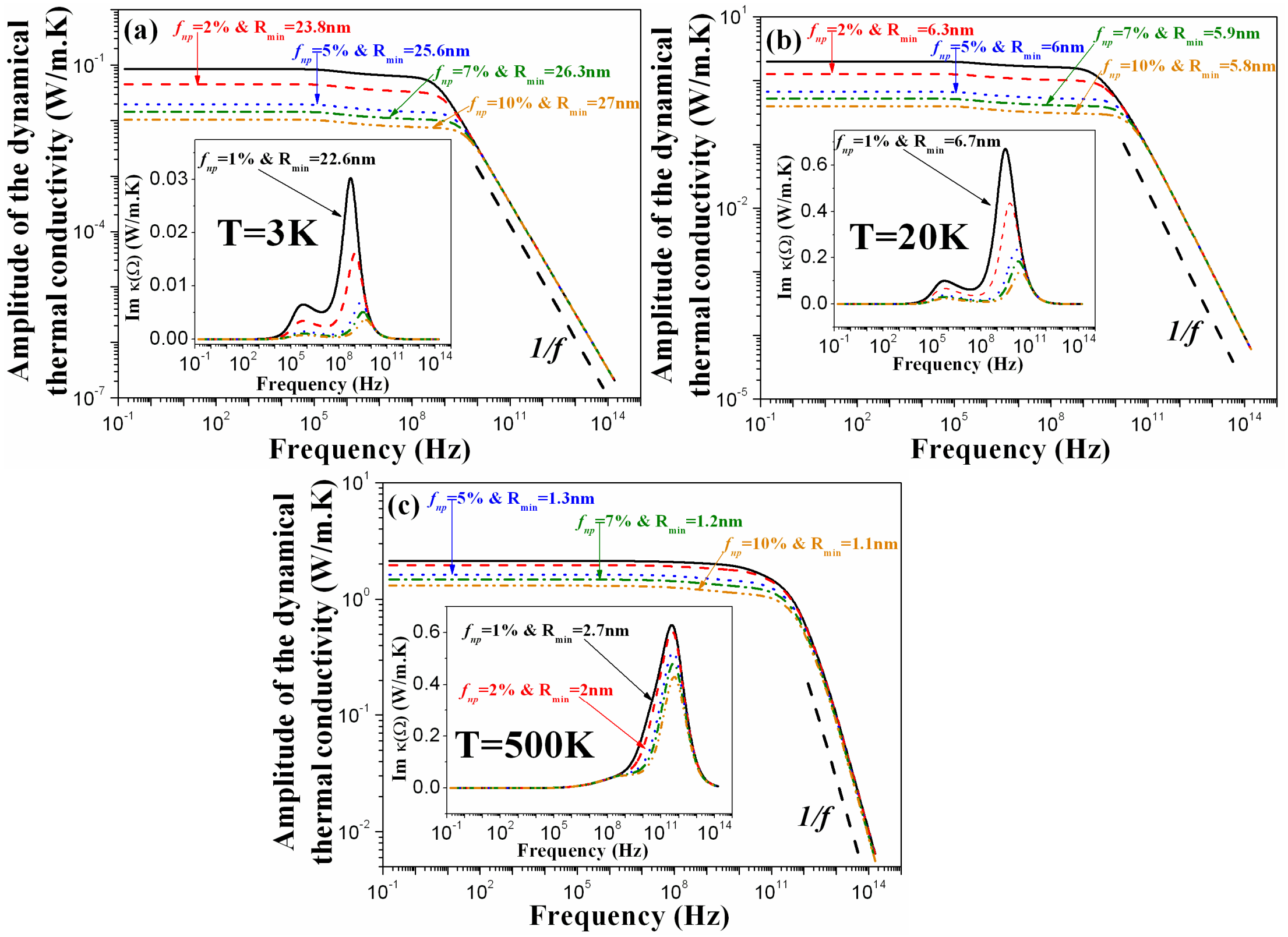

**Figure 10 :** Computed behaviors of the amplitude and imaginary part (inset) of $\kappa^{np}(\Omega)$ of $Si_{0.7}Ge_{0.3}$ SC alloy with embedded Ge nanoparticles as a function of frequency for different values of $f_{np}$ and corresponding $R_{min}$ at: (a) $T=3K$, (b) $T=20K$ and (c) $T=500K$.

The calculated dynamical behaviors of the amplitude and imaginary part of $\kappa^{np}(\Omega)$ of $Si_{0.7}Ge_{0.3}$ SC alloy matrix with embedded Ge nanoparticles are reported in Figs 10 (a), (b) and (c) at $T=3K$, $T=20K$ and $T=500K$, respectively, for different values of $f_{np}$ and corresponding values of $R_{min}$ that lead to $Min[\kappa^{np}(0)]$ [Figs 8 and 9]. While in the low frequency regime, the amplitude of $\kappa^{np}(\Omega)$ decreases by increasing $f_{np}$ as expected, all the curves seem to collapse in the high frequency regime where the $f^{-1}$ power law continues to be valid. The dynamical behavior of the imaginary part of $\kappa^{np}(\Omega)$ is reported in the insets of Figs 10. The imaginary part manifests a double resonance peaks; a primary one at the right and a secondary one at the left. In the high $T$ regime where scattering of phonons by the nanoparticles becomes less significant, the secondary peak disappears and only remains the primary one, the position of which shifts to the right by increasing $f_{np}$. In the low $T$ regime, the amplitude of both peaks decreases by increasing $f_{np}$, while their positions seem to shift to the right also.

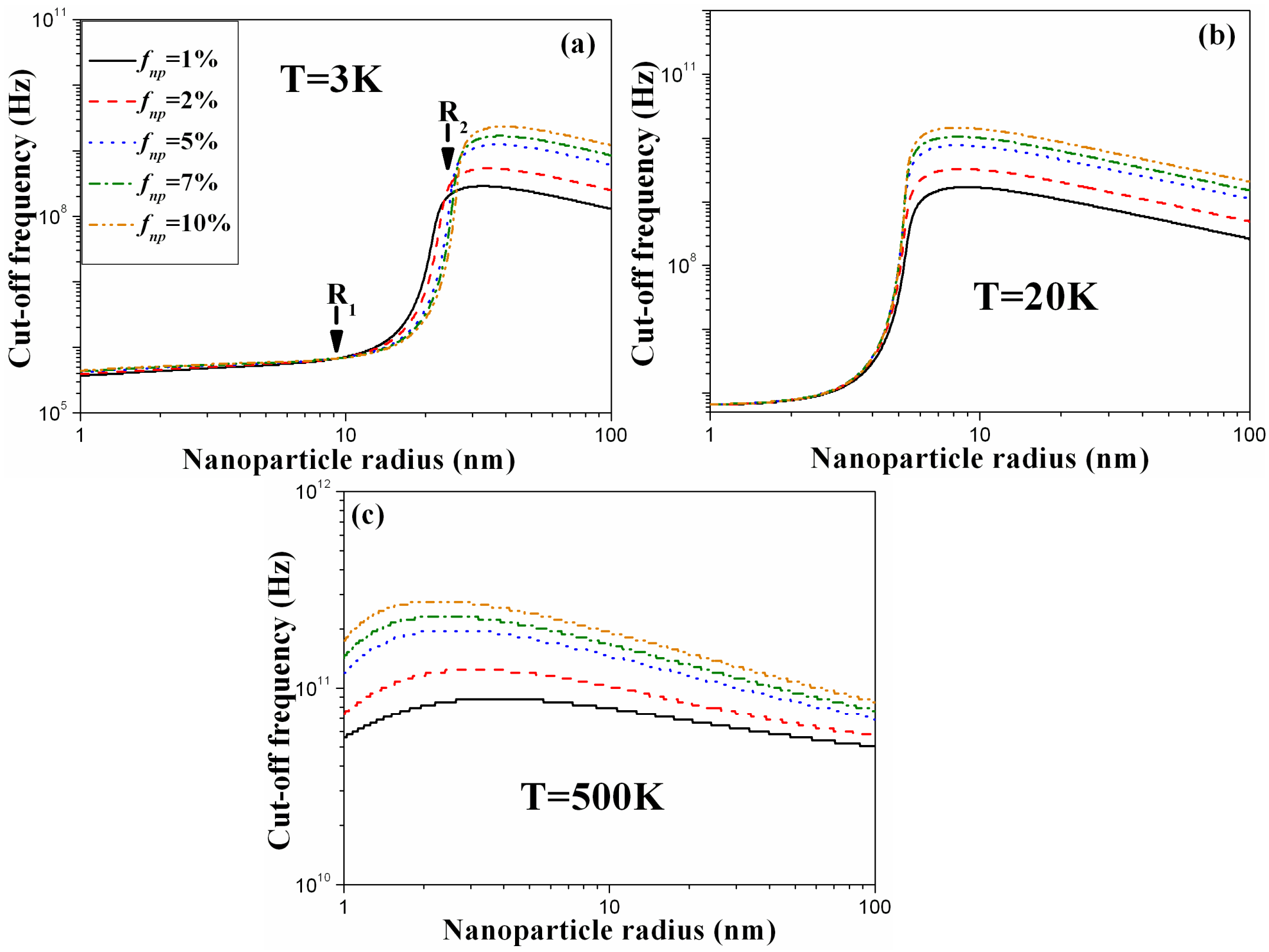

**Figure 11 :** Computed behaviors of $f_C$ of $\kappa^{np}(\Omega)$ of $Si_{0.7}Ge_{0.3}$ SC alloy with embedded Ge nanoparticles as a function of $R$ for different values of $f_{np}$ at: (a) $T=3K$, (b) $T=20K$ and (c) $T=500K$. The arrows on Fig 11(a) point the two threshold values of $R$ at which the behavior of $f_C$ as function of $f_{np}$ switches (see text for description).

The last figure relative to the effect of changing $f_{np}$ on the behavior of $\kappa^{np}(\Omega)$ of $Si_{0.7}Ge_{0.3}$ SC alloy matrix with embedded Ge nanoparticles is Fig 11. The latter reports the behavior of the cut-off frequency $f_C$ of $\kappa^{np}(\Omega)$ as a function of the nanoparticles radius $R$ for different values of $f_{np}$ and at different ambient temperatures $T=3K$ [Fig 11 (a)], $T=20K$ [Fig 11 (b)] and $T=500K$ [Fig 11 (c)]. These figures show the same interesting behaviors as the ones discussed in Fig 7 (d). In addition, we can see in the low $T$ regime [Fig 11(a)] that the behavior of $f_C$ as function of $f_{np}$ goes through three different regimes that correspond somehow to the previously discussed three regimes of the behavior of $f_C$ as a function of $R$ [Fig 7 (d)]. $f_C$ seems to increase a little by increasing $f_{np}$ for values of $R$ smaller than the lowest threshold value $R_1$, then the tendency reverses when $R$ becomes between $R_1$ and $R_2$. Once $R$ becomes higher than $R_2$, $f_C$ returns to its first tendency and starts to increase as $f_{np}$ increases. Analyzing Figs 11 shows also that by increasing $f_{np}$, the position of the highest threshold value $R_2$ shifts to the right at low $T$ while it tends to shift to the left at high $T$. This behavior is deeply connected to the interesting behavior of $R_{min}$ of steady-state $\kappa^{np}(0)$ as function of $f_{np}$ for different ambient temperatures.

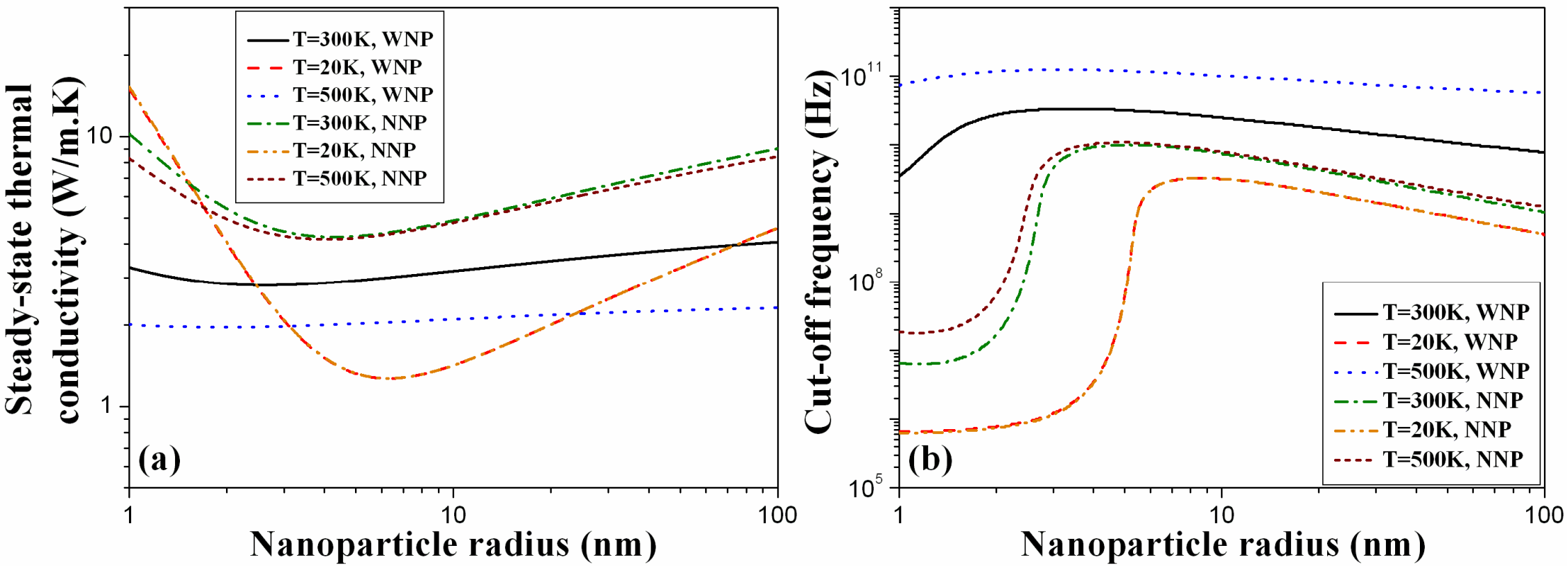

**Figure 12 :** (a) Computed behaviors of $\kappa^{np}(0)$ of $Si_{0.7}Ge_{0.3}$ SC alloy with embedded Ge nanoparticles as a function of the nanoparticles radius *R* for different *T*. (b) Computed behaviors of $f_C$ of $\kappa^{np}(\Omega)$ of $Si_{0.7}Ge_{0.3}$ SC alloy with embedded Ge nanoparticles as a function of *R* for different *T*. In both figures $f_{np}=2\%$ and we consider two cases; with (WNP) and without (NNP) three-phonon scattering N-processes.

This intriguing, yet very interesting behavior, of $f_C$ as function of embedded nanoparticles radius *R* and volume fraction $f_{np}$, shows clearly how sensitive the dynamical behavior of $\kappa^{np}(\Omega)$ could be to the nanoparticles size and concentration. It shows that a crucial and careful manipulation of the nanoparticles size and concentration is needed in order to adjust $f_C$ to the expected range, especially for nanoparticles engineered and grown with sizes that surround the optimal size leading to the minimum value of $\kappa^{np}(0)$. Generally speaking, while embedding nanoparticles in SC alloy matrices reduces the steady-state $\kappa^{np}(0)$, it however tends to increase $f_C$ of the dynamical $\kappa^{np}(\Omega)$ in comparison with intrinsic SC alloys. This opposite double effect could be very beneficial in many microelectronic and optoelectronic devices.

To end this subsection, we report respectively in Figs 12(a) and 12(b) a comparison between the calculated steady-state $\kappa^{np}(0)$ and the calculated $f_C$ of the dynamical $\kappa^{np}(\Omega)$ of $Si_{0.7}Ge_{0.3}$ SC alloy matrix as a function of the embedded Ge nanoparticles radius *R* at different ambient temperatures *T*, in two cases where three-phonon scattering N-processes are considered or disregarded. As these figures show, the effect of including N-processes is insignificant in the low *T* regime where phonon scattering is dominated by extrinsic processes, the calculated curves totally superimpose. On the other hand, in the high T regime, where anharmonic phonon-phonon scattering processes dominate, it is crucial to include N-processes. By disregarding N-processes, both values of the optimal $R_{min}$ and the minimum $\kappa^{np}(0)$ increase, while the value of the cut-off frequency $f_C$ of $\kappa^{np}(\Omega)$ decreases. Depending on the nanoparticle size, a change of few factors can be expected in $\kappa^{np}(0)$, while $f_C$ can vary by one to few orders of magnitude.

### 3. Koh and Cahill experimental results

In a very recent experiment, Koh and Cahill[53] published very interesting and intriguing experimental results of the frequency behavior of the thermal conductivity of SC alloy crystals at different ambient temperatures. As we mentioned in the introduction, Koh and Cahill's results over the frequency range used in their experiments [*0.6-10MHz*], show $\kappa(\Omega)$ of SC alloy crystals to have a cut-off frequency $f_C<10MHz$ while $\kappa(\Omega)$ of single SC crystals manifested a plateau at room temperature. The authors used BPTE in the steady-state regime to explain their results in which they translate *Koh and Cahill statement* (see introduction) as a boundary scattering process that phonons would undergo at a virtual interface. This virtual interface is actually the surface of a hemisphere whose radius is the thermal penetration depth $\delta(\Omega)$. The authors found a good and satisfactory agreement between experimental data and the results of this phenomenological approach.

Even though their model seems to agree with the experimental data, but it cannot be considered as a relevant and robust explanation of the behavior of $\kappa(\Omega)$ of SC crystals and that for two reasons: (i) the thermal penetration depth $\delta$ is a characteristic depth of the applied heat source at the surface of the crystal which is calculated based on the knowledge of the thermal diffusivity (the latter property itself is related to the thermal conductivity of the given crystal) and the frequency of the heat source, and then cannot be considered as a limit for the phonon MFP inside the crystal, this raises a consistency question. (ii) There is no physically plausible reason to consider scattering of phonons at a fictitious interface.

Koh and Cahill experimental results are very low compared to the predicted values of $f_C$ based on the above modeling of $\kappa(\Omega)$. Furthermore, a value of $f_C=10MHz$ at room temperature is equivalent to a relaxation time of the dominant phonon scattering process of $\tau\sim 16ns$, this is a very large value and it is hard to be accepted physically; besides, it is difficult to admit that the dynamical $\kappa(\Omega)$ of bulk SC alloys will undergo such a huge reduction on a small frequency range. The most reasonable explanation to the measured low value of $f_C$ in SC crystal alloys in Koh and Cahill's experiment[53] would be related to the physical meaning of the measured thermal conductivity $\kappa$ itself and the role of the cumulative effect of the laser train pulses in the experimental setup and thermal modeling assumptions used to extract $\kappa$ especially at high frequency of the excitation source. The recent experiments of Minnich et al,[72,73] take the same reasoning direction and come to support this conclusion, where the authors used again TDTR technique to study quasiballistic heat transport and to measure phonon MFP. The results of the authors show the measured thermal conductivity $\kappa$ to depend on the size of the heating laser pump source spot and it is independent of the modulation frequency of the latter.

Koh and Cahill experimental results are still intriguing and deserve much more investigation. It is more plausible that, in the experiment, the authors measured an *apparent thermal conductivity* $\kappa_{app}$ that matches the real intrinsic $\kappa$ at low frequency but deviates from it as the frequency of the excitation source gets higher due mainly to the cumulative effect of the TDTR experimental set-up and thermal modeling assumptions.[54-56] We believe these two factors need further study to separate their effect from the measured $\kappa_{app}$ in order to have access to the real intrinsic thermal conductivity of the studied dielectric crystal material.

### 4. Shastry's Sum Rule

In this last subsection, we will shed light on a very interesting result that has been recently introduced by Shastry,[51] namely the sum-rule for the real part of the dynamical thermal conductivity. Shastry derived this sum rule for several standard models of current interest in condensed matter. The sum rule is obtained using standard linear response theory and is expressed in terms of the expectation of an extensive object $\theta^{xx}$ that Shastry named "*thermal operator*" in his formalism.[51] As discussed by Shastry, the sum rule is closely related to the behavior of energy (heat) transport in the ballistic regime where the expectation of $\theta^{xx}$ is an equilibrium value that determines the magnitude of the ballistic force exerted by the applied temperature field.[52] Here we study the applicability of Shastry's sum rule (SSR) and we develop rather "*classical*" expressions for it as well as the expectation of $\theta^{xx}$ in the case of the dynamical lattice thermal conductivity $\kappa(\Omega)$ of bulk SC crystals as calculated based on BPTE for phonon transport.

According to the calculations developed earlier in the theory section, we can calculate the integral of the real part of $\kappa(\Omega)$. We will consider two cases of anharmonic phonon-phonon scattering processes; (i) both Normal and Umklapp processes are included and (ii) only Umklapp process is active. In both cases of course, in addition to the above mentioned phonons scattering processes, all other total phonon crystal momentum destroying processes are taken into account. The integral is taking over all frequencies from zero to infinity. In the first case [Eqs. (13-16)], one obtains:

$$\int_0^{+\infty}\kappa_r(\Omega)\,d\Omega=\sum_S\int\kappa_{q,S}^0\left[\int_0^{+\infty}\frac{d\Omega}{1+\left(\Omega\tau_{q,S}^C\right)^2}\right]d^3\boldsymbol{q}$$
$$=\frac{\pi}{2}\sum_S\int\frac{\kappa_{q,S}^0}{\tau_{q,S}^C}d^3\boldsymbol{q}=\frac{1}{48\pi^2}\sum_S\int\left[1+\frac{\beta_S}{\tau_{q,S}^N}\right]v_S^2C_{Ph}(\boldsymbol{q},S)\,d^3\boldsymbol{q}\qquad(37)$$

Since the resulting function depends on phonons scattering rates due to the term $\beta_S/\tau_{q,S}^N$ (effect of N-processes), it cannot be viewed as a sum rule; as explained by Shastry,[51] the thermal operator $\theta^{xx}$ does not contain any scattering rate. On the other hand, when N-processes are disregarded and only resistive processes are considered, we can easily show that the calculation of the integral of the real part of $\kappa(\Omega)$ gives:

$$\int_0^{+\infty}\kappa_r(\Omega)\,d\Omega=\frac{1}{8\pi^3}\sum_S\int\tau_{q,S}^R v_{S,\iota}^2C_{Ph}(\boldsymbol{q},S)\left[\int_0^{+\infty}\frac{d\Omega}{1+\left(\Omega\tau_{q,S}^R\right)^2}\right]d^3\boldsymbol{q}$$
$$=\frac{1}{16\pi^2}\sum_S\int v_{S,\iota}^2C_{Ph}(\boldsymbol{q},S)\,d^3\boldsymbol{q}=\frac{\pi}{6}\sum_S\int_0^{\omega_D^S}v_S^2C_{Ph}(\omega,S)\,g_S(\omega)\,d\omega\qquad(38)$$

where $g_S(\omega)=\dfrac{\omega^2}{2\pi^2v_S^3}$ is the Debye density of states of phonons in the acoustic polarization branch $S$.[59] This last expression shows that the resulting function is independent of any phonons scattering rate and hence can be viewed as a sum rule. Comparison of Eq. (38) to Eq. (17) from Shastry's paper[51] implies that the classical expression of the expectation of the thermal operator $\theta^{xx}$ can be written as:

$$\frac{1}{\hbar}\left\langle\theta^{xx}\right\rangle_{Classical}^{BPTE}=\frac{2T_0W}{\pi}\int_0^{+\infty}\kappa_r(\Omega)\,d\Omega=\frac{T_0W}{3}\sum_S\int_0^{\omega_D^S}v_S^2C_{Ph}(\omega,S)\,g_S(\omega)\,d\omega\qquad(39)$$

where $W$ denotes the total volume of the SC crystal material. Then using the usual change of variable $x=\hbar\omega/k_BT_0$, one can further write Eq. (39) in a more convenient form for numerical calculation, this gives:

$$\left\langle \theta^{xx} \right\rangle_{Classical}^{BPTE} = \frac{k_B^4}{\pi^2 \hbar^2} W T_0^4 \left\{ \frac{1}{6 v_L} \int_0^{\theta_D^L / T_0} D(x)\,dx + \frac{1}{3 v_T} \int_0^{\theta_D^T / T_0} D(x)\,dx \right\} \qquad (40)$$

where again *D(x)* denotes Debye function.

The expression of $\left\langle \theta^{xx} \right\rangle_{Classical}^{BPTE}$ as given by Eq. (40) presents the remarkable feature of vanishing at *T=0* (i.e in the true thermodynamic ground state). This finding corroborates the arguments of Shastry who discussed this fundamental behavior and showed its deep connection to the vanishing of the lattice specific heat.[51] As a matter of fact, starting from the definition of the latter thermodynamic property, in our case the lattice specific heat at constant volume $C_W$, it is straightforward to show that it has the following expression:[22,32,59]

$$\begin{cases} U = \frac{1}{W} \sum_{S,q} \hbar \omega_{q,S} \left( n_{q,S} + \frac{1}{2} \right) \\ C_W(T) = \left. \frac{\partial U}{\partial T} \right|_{W=cte} = \frac{1}{W} \sum_{S,q} \hbar \omega_{q,S} \frac{\partial n_{q,S}}{\partial T} \\ \frac{\partial n_{q,S}}{\partial T} \cong \frac{\partial n_{q,S}^0}{\partial T_0} \\ C_W(T_0) = \frac{1}{8\pi^3} \sum_S \int C_{Ph}(\boldsymbol{q},S)\, d^3\boldsymbol{q} = \sum_S \int_0^{\omega_D^S} C_{Ph}(\omega,S)\, g_S(\omega)\, d\omega \\ C_W(T_0) = \frac{k_B^4}{2\pi^2 \hbar^3} T_0^3 \left\{ \frac{1}{v_L^3} \int_0^{\theta_D^L / T_0} D(x)\,dx + \frac{2}{v_T^3} \int_0^{\theta_D^T / T_0} D(x)\,dx \right\} \end{cases} \qquad (41)$$

Comparison of Eqs. (40) and (41) shows clearly the connection between $\left\langle \theta^{xx} \right\rangle_{Classical}^{BPTE}$ and $C_W$. As a matter of fact, one can write:

$$\begin{cases} \frac{\left\langle \theta^{xx} \right\rangle_{Classical}^{BPTE}}{\hbar T_0 W} = C_W v_{eff}^2 \\ v_{eff}^2 = \frac{1}{3} \frac{\left\{ \frac{1}{v_L} \int_0^{\theta_D^L / T_0} D(x)\,dx + \frac{2}{v_T} \int_0^{\theta_D^T / T_0} D(x)\,dx \right\}}{\left\{ \frac{1}{v_L^3} \int_0^{\theta_D^L / T_0} D(x)\,dx + \frac{2}{v_T^3} \int_0^{\theta_D^T / T_0} D(x)\,dx \right\}} \end{cases} \qquad (42)$$

where $v_{eff}$ represents an effective phonon group velocity averaged over all phonon acoustic polarization branches.

The fact that $\left\langle \theta^{xx} \right\rangle_{Classical}^{BPTE}$ and $C_W$ can be related by such a compact formula [Eq. (42)] that is similar to the one suggested by Shastry [Eq. (91) in Ref 51] proves somehow the meaningfulness of the classical limit of the expectation of the thermal operator $\theta^{xx}$ and confirms the physical meaning of the latter variable in capturing the ballistic dynamics aspect in the energy (heat) transport phenomenon.

It is interesting to note that the expression of $v_{eff}$ as defined by Eq. (42) tends to the conventional driftless expression of the second sound in dielectric solids in the limit of very low temperature where the upper bound in both longitudinal and transverse integrals tends to infinity.[57] Therefore, Eq.(42) can be considered as a kind of generalization form of the second sound velocity in bulk dielectric and SC crystals. This consolidates even more the physical meaning of the expectation of the thermal operator $\theta^{xx}$.

## 5. Heuristic treatment of second sound in bulk SC crystals

Since the first speculations[74-77] regarding the occurrence of second sound phenomenon in anharmonic dielectric crystals, many authors have devoted considerable efforts attempting different theoretical approaches in order to study and understand the conditions of existence of the phenomenon and under which it can be observed experimentally.[45-47,57,78-83] Few years later the first experimental observations were made available and second sound phenomenon was definitely proven to be an intrinsic property of the dielectric solid independent of how the latter is excited.[84-87] Therefore second sound in dielectric solids was confirmed both theoretically and experimentally to be the same physical phenomenon that was first observed in liquid $He^2$.[45-47,57,74-87] As such, the term *"second sound"* is used to describe a hydrodynamic collective mode involving a coherent mixture of different phonons in an interacting phonon gas.[57] Alternatively, "second sound" could be looked at as describing the coherent propagation of energy (heat) fluctuations in a phonon gas in analogy to first sound which describes the coherent propagation of density (pressure) fluctuations in a particle gas.[45-47,78]

Very comprehensive reviews of the theory of second sound in dielectric crystals can be found in the review papers by Griffin[57] and Enz.[81] As mentioned by Hardy,[82] discussions about second sound lie in the framework of two general approaches. (i) The first one is Boltzmann Peierls Transport Equation (BPTE) based on which one usually couples two continuity equations for energy balance and crystal-momentum balance, respectively. The effect obtained has been called *"drifting"* second sound.[81] The second approach is based on linear theory and has led to a different effect called *"driftless"* second sound.[57] The theoretical derivations of these two types of second sound are quite convincing and well established which then was looked at as a real puzzle. Enz[81] emphasized the discussion about these two types of second sounds and came to the conclusion that the two types exist in principle but are encountered under different conditions distinguished by the frequency of excitation and the temperature. According to Enz,[81] *"drifting"* second sound must be associated with the neighboring of the steady-state thermal conductivity peak where *"nonballistic convection"* is the dominant heat transport whereas *"driftless"* second sound would have to be sought above the peak where *"conduction"* dominates. In his very interesting paper,[82] Hardy studied the conditions of existence of these two types of second sounds in dielectric solids and obtained the velocity of drifting and driftless second sounds, both from the same linearized BPTE.

In one of the first investigations of second sound in solids using BPTE, Guyer and Krumhansl[45] considered solving the latter equation within the framework of the single relaxation time approximation and using Callaway approximated form of the collision operator. Starting from an exact solution of the BPTE using the trajectory integral method of Chamber, the authors were able to obtain a dispersion relation for second sound that explicitly exhibits the need for a *"window"* in the relaxation time spectrum. The analysis of Guyer and Krumhansl was made in two different stages; in the first one, all phonon scattering relaxation times were taken constant independent of the phonon wave vector and in the second stage, the authors considered a sort of average relaxation times as described in the appendix of the paper. In both cases, the second sound dispersion relation was obtained with similar damping terms.[45]

We suggest here to use rather different simplified approach than Guyer and Krumhansl to deal with BPTE using Callaway collision operator in the framework of modified Debye-Callaway model in which both longitudinal and transverse phonon modes are explicitly included (see theory section). Our study focuses on bulk SC crystals for which we derive the dispersion relation of second sound using the self-consistency criterion of Griffin.[57,80] Another less justified approach using Hardy's method[82] will be detailed in appendix D. Surprisingly, this latter method gives similar results to the ones we present in the following.

The starting point of our theoretical approach is the expression of the dynamical lattice thermal conductivity $\kappa(\Omega)$ of bulk SC crystals that we just derived in the theory section (section 2). The possibility of unforced propagation of temperature waves is expressed according to Griffin's self-consistency criterion[57] which leads directly to the second sound dispersion relation that describes the way the frequency $\Omega$ of a temperature wave depends on its propagation vector $Q$.

$$\left(\frac{\Omega}{Q}\right)^2 = -j\Omega\left[\frac{\kappa(\Omega)}{C_W}\right] = \Omega\frac{\kappa_i(\Omega)}{C_W} - j\Omega\frac{\kappa_r(\Omega)}{C_W} \qquad (43)$$

where $\kappa_r$ and $\kappa_i$ are the real and imaginary parts of $\kappa(\Omega)$. The square of the propagation velocity of second sound is given by the real part of the dispersion relation.[57] Assuming the timing relation $\Omega\tau_{q,S}^C \gg 1$ to be valid for each normal phonon state $(\boldsymbol{q}, S)$ (according to Griffin, this relation is indeed the condition for quasi-particle approximation to be good in the first place.[57]) and using the expression of $\kappa_i(\Omega)$ as given by Eq. (16), one finds:

$$\left(v_{II}^G\right)^2 = \mathrm{Re}\left\{\left(\frac{\Omega}{Q}\right)^2\right\} = \frac{1}{C_W}\sum_S\int\frac{\kappa_{q,S}^0}{\tau_{q,S}^C}d^3\boldsymbol{q} \qquad (44)$$

A more manageable expression of $v_{II}^G$ can be obtained by taking into account above equations given $\kappa_{q,S}^0$ [Eq. (14)] and $C_W$ [Eq. (41)] and using the usual change of variable $x = \hbar\omega/k_BT_0$, one obtains the following expression:

$$\begin{cases}\left(v_{II}^G\right)^2 = \dfrac{1}{3}\dfrac{\sum_S \dfrac{1}{v_S}\left[a_S + \beta_S b_S\right]}{\sum_S \dfrac{1}{v_S^3}a_S} \\ a_S = \displaystyle\int_0^{\theta_D^S/T_0} D(x)\,dx;\; b_S = \int_0^{\theta_D^S/T_0}\frac{D(x)}{\tau_S^N(x)}dx\end{cases} \qquad (45)$$

The Callaway pseudo-relaxation time $\beta_S$ is as given by Eq. (8). The second sound relaxation time is given according to the relation $\tau_{II}^G = \kappa(0)\Big/\left[C_W\left(v_{II}^G\right)^2\right]$,[82] with $\kappa(0)$ representing the steady-state lattice thermal conductivity. This leads to:

$$\tau_{II}^{G}=\frac{\sum_{S}\int \kappa_{q,S}^{0}d^{3}\boldsymbol{q}}{\sum_{S}\int \frac{\kappa_{q,S}^{0}}{\tau_{q,S}^{C}}d^{3}\boldsymbol{q}} \qquad (46)$$

which after simplifications using the standard change of variable $x=\hbar\omega/k_{B}T_{0}$, one obtains:

$$\begin{cases} \tau_{II}^{G}=\dfrac{\sum_{S}\dfrac{1}{v_{S}}\left[c_{S}+\beta_{S}d_{S}\right]}{\sum_{S}\dfrac{1}{v_{S}}\left[a_{S}+\beta_{S}b_{S}\right]} \\ c_{S}=\displaystyle\int_{0}^{\theta_{D}^{S}/T_{0}}\tau_{S}^{C}(x)D(x)dx;\ d_{S}=\displaystyle\int_{0}^{\theta_{D}^{S}/T_{0}}\frac{\tau_{S}^{C}(x)}{\tau_{S}^{N}(x)}D(x)dx \end{cases} \qquad (47)$$

The square of second sound damping factor is given by the imaginary part of the dispersion relation [Eq. (43)]. Without considering the timing relation $\Omega\tau_{q,S}^{C}\gg 1$ at this stage and using the expression of $\kappa_{r}(\Omega)$ from Eq. (16), it follows:

$$\begin{cases} \left(\Gamma_{II}^{G}\right)^{2}=-\mathrm{Im}\left\{\left(\dfrac{\Omega}{Q}\right)^{2}\right\}=\dfrac{1}{C_{W}}\sum_{S}\displaystyle\int\frac{\Omega\kappa_{q,S}^{0}}{1+\left(\Omega\tau_{q,S}^{C}\right)^{2}}d^{3}\boldsymbol{q} \\ =\dfrac{1}{8\pi^{3}C_{W}}\sum_{S}\displaystyle\int\frac{\Omega\tau_{q,S}^{C}\left(1+\dfrac{\beta_{S}}{\tau_{q,S}^{N}}\right)}{1+\left(\Omega\tau_{q,S}^{C}\right)^{2}}v_{S,t}^{2}C_{Ph}(q,S)d^{3}\boldsymbol{q} \\ =\dfrac{1}{8\pi^{3}C_{W}}\sum_{S}\displaystyle\int\frac{v_{S,t}^{2}C_{Ph}(q,S)}{\dfrac{1}{\Omega\tau_{q,S}^{C}\left(1+\dfrac{\beta_{S}}{\tau_{q,S}^{N}}\right)}+\Omega\dfrac{\tau_{q,S}^{C}}{\left(1+\dfrac{\beta_{S}}{\tau_{q,S}^{N}}\right)}}d^{3}\boldsymbol{q} \end{cases} \qquad (48)$$

The second line in Eq. (48) is obtained using the expression of $\kappa_{q,S}^{0}$ [Eq. (14)].

As one can see, the damping term is a complicated function of the relaxation times of all phonon scattering processes; N-processes as well as all resistive processes for each normal phonon state *(**q**, S)*. In addition, one can even guess the existence of a window in the relaxation time spectrum. One can obtain an irrefutable indication about it by assuming the grey spectrum approximation (GSA). In fact, under this approximation, Eq. (48) becomes:

$$\begin{cases} \left(\Gamma_{II}^{G}\right)^2 = \dfrac{\sum_{S}\int v_S^2 C_{Ph}(q,S)\, d^3\boldsymbol{q}}{24\pi^3 C_W} \times \dfrac{\Omega\tau_R}{1+\left(\Omega\tau_C\right)^2} \\ \dfrac{\Omega\tau_R}{1+\left(\Omega\tau_C\right)^2} = \dfrac{1}{\dfrac{1}{\Omega\tau_R}+\Omega\dfrac{\tau_C^2}{\tau_R}} = \left[\dfrac{1}{\Omega\tau_R}+\Omega\dfrac{\tau_N^2\tau_R}{\left(\tau_N+\tau_R\right)^2}\right]^{-1} \end{cases} \quad (49)$$

To obtain Eq. (49), we used as before, the isotropy of the group velocity in the real and reciprocal spaces $v_{S,q_x}^2 = v_{S,q_y}^2 = v_{S,q_z}^2 = \frac{1}{3} v_S^2$ and the expression of the effective wave vector independent combined relaxation time $\tau_C^{-1} = \tau_R^{-1} + \tau_N^{-1}$. Eq. (49) is very similar to the expression obtained by Guyer and Krumhansl (see the result after Eq. (62) in Ref [46]) with the difference that we have a term $\Omega\tau_N^2\tau_R / \left(\tau_N+\tau_R\right)^2$ instead of just $\Omega\tau_N$. Following the discussion of Guyer and Krumhansl,[46] this will indeed constitutes a very solid argument to assert of the existence of a window in the relaxation time spectrum for the damping factor $\Gamma_{II}^{G}$, as given by Eq. (48), to be small and as such for the second sound to exist and propagate into the bulk SC crystal.

As a matter of fact, the existence of this time window has very fundamental physical roots as it describes the transition from ballistic heat transfer to second sound to diffusive heat transfer. The experimental observation of this transition has been made early in the 1970 especially in the work of Narayanamurti, Varma and Dynes.[85-87] The time window can be established based on very fundamental physical grounds by taking into consideration the condition of establishment of the local thermal equilibrium necessary for the concept of temperature and as such second sound, as this describes temperature wave propagation, to have a meaning. The thermalization process of the phonon system in a dielectric solid is intimately linked to the relation between anharmonic normal scattering processes and all resistive (R) scattering processes. Whether this thermalization is reached due to dominant N-processes (momentum conserving processes) or R-processes (momentum destroying processes) will lead totally to different thermal phenomena.[46,47] As we have seen, through their role of shuffling the total phonon crystal momentum between different phonon states, N-processes contribute implicitly to the thermal conductivity, but can't by themselves lead to a finite thermal conductivity.[21,24] When N-processes characterizes the thermalization of the phonon system, one then can speak of a temperature after a time $\tau_N$ (average relaxation time due to N-processes over all phonons wave vectors and polarization branches), the total phonon crystal momentum is redistributed between different phonon states, and the phonon gas can follow the input temperature disturbance and therefore the possibility of a wave-like propagation of temperature in the heat transport regime before the latter transition to a fully diffusive regime when R-processes become dominant. On the other hand, if R-processes characterizes the thermalization of the phonon system, after a time $\tau_R$ (average relaxation time due to R-processes over all phonons wave vectors and polarization branches), the total phonon crystal momentum starts to relax and the heat transport flux to decay, which leads directly to a diffusive regime of heat transport and there can't be any wave-like propagation of temperature. As a consequence, second sound is expected to occur more likely when N-processes dominant R-processes.

In the GSA, one can formally write the time window as:

$$\tau_R^{-1} < \Omega < \left(\tau_N^{eff}\right)^{-1} = \left(\tau_N + \tau_R\right)^2 / \tau_N^2 \tau_R \qquad (50)$$

We should note here that this form would still be correct if instead of the GSA, one considers averaged expressions of $\tau_N^{eff}$ and $\tau_R^{eff}$ over all phonon wave vectors and polarization branches in a Mathienssen's like fashion. In fact, starting from Eq. (48) and in order to preserve the limit of the GSA, one might be tempted to write the time window in a more general conceptual form as:

$$\begin{cases} \left(\tau_R^{eff}\right)^{-1} < \Omega < \left(\tau_N^{eff}\right)^{-1} \\ \tau_R^{eff} = \dfrac{\sum_S \int \tau_{q,S}^C \left(1 + \dfrac{\beta_S}{\tau_{q,S}^N}\right) v_S^2 C_{Ph}\left(q,S\right) d^3\boldsymbol{q}}{\sum_S \int v_S^2 C_{Ph}\left(q,S\right) d^3\boldsymbol{q}} \\ \left(\tau_N^{eff}\right)^{-1} = \dfrac{\sum_S \int \dfrac{\left(1 + \dfrac{\beta_S}{\tau_{q,S}^N}\right)}{\tau_{q,S}^C} v_S^2 C_{Ph}\left(q,S\right) d^3\boldsymbol{q}}{\sum_S \int v_S^2 C_{Ph}\left(q,S\right) d^3\boldsymbol{q}} \end{cases} \qquad (51)$$

One can easily check that the time window as given by Eq. (51) tends to the one given by Eq. (50) in the GSA limit. This might be not the only way to define such averaging expressions for the second sound time window. But the latter form has the advantage to capture the interplay between anharmonic phonon-phonon N-processes and all resistive processes in a very elegant and simple fashion. Furthermore, this expression does not present any contradiction with the first timing relation $\Omega\tau_{q,S}^{C} \gg \mathbf{1}$.

Eqs. (50) and (51) state the need for a critical frequency for the onset of second sound thermal phenomenon. This frequency has to be greater than the highest frequency for the onset of thermal diffusion, but smaller than the lowest frequency for the establishment of local thermal equilibrium.[77] These equations show also that both N and R processes would contribute to the establishment of local thermal equilibrium. In addition, one can easily check that the condition $\tau_R^{-1} < \left(\tau_N + \tau_R\right)^2 / \tau_N^2 \tau_R$ in the GSA and generally the condition $\left(\tau_R^{eff}\right)^{-1} < \left(\tau_N^{eff}\right)^{-1}$ are always true, but regarding the needed averaging processes to obtain $\tau_N^{eff}$ and $\tau_R^{eff}$, Eqs. (50) and (51) are more likely to be legitimate in the low temperature regime.[45-47,78]

As a matter of fact, the expressions of $\left(\tau_R^{eff}\right)^{-1}$ and $\left(\tau_N^{eff}\right)^{-1}$ can be further simplified using the standard change of variable $x = \hbar\omega / k_B T_0$. By doing so, we obtain:

$$\begin{cases} \tau_R^{eff} = \dfrac{\sum_S \dfrac{\mathbf{1}}{v_S}\left[c_S + \beta_S d_S\right]}{\sum_S \dfrac{\mathbf{1}}{v_S} a_S} \\ \tau_N^{eff} = \dfrac{\sum_S \dfrac{\mathbf{1}}{v_S} a_S}{\sum_S \dfrac{\mathbf{1}}{v_S}\left[e_S + \beta_S f_S\right]} \\ e_S = \displaystyle\int_{\mathbf{0}}^{\theta_D^S / T_\mathbf{0}} \frac{D(x)}{\tau_S^C(x)} dx;\ f_S = \int_{\mathbf{0}}^{\theta_D^S / T_\mathbf{0}} \frac{D(x)}{\tau_S^C(x)\tau_S^N(x)} dx \end{cases} \qquad (52)$$

Now, using the timing relation $\Omega\tau_{q,S}^{C} \gg \mathbf{1}$, Eq. (48) becomes:

$$\left(\Gamma_{II}^G\right)^2 = \frac{\mathbf{1}}{\Omega C_W} \sum_S \int \frac{\kappa_{q,S}^{\mathbf{0}}}{\left(\tau_{q,S}^C\right)^2} d^{\mathbf{3}}\boldsymbol{q} \qquad (53)$$

which after simplification using the change of variable $x = \hbar\omega / k_B T_0$, we obtain the following expression:

$$\left(\Gamma_{II}^{G}\right)^2 = \frac{1}{3\Omega}\frac{\sum_S \frac{1}{v_S}\left[e_S + \beta_S f_S\right]}{\sum_S \frac{1}{v_S^3} a_S} \quad (54)$$

The second sound properties in terms of velocity, relaxation time and damping factor as formulated in Eqs. (45), (47) and (48-54) exhibit more general forms of the driftless second sound compared to the conventional formula first obtained by Griffin.[57] These general forms show additional weighting functions that capture the fundamentally intertwining aspect between anharmonic Normal and Umklapp phonon-phonon scattering processes as well as the combined effect of these processes with other extrinsic phonon scattering processes. It is very interesting to note that the introduction of Callaway pseudo-relaxation time $\beta_S$ starting from its relation to the drifting velocity [Eq. (4)],[21,24] excludes the possibility of a drifting second sound type as was derived by Hardy in his analysis when he kept the drifting velocity as a fundamental property of the phonon gas when the latter can be characterized by an intermediate drifting distribution.[82]

To have a close insight onto the behavior of second sound properties as a function of ambient temperature as well as different intrinsic and extrinsic parameters including the effect of embedding SC nanoparticles as extrinsic phonon scattering centers, we consider again $Si_{0.7}Ge_{0.3}$ SC crystal alloy as a test bulk material with embedded Ge spherical nanoparticles.

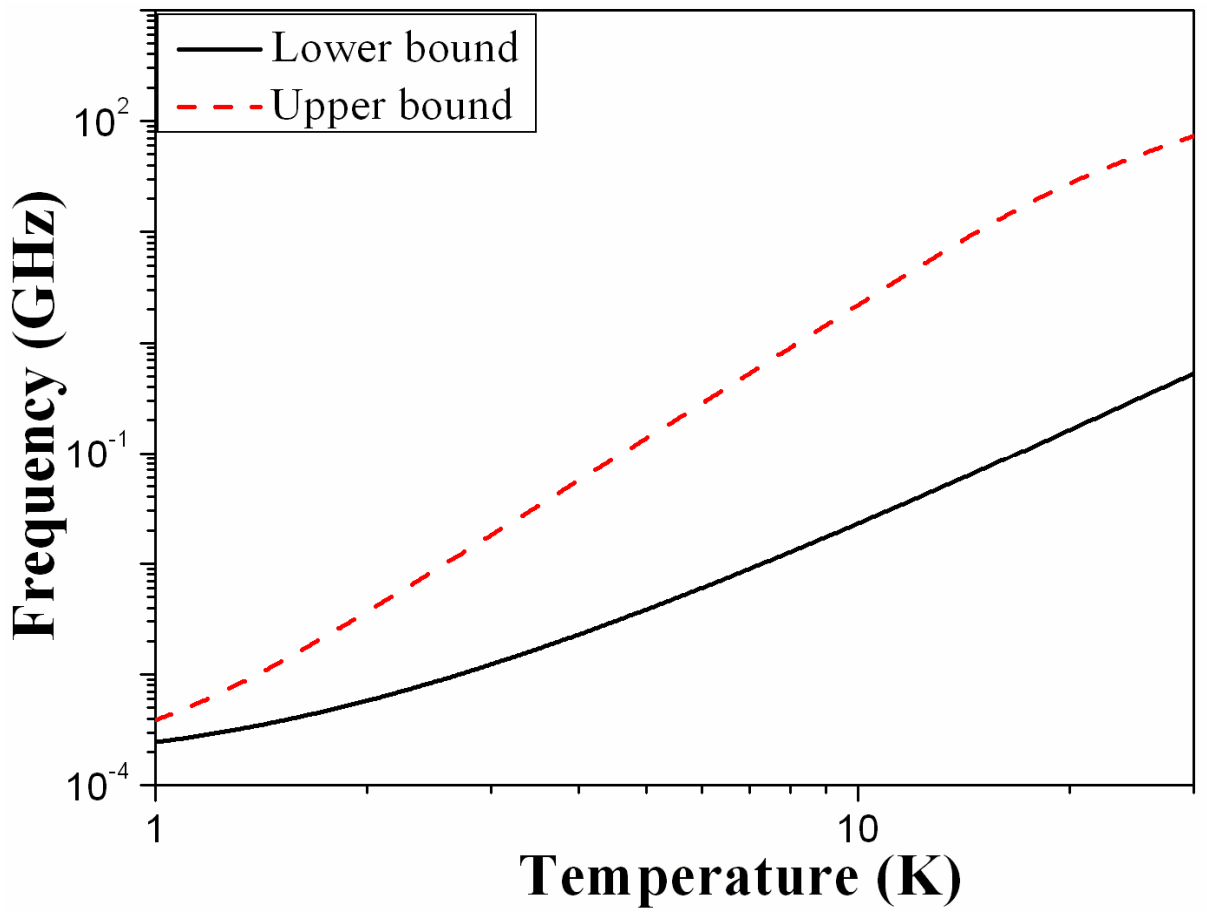

**Figure 13 :** Computed behavior of the time window boundaries as a function of temperature in bulk $Si_{0.7}Ge_{0.3}$ SC crystal alloy with no embedded Ge nanoparticles.

Figure 13 illustrates the temperature behavior of the time window boundaries as given by Eq. (52). As one can see, depending on the temperature in the low temperature regime, where one can expect Eq. (51) to be valid, the frequency of the temperature disturbance for the occurrence of second sound can vary from few MHz to few tens of GHz.

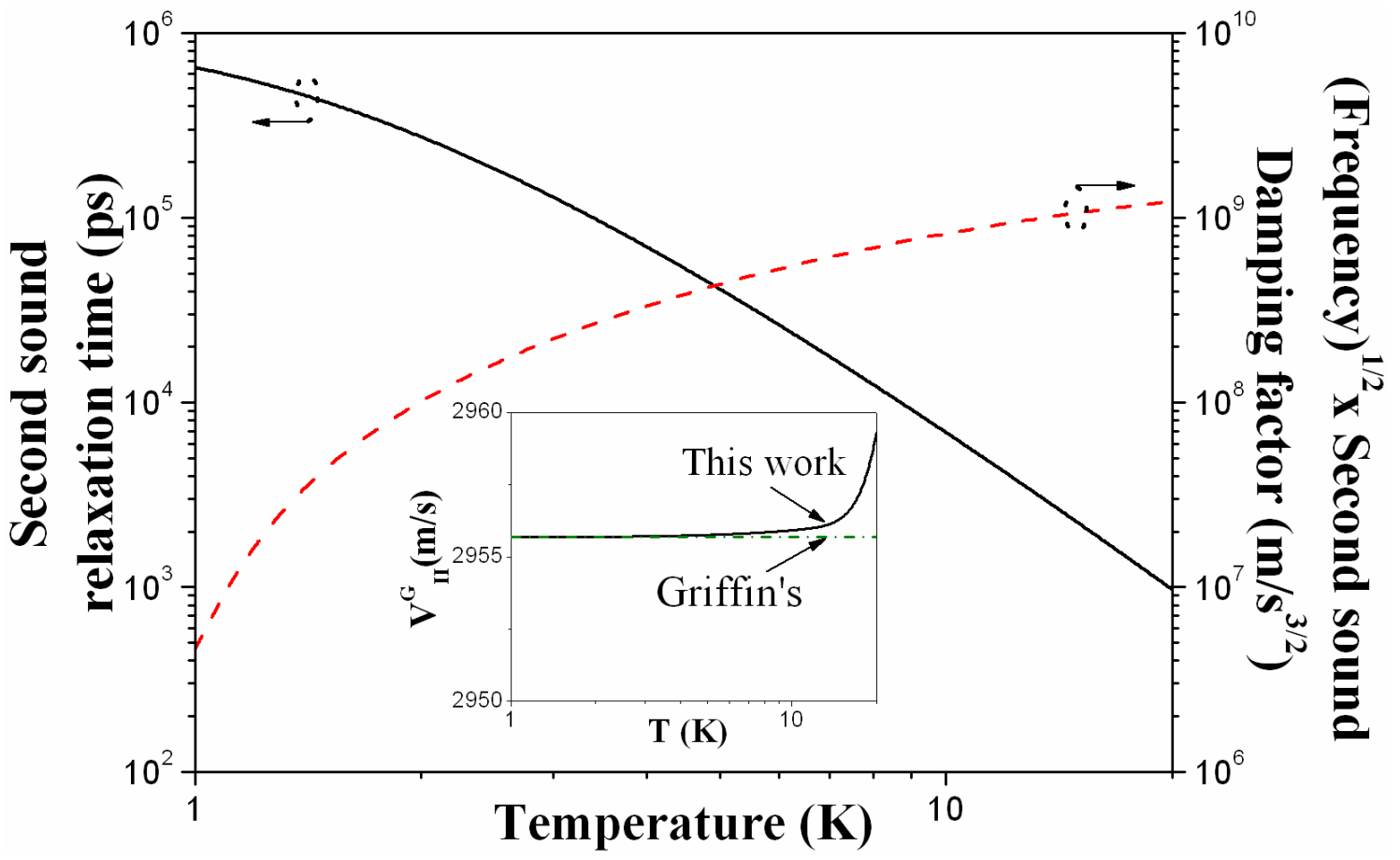

**Figure 14 :** Computed behaviors of second sound relaxation time (solid line), damping factor (dashed line) and velocity (inset) as a function of temperature in bulk $Si_{0.7}Ge_{0.3}$ SC crystal alloy with no embedded Ge nanoparticles. The inset shows a comparison of second sound velocity between classical Griffin's result[57] and this work's result.

In Fig 14, we report the computed temperature behaviors of second sound velocity $v_{II}^G$, relaxation time $\tau_{II}^G$ and damping factor $\Gamma_{II}^G$ in bulk $Si_{0.7}Ge_{0.3}$ SC crystal alloy over a temperature interval *[1-20K]*. The temperature interval is chosen to vary from low temperatures up to the optimum temperature at which the steady-state lattice thermal conductivity $\kappa(0)$ of bulk $Si_{0.7}Ge_{0.3}$ SC crystal alloy reaches a maximum [Fig 1]. This temperature interval is way below both transverse $\theta_D^T$ and longitudinal $\theta_D^L$ Debye temperatures.

For second sound propagation to occur, two conditions have to be fulfilled; (i) the time window [Eqs. (50) or (51)] and (ii) the weakness of the damping factor $\Gamma_{II}^G$. One expects second sound phenomena to occur more likely at the vicinity of the peak of $\kappa(0)$,[81] which in our case corresponds to the vicinity of the upper bound of the $T$ interval.

As one can see in Fig 14, the behaviors of $\tau_{II}^G$ and $\Gamma_{II}^G$ are very straightforward; by increasing the temperature $T$, $\tau_{II}^G$ decreases while $\Gamma_{II}^G$ increases. It seems however, that the rate of increasing of $\Gamma_{II}^G$ slows down a little by increasing $T$ in contrast to $\tau_{II}^G$, the decreasing of which continues almost in the same rate.

The inset, on the other hand, shows the behavior of $v_{II}^{G}$ calculated based on the present work [Eq. (45)] in comparison to the classical expression obtained by Griffin,[57] this one takes the form:

$$\left(v_{II}^{G}\right)_{Classical}^{2} = \frac{1}{3}\frac{\sum_{S}\frac{1}{v_S}}{\sum_{S}\frac{1}{v_S^3}} \qquad (55)$$

$v_{II}^{G}$ is almost independent of $T$ up to about $6K$ and as such superimposes with $\left.v_{II}^{G}\right)_{Classical}$ over this $T$ interval. Above 6K, $v_{II}^{G}$ starts increasing very slightly due to the effects of the different weighting functions that include the different phonon scattering $N$ and $R$ processes. An estimation of $\left.v_{II}^{G}\right)_{Classical}$ gives $\left.v_{II}^{G}\right)_{Classical} \approx \mathbf{2956}\text{m/s}$ . This value is closer to the transverse rather than to the longitudinal first sound velocity in $Si_{0.7}Ge_{0.3}$ SC crystal alloy, which proves the extra role and weight transverse phonon polarization branches have in the onset of the coherent phonon density wave propagation (second sound). As noticed by Narayanamurti and Dynes, experimental observations indicate that second sound consist of a true thermodynamic mixture of all the modes of the system; in this mixture transverse modes play the key role.[86,87] The close comparison between $v_{II}^{G}$ and $\left.v_{II}^{G}\right)_{Classical}$ shows that the influence of the different phonon scattering processes on the $T$-behavior of second sound velocity appears to be insignificant.

Also we can note that by comparing Eqs. (42) and (45), the expression of $v_{II}^{G}$ reduces to the one of $v_{eff}$ in the case we neglect phonon-phonon scattering N-processes ($\beta_S \rightarrow \mathbf{0}$).

Similarly we checked the effect of changing transverse $\gamma_T$ and longitudinal $\gamma_L$ Grüneisen parameters as well as mass-difference fluctuation parameter $\Gamma$ (not to be confused with the second sound damping factor as noted here $\Gamma_{II}^{G}$) on the behaviors of second sound properties at a fixed ambient temperature within the above $T$ interval, and the results showed an almost insignificant influence of these parameters; second sound properties remain almost insensitive to the changes of $0.2<\gamma_T<1.2$, $0.5<\gamma_T<1.5$ and $0.24<\Gamma<0.32$. As we have seen previously, the study of the cut-off frequency $f_C$ of the dynamical lattice thermal conductivity $\kappa(\Omega)$ led to the same conclusion regarding the effect of these parameters on a very low $T$ interval; $f_C$ is almost insensitive to these parameters.

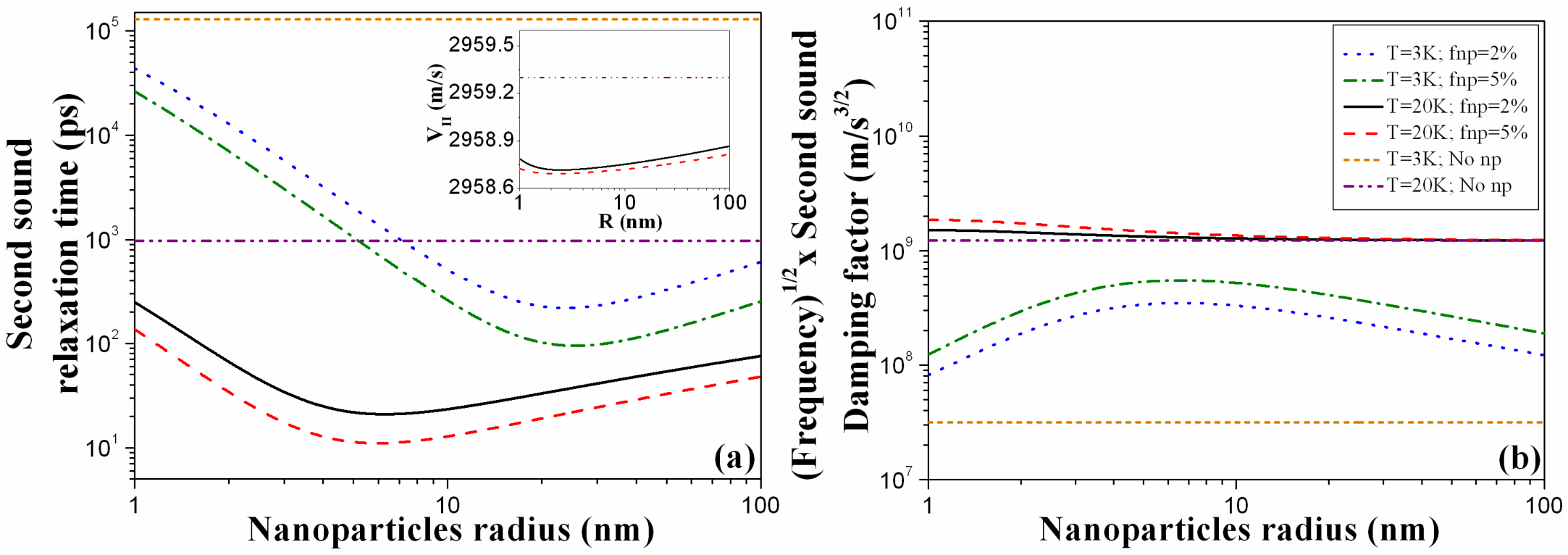

**Figure 15 :** Computed behaviors of second sound relaxation time (a), damping factor (b) and velocity (inset in (a)) in bulk $Si_{0.7}Ge_{0.3}$ SC crystal alloy with embedded Ge nanoparticles as a function of the nanoparticles radius $R$ for two values of the nanoparticles volume fraction $f_{np}$ and at two ambient temperatures $T$, also are added the computed behaviors with no nanoparticles.

On the other hand, very interesting and intriguing behaviors occur when one considers embedding Ge nanoparticles as extrinsic phonon scattering centers within bulk $Si_{0.7}Ge_{0.3}$ SC crystal alloy host matrix. Figs 15 (a) and (b) illustrate the behaviors of second sound properties $\tau_{II}^{G}$, $\Gamma_{II}^{G}$ and $v_{II}^{G}$ in $Si_{0.7}Ge_{0.3}$ SC crystal alloy with embedded Ge nanoparticles as a function of the nanoparticles radius $R$ for two values of the nanoparticles volume fraction $f_{np}=2\%$ and $f_{np}=5\%$ and at two ambient temperatures $T=3K$ and $T=20K$. The computations show $\tau_{II}^{G}$ to manifest a minimum at an optimal radius $R_{min}$, the position of which decreases by increasing $T$ at a fixed $f_{np}$ and seems to shift slightly to the right by increasing $f_{np}$ at a fixed $T$. We also see that at a fixed $T$, $\tau_{II}^{G}$ decreases by increasing $f_{np}$; this effect is expected since increasing $f_{np}$ will increase the corresponding scattering rate of phonons by the embedded nanoparticles. On the other hand, while $\Gamma_{II}^{G}$ manifests a maximum at $T=3K$, its trend seems to reverse at $T=20K$ where $\Gamma_{II}^{G}$ decays almost monotonically to saturate for high values of $R$. The position of $R_{max}$ for $\Gamma_{II}^{G}$ seems to be insensitive to the value of $f_{np}$, but $\Gamma_{II}^{G}$ increases by increasing $f_{np}$, the rate of increasing seems however to get slower by increasing $T$. At both $T=3K$ and $T=20K$, the computations show $v_{II}^{G}$ to be a constant almost independent of $R$ and $f_{np}$. The example of computations of the behavior of $v_{II}^{G}$ at $T=20K$ shown in the inset of Fig 15(a) is very illustrative and the apparent minimum may be misleading if one considers the accuracy of numerical integrations. We also added to these figures the behaviors of second sound properties in the case where no nanoparticles were added. As expected at a fixed ambient temperature, embedding nanoparticles reduces second sound $v_{II}^{G}$ and $\tau_{II}^{G}$ while it increases $\Gamma_{II}^{G}$.

It is really worthwhile to mention that the characteristic features in the behaviors of second sound properties as a function of the embedded nanoparticles radius $R$ do not occur for the same value of $R$ at fixed ambient $T$ and $f_{np}$; this is a consequence of the different expressions of the weighting functions in the general formulas of these properties as given by Eqs. (45), (47) and (53). These features result from the interplay between long and short wavelength scattering regimes of phonons by the embedded nanoparticles [Eqs. (33) and (34)]. We should remind here that the scattering of phonons by the embedded nanoparticles is assumed to be in the linear regime that is valid for small values of $f_{np}$, so that multiple scattering effects are neglected. While so far in second sound treatments, second sound relaxation time $\tau_{II}^{G}$ and damping factor $\Gamma_{II}^{G}$ have been shown to behave in an opposite way, meaning that when one decreases, the other increases and vice versa, the actual computed behaviors of these properties as a function of the embedded nanoparticles radius $R$ comes to break a little this result tendency, since we can have both of them decreasing at the same time when one changes $R$ at a fixed ambient $T$ (See the computed behaviors of $\tau_{II}^{G}$ and $\Gamma_{II}^{G}$ at $T=3K$ for $R$ ranging from *~10nm* to *~20nm*). This very intriguing result is not trivial and will require further investigation.

While $v_{II}^{G}$ remains almost insensitive and $\Gamma_{II}^{G}$ tends to saturate as a function of $R$ at a fixed ambient $T$ (especially for high $T$ within the allowed $T$ interval), our heuristic treatment shows $\tau_{II}^{G}$ to be second sound property that is the most affected by embedding nanoparticles and can exhibit a huge variation of few orders of magnitudes when $R$ varies over a decade.

A comparison of two experimental situations, with and without embedded nanoparticles of different sizes, would be a proof of a direct transition from the ballistic regime to the diffusive regime of heat transport at a fixed temperature in the low $T$ regime. As a matter of fact, by increasing the size of the nanoparticles, the total scattering rate of R-processes can get significantly increased; this leads to a reduction of the time window necessary for the occurrence of second sound, hence the more likely direct transition between the aforementioned two heat transport regimes if one keeps the same value of the excitation frequency of the temperature disturbance.

Based on this analysis of the behavior of $\kappa(\Omega)$ of SC crystals within the frame work of Boltzmann-Peierls theory of phonon transport using the single relaxation time approximation with Callaway approximated form of the collision operator, anharmonic phonon-phonon N-processes appear to play a very fundamental role in capturing the dynamics of energy (heat) transport in SC crystals. Through their role of shuffling the total phonon crystal momentum between different phonon states, N-processes have always to be considered in any study of phonon transport. The fact that taking them into account within the above framework with time independent Callaway parameter, leads to a breakdown of Shastry's Sum Rule of the real part of $\kappa(\Omega)$, constitutes a very interesting finding regarding phonon transport phenomena in SC crystals that needs to be checked using more sophisticated and complete modeling of energy and heat transport in these dielectric materials using first principles calculations and atomistic *ab-initio* Green's function approaches.[88,89]

## IV. SUMMARY AND CONCLUDING REMARKS

We developed in this paper an approach based on solving Boltzmann-Peierls Phonon Transport Equation (BPTE) in the frequency domain to analyze the dynamical behavior of the lattice thermal conductivity $\kappa(\Omega)$ of bulk semiconductor (SC) crystal materials as a response to an applied dynamical temperature gradient. The approach uses modified Debye-Callaway model within the framework of the single relaxation time approximation in which use has been made of Callaway approximated form of the collision operator with a time independent Callaway parameter. We were able to develop a compact expression for $\kappa(\Omega)$ that captures the leading behavior and the essential features of the dynamical thermal conduction by phonons. This expression fulfills the causality requirement and leads to a convolution type relationship between the heat flux density current and the temperature gradient in the real space-time domain in agreement with Gurtin-Pipkin theory. We considered the study of the effect of ambient temperature as well as different intrinsic and extrinsic parameters. Our calculations confirm previous theoretical studies regarding the order of magnitude of the cut-off frequency $f_C$ of $\kappa(\Omega)$ and further show $f_C$ to be very sensitive to the variation of temperature, Grüneisen parameter as well as embedded SC nanoparticles sizes and concentration. On the other hand, varying the mass-difference fluctuation parameter seems to have no effect on $f_C$. Embedding SC nanoparticles seems to considerably affect the dynamics of phonon heat transport in the host SC crystal alloy matrix depending on temperature, size and concentration of the nanoparticles. While embedding nanoparticles in SC alloy matrices reduces the steady-state $\kappa^{np}(0)$, it however tends to increase $f_C$ of the dynamical $\kappa^{np}(\Omega)$ in comparison with intrinsic SC alloys. This opposite double effect could be very beneficial in many microelectronic and optoelectronic device applications. The behavior of $\kappa(0)$ and $f_C$ of $\kappa^{np}(\Omega)$ as a function of the nanoparticles size and concentration revealed very interesting trends that show how important is a careful manipulation of the nanoparticles size and concentration for a better control of phonon heat transport in SC crystal alloys. The study showed also the N-processes to be indispensible and play a very fundamental role in both the steady-state and dynamical phonon transport regimes.

Low values of the cut-off frequency $f_C$ in the low temperature regime, is another manifestation of the ballistic phonon transport regime in which the intertwining between anharmonic phonon-phonon scattering N-processes and U-processes plays a key role.

Our model is unable however to explain Koh and Cahill experimental results and one needs to conduct more experiments and sophisticated modeling in order to check the relevance of the calculations results and shed light on the eventual discrepancies.

The applicability of Shastry's Sum Rule (SSR) to $\kappa(\Omega)$ revealed anharmonic phonon-phonon N-processes to play a very fundamental role in capturing the dynamics of energy (heat) transport in SC crystals. We found that SSR holds only when these phonon scattering processes are disregarded and only resistive phonon scattering processes are considered. In this latter case, we were able to extract a classical expression to the expectation of the thermal operator $\theta^{xx}$ introduced by Shastry. This expression preserves the deep connection linking the expectation of this operator to the lattice specific heat, namely the vanishing in the true thermodynamic ground state (i.e at $T=0$), as already discussed by Shastry. As given by the product of the lattice specific heat and the square of an effective group velocity that can be viewed as a generalized form of the driftless velocity of second sound in the bulk SC crystal, this classical expression confirms the physical meaning of $\theta^{xx}$ variable in capturing the ballistic dynamics aspect in the energy (heat) transport phenomenon.

We discussed also the possibility of existence and propagation of second sound in bulk SC crystals and general forms of driftless second sound properties were derived using a heuristic treatment based on Griffin's self-consistency criterion to study the dispersion relation of second sound in terms of its velocity, relaxation time and damping factor. Our approach exhibits also the need for a time window in the relaxation time spectrum in a way similar to Guyer and Krumhansl analysis These general forms captures elegantly the interplay aspect between anharmonic phonon-phonon scattering N-processes and U-processes as well as the combined effect of these processes with other extrinsic phonon scattering processes. Nevertheless, we found that over the temperature interval where one expects the timing window for the occurrence of second sound phenomenon, second sound properties seem to be almost insensitive to the change in temperature, Grüneissen parameter as well as mass-difference fluctuation parameter; the influence of the different phonon scattering processes seem to cancel out each other which does not hamper the coherent propagation of energy (heat) fluctuations in the phonon gas. On the other hand, embedding SC nanoparticles in the bulk SC alloy crystal host matrix lead to very interesting and intriguing behaviors of second sound properties as a function of the nanoparticles radius $R$. These characteristic features do not occur at the same value of $R$ for a fixed temperature and nanoparticles concentration and are a manifestation of the interplay between long and short wavelength scattering regimes of phonons by the embedded nanoparticles. The particular behavior of second sound relaxation time as a function of $R$ is very interesting and will surely need further investigations both experimentally and theoretically using more sophisticated and complete modeling for a better understanding of the effect of embedding nanoparticles on the behavior of the propagation of second sound in dielectric and SC bulk crystals.

The treatment outlined forth in the theory section highlights the leading dynamical behavior of $\kappa(\Omega)$, it allows us to have a consistent and meaningful classical limit of SSR and leads us to obtain general forms of driftless second sound properties in bulk SC crystals. Nevertheless this treatment of BPTE within the framework of the single relaxation time can't be considered as fully rigorous as it depends on few assumptions that can eventually be relaxed. Therefore many improvements can be contemplated in order to check the relevance and consistency of some of the calculations results we presented above, regarding particularly the behavior of the cut-off frequency $f_C$ of the dynamical thermal conductivity $\kappa(\Omega)$ as a function of temperature $T$ as well as the behavior of the embedded nanoparticle optimal size $R_{min}$ as a function of $T$ and nanoparticle volume fraction $f_{np}$ and generally the effect of embedded nanoparticles on the behavior of $f_C$ of $\kappa^{np}(\Omega)$. The improvements would be related to:

i. the form of the collision operator including the separate effect of anharmonic phonon-phonon N-processes and U-processes.

ii. the effect of specularity in phonon boundary scattering especially in the low temperature regime. We have considered a fixed characteristic length $L_C$ in our analysis; changing $L_C$ would affect the behavior of $\kappa(0)$ and $\kappa(\Omega)$ enormously in this low $T$ regime.

iii. the phonon wave vector dependence of the expressions of all phonon scattering rates, particularly anharmonic phonon-phonon N-processes and U-processes. In this case, a complete modeling using first principles calculations and atomistic *ab-initio* Green's function approaches will be very helpful and will allow to treat intrinsic and extrinsic phonon scattering mechanisms more respectfully including any eventual contribution from optical phonons especially in the high $T$ regime as well as any eventual nonlinearity as the one that might arise for scattering of phonon by embedded nanoparticles with high $f_{np}$.

iv. the explicit time dependence of Callaway parameter $\beta$ when approximated Callaway form of the collision operator is used as we did in our analysis except that we assumed $\beta$=*cte* independent of time in our treatment. In this case of time dependent $\beta$, the mathematical treatment of BPTE would be a bit tedious in which it would be easier to solve the problem directly in the time domain following the approach of Guyer and Krumhansl and using the trajectory integral method. We plan to investigate this approach in a future work in order to study the impact on the behavior of $f_C$ of $\kappa(\Omega)$ as function of $T$ (it is more likely that $f_C$ would be a little bit lower than what we expected above especially in the low $T$ regime) as well as the applicability of SSR and the causality requirement when anharmonic phonon-phonon N-processes are turned on.

# ACKNOWLEDGMENT

The authors would like to thank Professor Ali Shakouri for his numerous insights, fruitful and stimulating discussions regarding this work.

## APPENDIX A: Callaway pseudo-relaxation time $\beta_S$

The condition of conservation of the total crystal momentum of the phonon system after a collision involving anharmonic phonon-phonon N-processes is:[21,24]

$$\int \left( \frac{\partial n_{q,S}}{\partial t} \right)_N \boldsymbol{q} d^3\boldsymbol{q} \equiv \int \left( \frac{n_{\boldsymbol{q},S}(\boldsymbol{\lambda}_S) - n_{\boldsymbol{q},S}}{\tau_{q,S}^N} \right) \boldsymbol{q} d^3\boldsymbol{q} = 0; \text{ all polarizations S} \qquad (A.1)$$

The effect of resistive or destroying processes represented by $\tau_{q,S}^R$ due to anharmonic phonon-phonon U-processes and all extrinsic processes, is contained in $n_{\boldsymbol{q},S}$.

In the steady-state case $\left( \partial n_{\boldsymbol{q},S} / \partial t = 0 \right)$ the combination of Eqs. (3), (5) and (6) in combination with the approximation $\frac{dn_{\boldsymbol{q},S}}{dT} \cong \frac{dn_{\boldsymbol{q},S}^0}{dT_0}$ allows us then to write:

$$n_{\boldsymbol{q},S}(\boldsymbol{\lambda}_S) - n_{\boldsymbol{q},S} = \boldsymbol{V}_{\boldsymbol{q},S} \left[ \tau_{q,S}^{eff} - \beta_S \right] \frac{dn_{\boldsymbol{q},S}^0}{dT_0} \boldsymbol{\nabla} T \qquad (A.2)$$

On using Eq. (A.2), Eq. (A.1) becomes:

$$\int \left( \frac{\partial n_{\boldsymbol{q},S}}{\partial t} \right)_N \boldsymbol{q} d^3\boldsymbol{q} \equiv \int \left( \frac{\tau_{q,S}^{eff} - \beta_S}{\tau_{q,S}^N} \right) \frac{dn_{\boldsymbol{q},S}^0}{dT_0} \boldsymbol{V}_{\boldsymbol{q},S} . \boldsymbol{\nabla} T \boldsymbol{q} d^3\boldsymbol{q} = 0$$

$$\Rightarrow \int_0^{\omega_D^S} \left( \frac{\tau_S^{eff}(\omega) - \beta_S}{\tau_S^N(\omega)} \right) \frac{\hbar\omega}{k_B T_0^2} \frac{e^{\frac{\hbar\omega}{k_B T_0}}}{\left( e^{\frac{\hbar\omega}{k_B T_0}} - 1 \right)^2} \omega^3 d\omega = 0 \qquad (A.3)$$

To obtain the second line in Eq. (A.3), we used Debye linear relation (acoustic approximation) ($\omega_{\boldsymbol{q},S} = v_{S,\iota} |\boldsymbol{q}|$) and the fact that the system is an isotropic elastic medium. $\omega_D^S$ is Debye cut-off frequency of the acoustic polarization branch *S*.[59] By using the standard change of variable $x = \hbar\omega / k_B T_0$, replacing Eq. (7) into Eq. (A.3) and solving for the constant $\beta_S$ which is independent of *x*, one obtains:

$$\beta_S = \frac{\int_0^{\theta_D^S/T_0} \frac{\tau_S^C(x)}{\tau_S^N(x)} D(x)\,dx}{\int_0^{\theta_D^S/T_0} \frac{1}{\tau_S^N(x)}\left[1-\frac{\tau_S^C(x)}{\tau_S^N(x)}\right] D(x)\,dx} = \frac{\int_0^{\theta_D^S/T_0} \frac{\tau_S^C(x)}{\tau_S^N(x)} D(x)\,dx}{\int_0^{\theta_D^S/T_0} \frac{\tau_S^C(x)}{\tau_S^N(x)\tau_S^R(x)} D(x)\,dx} \qquad (A.4)$$

where $D(x) = x^4 e^x \big/ \left(e^x - 1\right)^2$ is Debye function and $\theta_D^S$ is Debye temperature of the acoustic polarization branch S.[59]

## APPENDIX B: Simplification of the expressions of $\kappa_r$ and $\kappa_i$ as given in Eqs. (18) and (19)

Starting from Eqs. (13-17) and using the isotropy of the group velocity in the real and reciprocal spaces $v_{S,q_x}^2 = v_{S,q_y}^2 = v_{S,q_z}^2 = \frac{1}{3} v_S^2$ (cubic symmetry) one can write the dynamical lattice thermal conductivity of each phonon polarization branch S as (*S=L, T*):

$$\kappa_S(\Omega) = \frac{1}{3}\frac{1}{8\pi^3}\int \frac{\tau_{q,S}^C\left[1+\frac{\beta_S}{\tau_{q,S}^N}\right]}{1-j\Omega\tau_{q,S}^C} v_S^2 C_{Ph}(\mathbf{q},S)\,d^3\mathbf{q} = \kappa_{S1}(\Omega)+\kappa_{S2}(\Omega)$$

$$\begin{cases} \kappa_{S1}(\Omega) = \frac{1}{3}\frac{1}{8\pi^3}\int \frac{\tau_{q,S}^C}{1-j\Omega\tau_{q,S}^C} v_S^2 C_{Ph}(\mathbf{q},S)\,d^3\mathbf{q} = \frac{1}{3}\int_0^{\omega_D^S} \frac{\tau_S^C(\omega)}{1-j\Omega\tau_S^C(\omega)} v_S^2 C_{Ph}(\omega,S)\, g_S(\omega)\,d\omega \\ \kappa_{S2}(\Omega) = \frac{1}{3}\frac{1}{8\pi^3}\beta_S\int \frac{\frac{\tau_{q,S}^C}{\tau_{q,S}^N}}{1-j\Omega\tau_{q,S}^C} v_S^2 C_{Ph}(\mathbf{q},S)\,d^3\mathbf{q} = \frac{1}{3}\beta_S\int_0^{\omega_D^S} \frac{\frac{\tau_S^C(\omega)}{\tau_S^N(\omega)}}{1-j\Omega\tau_S^C(\omega)} v_S^2 C_{Ph}(\omega,S)\, g_S(\omega)\,d\omega \\ C_{Ph}(\omega,S) = \hbar\omega\frac{dn_{q,S}^0}{dT_0} = k_B\left(\frac{\hbar\omega}{k_B T_0}\right)^2 \frac{e^{\frac{\hbar\omega}{k_B T_0}}}{\left(e^{\frac{\hbar\omega}{k_B T_0}}-1\right)^2} \end{cases} \qquad (B.1)$$

The partial conductivities $\kappa_{S1}$ and $\kappa_{S2}$ are the usual Debye-Callaway terms, where $g_S(\omega) = \frac{\omega^2}{2\pi^2 v_S^3}$ is the Debye phonon density of states in the acoustic polarization branch *S*.[59]

Then we use as usual the change of variable $x = \hbar\omega / k_B T_0$ , it is straightforward to show that $\kappa_{S1}$ and $\kappa_{S2}$ can be written as:

$$\begin{cases} \kappa_{S1}(\Omega) = \frac{1}{3} C_S T_0^3 \int_0^{\theta_D^S/T_0} \frac{\tau_S^C(x)}{1 - j\Omega\tau_S^C(x)} D(x)\,dx \\ \kappa_{S2}(\Omega) = \frac{1}{3} C_S T_0^3 \beta_S \int_0^{\theta_D^S/T_0} \frac{\frac{\tau_S^C(x)}{\tau_S^N(x)}}{1 - j\Omega\tau_S^C(x)} D(x)\,dx \\ C_S = \frac{k_B^4}{2\pi^2\hbar^3 v_S} \end{cases} \qquad (\mathrm{B.2})$$

where *D(x)* is Debye function.[59]

As it is customary in the modified Debye-Callaway model, we express the total dynamical lattice thermal conductivity $\kappa(\Omega)$ of the SC crystal as the sum over one longitudinal ($\kappa_L$) and two degenerate transverse ($\kappa_T$) phonon polarization branches:[34]

$$\kappa(\Omega) = \kappa_L(\Omega) + 2\kappa_T(\Omega) \qquad (B.3)$$

We can then write the real part $\kappa_r(\Omega)$ and the imaginary $\kappa_i(\Omega)$ part as:

$$\begin{cases} \kappa_r(\Omega) = \kappa_L^r(\Omega) + 2\kappa_T^r(\Omega) \\ \kappa_S^r(\Omega) = \kappa_{S1}^r(\Omega) + \kappa_{S2}^r(\Omega) \\ \kappa_{S1}^r(\Omega) = \frac{1}{3} C_S T_0^3 \int_0^{\theta_D^S/T_0} \frac{\tau_S^C(x)}{1 + \left[\Omega\tau_S^C(x)\right]^2} D(x)\,dx \\ \kappa_{S2}^r(\Omega) = \frac{1}{3} C_S T_0^3 \beta_S \int_0^{\theta_D^S/T_0} \frac{\frac{\tau_S^C(x)}{\tau_S^N(x)}}{1 + \left[\Omega\tau_S^C(x)\right]^2} D(x)\,dx \end{cases} \qquad (\mathrm{B.4})$$

$$\begin{cases} \kappa_i(\Omega) = \kappa_L^i(\Omega) + 2\kappa_T^i(\Omega) \\ \kappa_S^i(\Omega) = \kappa_{S1}^i(\Omega) + \kappa_{S2}^i(\Omega) \\ \kappa_{S1}^i(\Omega) = \dfrac{1}{3} C_S T_0^3 \displaystyle\int_0^{\theta_D^S/T_0} \dfrac{\Omega\left[\tau_S^C(x)\right]^2}{1+\left[\Omega\tau_S^C(x)\right]^2} D(x)\,dx \\ \kappa_{S2}^i(\Omega) = \dfrac{1}{3} C_S T_0^3 \beta_S \displaystyle\int_0^{\theta_D^S/T_0} \dfrac{\Omega \dfrac{\left[\tau_S^C(x)\right]^2}{\tau_S^N(x)}}{1+\left[\Omega\tau_S^C(x)\right]^2} D(x)\,dx \end{cases} \qquad (\text{B.5})$$

## APPENDIX C: Hilbert Transforms and Kramers-Kronig relations applicability to the dynamical lattice thermal conductivity $\kappa(\Omega)$

Kramers-Kronig relations are a consequence of the causality requirement and as such they are general and apply to a variety of dynamical susceptibilities (*after-effect functions*). In this appendix, we show that the expression of the dynamical lattice thermal conductivity $\kappa(\Omega)$ that we derived in this paper, verifies these relations as well.

$\kappa(\Omega)$ as given by Eq. (15) has all its poles (singularities) lying in the lower half complex plane. It is an holomorphic (analytic) function on the upper half complex plane. Therefore:

$$P\int_{-\infty}^{+\infty} \frac{\kappa(\Omega)}{\Omega-\Omega_0} d\Omega = \pi j \kappa(\Omega_0) \qquad (C.1)$$

where $P$ represents the Cauchy principal value. In terms of the real and imaginary parts $\kappa_r(\Omega)$ and $\kappa_i(\Omega)$ of $\kappa(\Omega)$, Eq. (C.1) leads to:

$$\begin{cases} \kappa_r(\Omega_0) = \dfrac{1}{\pi} P\displaystyle\int_{-\infty}^{+\infty} \dfrac{\kappa_i(\Omega)}{\Omega-\Omega_0} d\Omega & (a) \\ \kappa_i(\Omega_0) = -\dfrac{1}{\pi} P\displaystyle\int_{-\infty}^{+\infty} \dfrac{\kappa_r(\Omega)}{\Omega-\Omega_0} d\Omega & (b) \end{cases} \qquad (C.2)$$

$\kappa_r(\Omega)$ and $\kappa_i(\Omega)$ are then said to be Hilbert transforms of each other.

Furthermore, we can easily check that $\kappa(\Omega)$ fulfills the Hermitian symmetry property, namely:

$$\kappa(-\Omega) = \kappa^*(\Omega) \qquad (C.3)$$

where $\kappa^*(\Omega)$ denotes the complex conjugate of $\kappa(\Omega)$. Eq. (C.3) describes the evenness and the oddness of $\kappa_r(\Omega)$ and $\kappa_i(\Omega)$ respectively:

$$\begin{cases} \kappa_r(-\Omega) = \kappa_r(\Omega) \\ \kappa_i(-\Omega) = -\kappa_i(\Omega) \end{cases} \qquad (C.\mathbf{4})$$

Using some very simple algebra, it is easy to show that via Eqs. (C.4), Eqs. (C.2) become:

$$\begin{cases} \kappa_r(\Omega_0) = \dfrac{2}{\pi} P \displaystyle\int_0^{+\infty} \dfrac{\Omega}{\Omega^2 - \Omega_0^2} \kappa_i(\Omega) d\Omega \\ \kappa_i(\Omega_0) = -\dfrac{2}{\pi} P \displaystyle\int_0^{+\infty} \dfrac{\Omega_0}{\Omega^2 - \Omega_0^2} \kappa_r(\Omega) d\Omega \end{cases} \qquad (C.\mathbf{5})$$

These constitute the standard forms of Kramers-Kronig relations mostly used in literature that are related to any dynamical susceptibility since the dependent variable is frequency which is positive.

In order to check the applicability of Kramers-Kronig relations to the expressions we obtained in the theory section for $\kappa_r(\Omega)$ and $\kappa_i(\Omega)$, it is easier to start with Eqs. (C.2).

Let's consider Eq. (C.2a). Using the expression of $\kappa_i(\Omega)$ as given by Eq. (16), one can write:

$$\begin{cases} P\displaystyle\int_{-\infty}^{+\infty} \dfrac{\kappa_i(\Omega)}{\Omega - \Omega_0} d\Omega = \lim_{\substack{\varepsilon \to 0 \\ R \to \infty}} \left\{ \displaystyle\int_{-R}^{\Omega_0 - \varepsilon} \dfrac{\kappa_i(\Omega)}{\Omega - \Omega_0} d\Omega + \displaystyle\int_{\Omega_0 + \varepsilon}^{R} \dfrac{\kappa_i(\Omega)}{\Omega - \Omega_0} d\Omega \right\} \\ \qquad = \displaystyle\sum_S \int_q \kappa_{q,S}^0 \lim_{\substack{\varepsilon \to 0 \\ R \to \infty}} \left\{ I_{q,S}(\varepsilon, R) + J_{q,S}(\varepsilon, R) \right\} \\ I_{q,S}(\varepsilon, R) = \displaystyle\int_{-R}^{\Omega_0 - \varepsilon} \dfrac{\Omega \tau_{q,S}^C}{(\Omega - \Omega_0)\left(1 + \left(\Omega \tau_{q,S}^C\right)^2\right)} d\Omega \\ J_{q,S}(\varepsilon, R) = \displaystyle\int_{\Omega_0 + \varepsilon}^{R} \dfrac{\Omega \tau_{q,S}^C}{(\Omega - \Omega_0)\left(1 + \left(\Omega \tau_{q,S}^C\right)^2\right)} d\Omega \end{cases} \qquad (C.\mathbf{6})$$

The integrand in both functions $I_{q,S}(\varepsilon,R)$ and $J_{q,S}(\varepsilon,R)$ can be simplified using a fractional decomposition. It follows then:

$$\begin{cases} \dfrac{\Omega\tau_{q,S}^{C}}{(\Omega-\Omega_{0})\left(1+\left(\Omega\tau_{q,S}^{C}\right)^{2}\right)} = \dfrac{f}{\Omega-\Omega_{0}} + \dfrac{g\Omega+h}{1+\left(\Omega\tau_{q,S}^{C}\right)^{2}} \\ f = \dfrac{\Omega_{0}\tau_{q,S}^{C}}{1+\left(\Omega\tau_{q,S}^{C}\right)^{2}};\ g = -\dfrac{\Omega_{0}\left(\tau_{q,S}^{C}\right)^{3}}{1+\left(\Omega\tau_{q,S}^{C}\right)^{2}} \text{ and } h = \dfrac{\tau_{q,S}^{C}}{1+\left(\Omega\tau_{q,S}^{C}\right)^{2}} \end{cases} \qquad (C.7)$$

$I_{q,S}(\varepsilon,R)$ and $J_{q,S}(\varepsilon,R)$ can then be calculated analytically and the result is:

$$\begin{cases} I_{q,S}(\varepsilon,R) = f\ln\left|\dfrac{\varepsilon}{R+\Omega_{0}}\right| + \dfrac{g}{2\left(\tau_{q,S}^{C}\right)^{2}}\ln\left|\dfrac{1+\left[(\Omega_{0}-\varepsilon)\tau_{q,S}^{C}\right]^{2}}{1+\left[R\tau_{q,S}^{C}\right]^{2}}\right| \\ \qquad + \dfrac{h}{\tau_{q,S}^{C}}\left\{Arc\tan\left[(\Omega_{0}-\varepsilon)\tau_{q,S}^{C}\right] + Arc\tan\left[R\tau_{q,S}^{C}\right]\right\} \\ J_{q,S}(\varepsilon,R) = f\ln\left|\dfrac{R-\Omega_{0}}{\varepsilon}\right| + \dfrac{g}{2\left(\tau_{q,S}^{C}\right)^{2}}\ln\left|\dfrac{1+\left[R\tau_{q,S}^{C}\right]^{2}}{1+\left[(\Omega_{0}+\varepsilon)\tau_{q,S}^{C}\right]^{2}}\right| \\ \qquad + \dfrac{h}{\tau_{q,S}^{C}}\left\{Arc\tan\left[R\tau_{q,S}^{C}\right] - Arc\tan\left[(\Omega_{0}+\varepsilon)\tau_{q,S}^{C}\right]\right\} \end{cases} \qquad (C.8)$$

The sum of $I_{q,S}(\varepsilon,R)$ and $J_{q,S}(\varepsilon,R)$ leads to:

$$I_{q,S}(\varepsilon,R) + J_{q,S}(\varepsilon,R) = f\ln\left|\dfrac{R-\Omega_{0}}{R+\Omega_{0}}\right| + \dfrac{g}{2\left(\tau_{q,S}^{C}\right)^{2}}\ln\left|\dfrac{1+\left[(\Omega_{0}-\varepsilon)\tau_{q,S}^{C}\right]^{2}}{1+\left[(\Omega_{0}+\varepsilon)\tau_{q,S}^{C}\right]^{2}}\right|$$

$$+\dfrac{h}{\tau_{q,S}^{C}}\left\{2Arc\tan\left[R\tau_{q,S}^{C}\right] + Arc\tan\left[(\Omega_{0}-\varepsilon)\tau_{q,S}^{C}\right] - Arc\tan\left[(\Omega_{0}+\varepsilon)\tau_{q,S}^{C}\right]\right\} \qquad (C.9)$$

then by taking the limit when $\varepsilon\to 0$ and $R\to\infty$, we obtain:

$$\lim_{\substack{\varepsilon\to 0\\ R\to\infty}}\left\{I_{q,S}(\varepsilon,R) + J_{q,S}(\varepsilon,R)\right\} = \pi\dfrac{h}{\tau_{q,S}^{C}} = \dfrac{\pi}{1+\left(\Omega_{0}\tau_{q,S}^{C}\right)^{2}} \qquad (C.10)$$

By replacing Eq. (C.10) in Eq. (C.6) we get finally:

$$\begin{cases} P\displaystyle\int_{-\infty}^{+\infty}\dfrac{\kappa_{i}(\Omega)}{\Omega-\Omega_{0}}d\Omega = \lim_{\substack{\varepsilon\to 0\\ R\to\infty}}\left\{\int_{-R}^{\Omega_{0}-\varepsilon}\dfrac{\kappa_{i}(\Omega)}{\Omega-\Omega_{0}}d\Omega + \int_{\Omega_{0}+\varepsilon}^{R}\dfrac{\kappa_{i}(\Omega)}{\Omega-\Omega_{0}}d\Omega\right\} \\ \qquad = \displaystyle\sum_{S}\int_{q}\kappa_{q,S}^{0}\lim_{\substack{\varepsilon\to 0\\ R\to\infty}}\left\{I_{q,S}(\varepsilon,R) + J_{q,S}(\varepsilon,R)\right\}dq \\ \qquad = \pi\displaystyle\sum_{S}\int_{q}\dfrac{\kappa_{q,S}^{0}}{1+\left(\Omega_{0}\tau_{q,S}^{C}\right)^{2}}dq = \pi\kappa_{r}(\Omega_{0}) \end{cases} \qquad (C11)$$

which ends the proof of Eq. (C.2a).

The verification of the second relation [Eq. (C.2b)] follows exactly the same mathematical steps; hence there is no need to detail the proof.

## APPENDIX D: Hardy's like approach to second sound in bulk cubic SC and dielectric crystals

The basic idea in Hardy's method[82] is that second sound phenomenon in a dielectric solid is to be considered to exist when an accurate description of the variations of the local temperature $T(x,t)$ requires the use of a damped wave equation of the form:

$$\frac{\partial^2 T}{\partial t^2}+\frac{\mathbf{1}}{\tau_{II}^{H}}\frac{\partial T}{\partial t}-\left(v_{II}^{H}\right)^{2}\nabla^2 T=\mathbf{0} \qquad (D.\mathbf{1})$$

where $\tau_{II}^{H}$ and $v_{II}^{H}$ are the relaxation time and the propagation velocity of the second sound.

Let's set $J_{q,S}(x,t)=W^{-\mathbf{1}}\hbar\omega_{q,S}n_{q,S}(x,t)\boldsymbol{V}_{\boldsymbol{q},S}$ to be the heat flux density current associated with a phonon wave packet at state ***(q**, S)* with a group velocity $\boldsymbol{V}_{\boldsymbol{q},S}$. The total heat flux density current may then be expressed as:

$$J_Q(x,t)=\sum_{q,S}J_{q,S}(x,t)=\frac{W}{\mathbf{8}\pi^{\mathbf{3}}}\sum_{S}\int J_{q,S}(x,t)\,d^3\boldsymbol{q} \qquad (D.\mathbf{2})$$

where "*W*" denotes the bulk SC crystal volume and we used the standard relation $\sum_{q,S}\to W/8\pi^3\sum_{S}\int d^3\boldsymbol{q}$ .

By Fourier-transforming $J_{q,S}(x,t)$ and using the expression of $\overline{n_{q,S}}(x,\Omega)$ as given by Eq. (9) with the fact that $n^{0}_{\boldsymbol{q},S}$ does not contribute to any energy (heat) transport, one obtains:

$$J_{q,S}(x,\Omega)=-\frac{\mathbf{8}\pi^{\mathbf{3}}}{W}\frac{\kappa^{\mathbf{0}}_{q,S}}{\mathbf{1}-j\Omega\tau^{C}_{q,S}}\overline{\nabla T}(x,\Omega) \qquad (D.\mathbf{3})$$

where $\kappa^{\mathbf{0}}_{q,S}$ is given by Eq. (14). After rearranging both sides of Eq. (D.3), then taking the inverse Fourier transform, this leads to:

$$\tau^{C}_{q,S}\frac{\partial J_{q,S}}{\partial t}+J_{q,S}=-\frac{\mathbf{8}\pi^{\mathbf{3}}}{W}\kappa^{\mathbf{0}}_{q,S}\nabla T(x,t) \qquad (D.\mathbf{4})$$

Eq. (D.4) resembles a Cattaneo's like model for the heat flux density current associated with each phonon normal mode (or wave packet) $(\boldsymbol{q}, S)$ and characterized by a its own relaxation time $\tau^{C}_{q,S}$.

Now, let's take the sum of each term in Eq. (D.4) over all phonon states (all wave vectors and polarizations). One gets:

$$\frac{\partial}{\partial t}\left[\sum_{q,S}\tau^{C}_{q,S}J_{q,S}\right]+J_Q=-\kappa(\mathbf{0})\nabla T(x,t) \qquad (D.\mathbf{5})$$

where $\kappa(0)$ is the steady-state thermal conductivity of the bulk SC crystal.

To have a Cattaneo's like model for the total heat flux density current $J_Q$, we define an effective relaxation time $\tau^{eff}$ as:

$$\tau^{eff}=\frac{\sum_{q,S}\tau^{C}_{q,S}J_{q,S}}{J_Q}=\frac{\sum_{q,S}\tau^{C}_{q,S}J_{q,S}}{\sum_{q,S}J_{q,S}} \qquad (D.\mathbf{6})$$

Similarly to all other relaxation times considered in our study of $\kappa(\Omega)$ (see the theory section), the effective relaxation time $\tau^{eff}$ is supposed to be time or frequency independent. Based on this, one can derive the expression of $\tau^{eff}$ by taking the steady-state limits of both $J_{q,S}$ and $J_Q$. It follows then:

$$\tau^{eff}=\lim_{\Omega\to\mathbf{0}}\frac{\sum_{q,S}\tau^{C}_{q,S}J_{q,S}}{\sum_{q,S}J_{q,S}}=\frac{\sum_{S}\int\tau^{C}_{q,S}\kappa^{\mathbf{0}}_{q,S}d^{3}\boldsymbol{q}}{\sum_{S}\int\kappa^{\mathbf{0}}_{q,S}d^{3}\boldsymbol{q}} \qquad (D.7)$$

where the dominator in Eq. (D.7) is exactly the steady-state thermal conductivity $\kappa(0)$. One final equation of need at this stage is the energy continuity equation, which for small deviation from equilibrium takes the form:[82]

$$C_W\frac{\partial T(x,t)}{\partial t}+\nabla.J_Q(x,t)=\mathbf{0} \qquad (D.\mathbf{8})$$

Combining Eqs. (D.5-D.8) results in a damped wave equation for the local temperature $T(x,t)$ given by:

$$\frac{\partial^2 T}{\partial t^2}+\frac{\mathbf{1}}{\tau^{eff}}\frac{\partial T}{\partial t}-\frac{\kappa(\mathbf{0})}{C_W\tau^{eff}}\nabla^2 T=\mathbf{0} \qquad (D.\mathbf{9})$$

where $C_W$ denotes the specific heat per unit volume at a constant volume of the bulk SC crystal. Eqs. (D.9) and (D.1) are analogous with $\tau_{II}^{H}$ and $\left(v_{II}^{H}\right)^2$ being, respectively identified to $\tau^{eff}$ and $\kappa(\mathbf{0}) / C_W \tau^{eff}$.

A more manageable expression of the second sound relaxation time $\tau_{II}^{H}$ can easily be obtained using the expression of $\kappa_{q,S}^{\mathbf{0}}$ as given by Eq. (14) combined to the usual change of variable $x = \hbar\omega / k_B T_0$. One obtains the following expression:

$$\begin{cases} \tau_{II}^{H} = \dfrac{\sum_S \dfrac{\mathbf{1}}{v_S}\left[g_S + \beta_S h_S\right]}{\sum_S \dfrac{\mathbf{1}}{v_S}\left[c_S + \beta_S d_S\right]} \\ g_S = \displaystyle\int_{\mathbf{0}}^{\theta_D^S / T_{\mathbf{0}}} \left[\tau_S^C(x)\right]^2 D(x)\,dx;\; h_S = \int_{\mathbf{0}}^{\theta_D^S / T_{\mathbf{0}}} \frac{\left[\tau_S^C(x)\right]^2}{\tau_S^N(x)} D(x)\,dx \end{cases} \qquad (D.\mathbf{10})$$

where, $D(x)$ is Debye function and $\theta_D^S$ is Debye temperature of the acoustic polarization branch S.[59] The expression of Callaway pseudo-relaxation time $\beta_S$ is as calculated in the steady-state regime [Eq. (A.4)] while the expressions of $c_S$ and $d_S$ are as given by Eq. (47).

The expression of the second sound velocity $v_{II}^{H}$ is readily obtained based on Eqs. (D.7), and (D.10) combined to the equation given $C_W$ ($C_W = W^{-\mathbf{1}} \sum_{q,S} C_{Ph}(q,S)$). This leads to:

$$\left(v_{II}^{H}\right)^{\mathbf{2}} = \frac{\mathbf{1}}{\mathbf{3}} \frac{\left\{\sum_S \dfrac{\mathbf{1}}{v_S}\left[c_S + \beta_S d_S\right]\right\}^{\mathbf{2}}}{\left\{\sum_S \dfrac{\mathbf{1}}{v_S^{\mathbf{3}}} a_S\right\} \times \left\{\sum_S \dfrac{\mathbf{1}}{v_S}\left[g_S + \beta_S h_S\right]\right\}} \qquad (D.\mathbf{11})$$

where $a_S$ is as given by Eq. (45).

In order to obtain the second sound dispersion relation describing the way the frequency $\Omega$ of a temperature wave depends on the its propagation vector $Q$ in Hardy's method, one substitutes a wave-plane solution of the form $T(x,t) = T_{\mathbf{0}}\, e^{j(Qx - \Omega t)}$ into the damped wave equation for the local temperature [Eq. (D.9)]. One obtains:

$$\left(\frac{\Omega}{Q}\right)^2 = \left(v_{II}^H\right)^2 - j\frac{\Omega}{Q^2\tau_{II}^H} \qquad (D.\mathbf{12})$$

As can be seen by comparing Eqs. (45) and (47) to Eqs. (D.10) and (D.11), respectively, the expressions of second sound relaxation time and velocity as obtained based on the herein presented Hardy's generalized approach have forms that are very close if not similar to the ones obtained using Griffin's self consistency criterion, but with different weighting functions. On the other hand, second sound damping factor has a totally different expression as it depends on the temperature disturbance propagation vector $Q$ in Hardy's approach while it does not in Griffin's. Also the damping factor expression in Hardy's approach does not exhibit any time window in the relaxation time spectrum for the occurrence of second sound. The weakness of the herein presented Hardy's approach lies in the difficulty to rigorously justify the definition of the effective relaxation time as given by Eqs. (D.6) and (D.7), even though one can be less rigidly precise in saying that this definition might be justified by the final results obtained for both second sound relaxation time and velocity.

## REFERENCES

1. Y. Ezzahri and K. Joulain, J. Appl. Phys, **112**, 083515 (2012).
2. Y. Ezzahri and K. Joulain, J. Appl. Phys, **113**, 043510 (2013).
3. M. D. Row, "*Handbook of Thermoelectrics*", (CRC, Boca Raton, FL, 1995).
4. G. S. Nolas, J. Sharp and J. Goldsmid, "*Thermoelectrics: Basic Principles and New Materials Developments*", (Springer-Verlag, Berlin, Heidelberg, 2002).
5. R. Venkatasubramanian, E. Silvola, T. Colpitts and B. O'Quinn, Nature. **413**, 597 (2001).
6. T. C. Harman, P. J. Taylor, M.P. Walsh and B. E. Laforge, Science. **297**, 2229 (2002).
7. W. Kim, J. Zide, A. Gossard, D. Klenov, S. Stemmer, A. Shakouri and A. Majumdar, Phys. Rev. Lett. **96**, 045901 (2006).
8. B. Poudel, Q. Hao, Y. Ma, Y. C. Lan, A. Minnich, B. Yu, X. Yan, D. Z. Wang, A. Muto, D. Vashaee, X. Y. Chen, J. M. Liu, M. S. Dresselhaus, G. Chen and Z. F. Ren, Science. **320**, 63 (2008).
9. A. I. Hochbaum, R. Chen, R. D. Delgado, W. Liang, E. C. Garnett, M. Najarian, A. Majumdar and P. Yang, Nature. **451**, 163 (2008).
10. A. Shakouri, Annu. Rev. Mater. Res. **41**, 399 (2011).
11. M. Zebarjadi, K. Esfarjani, M. S. Dresselhaus, Z. F. Ren and G. Chen, Energy Environ. Sci. **5**, 514 (2012).
12. W. S. Capinski, H. J. Maris, T. Ruf, M. Cardona, K. Ploog and D. S. Katzer, Phys. Rev. B **59**, 8105 (1999).
13. R. Venkatasubramanian, Phys. Rev. B **61**, 3091 (2000).
14. D. Li, Y. Wu, P. Kim, L. Shi, P. Yang and A. Majumdar, Appl. Phys. Lett. **83**, 2934 (2003).
15. M. L. Lee and R. Venkatasubramanian, Appl. Phys. Lett. **92**, 053112 (2008).
16. N. Mingo, D. Hauser, N. P. Kobayashi, M. Plissonier and A. Shakouri, Nano. Lett. **9**, 711 (2009).
17. R. E. Peierls, Ann. Phys. (Leipzig) **3**, 1055 (1929).
18. P. G. Klemens, Proc. Phys. Soc. A **208**, 108 (1951).
19. C. Herring, Phys. Rev. **95**, 954 (1954).
20. P. G. Klemens, Proc. Phys. Soc. A **68**, 1113 (1955).
21. J. Callaway, Phys. Rev. **113**, 1046 (1959).

22. J. M. Ziman, “*Electron and Phonons*”, (Oxford university press, New York, 1960).
23. J. Callaway, Phys. Rev. **122**, 787 (1961).
24. P. Carruthers, Rev. Mod. Phys. **33**, 92 (1961).
25. B. Abeles, Phys. Rev. **131**, 1906 (1963).
26. R. E. Nettleton, Phys. Rev. **132**, 2032 (1963).
27. M. G. Holland, Phys. Rev. **132**, 2461 (1963).
28. G. A. Slack and S. Galginaitis, Phys. Rev. **133**, A253 (1964).
29. G. A. Slack and C. J. Glassbrenner, Phys. Rev. **120**, 782 (1960).
30. C. J. Glassbrenner and G. A. Slack, Phys. Rev. **134**, A1058 (1964).
31. J. E. Parrott, Phys. Status. Solidis B **48**, K159 (1971).
32. G. P. Srivastava, *"The Physics of Phonons"*, (Adam Hilger, Bristol, 1990).
33. M. Asen-Palmer, K. Bartkowski, E. Gmelin, M. Cardona, A. P. Zhemov, A. V. Inyushkin, A. Taldenkov, V. I. Ozhogin, K. M. Itoh and E. E. Haller, Phys. Rev. B **56**, 9431 (1997).
34. D. T. Morelli, J. P. Heremans and G. A. Slack, Phys. Rev. B **66**,195304 (2002).
35. Y. Ezzahri and A. Shakouri, Phys. Rev. B **79**, 184303 (2009).
36. Y. Ezzahri, K. Joulain and A. Shakouri, J. Heat. Transfer. **133**, 072401 (2011).
37. G. D. Mahan and F. Claro, Phys. Rev. B **38**, 1963 (1988).
38. D. D. Joseph and L. Preziosi, Rev. Mod. Phys. **61**, 41 (1989).
39. A. A. Joshi and A Majumdar, J. Appl. Phys. **74**, 31 (1993).
40. G. Chen, Phys. Rev. Lett. **86**, 2297 (2001).
41. F. X. Alvarez and D. Jou, Appl. Phys. Lett. **90**, 083109 (2007).
42. F. X. Alvarez and D. Jou, J. Appl. Phys. **103**, 094321 (2008).
43. M. E. Siemensi, Q. Li, R. Yang, K. A. Nelson, E. H. Anderson, M. M. Murnane and H. C. Kapteyn, Nature. **9**, 26 (2010).
44. H. Iwai, Microelectronic Engineering. **86**, 1520 (2009).
45. R. A. Guyer and J. A. Krumhansl, Phys. Rev. **133**, A1411 (1964).
46. R. A. Guyer and J. A. Krumhansl, Phys. Rev. **148**, 766 (1966).
47. R. A. Guyer and J. A. Krumhansl, Phys. Rev. **148**, 778 (1966).
48. S. G. Volz and G. Chen, Phys. Rev. B **61**, 2651 (2000).
49. S. G. Volz, Phys. Rev. Lett. **87**, 074301 (2001).
50. B. Hüttner, Physica Status Solidi (b). **245**, 2786 (2008).
51. B. S. Shastry, Phys. Rev. B **73**, 085117 (2006).
52. B. S. Shastry, Rep. Prog. Phys. **72**, 016501 (2009).

53. Y. K. Koh and D. G. Cahill, Phys. Rev. B **76**, 075207 (2007).
54. D. G. Cahill, Rev. Sci. Instrum. **75**, 5119 (2004).
55. A. J. Schmidt, X. Chen and G. Chen, Rev. Sci. Instrum. **79**, 114902 (2008).
56. S. Dilhaire, G. Pernot, G. Calbris, J. M. Rampnoux and S. Grauby, J. Appl. Phys. **110**, 114314 (2011).
57. A. Griffin, Rev. Mod. Phys. **40**, 167, (1968).
58. As P. Carruthers pointed it out in his review paper (see reference 24), the terminology of "phonon gas" often used by authors in the field of condensed matter and transport phenomena, can be misleading if one recognizes the fundamental difference in the onset of thermal resistance between a dielectric crystal and a fluid. While in the latter, thermal resistance is due to intermolecular forces and the randomness of the molecular distribution, in the former, the discrete character of the lattice and the anharmonic forces are both essential to have a noninfinite thermal conductivity. As this terminology is generally accepted, we continue to use it throughout this paper.
59. N. W. Ashcroft and N. D. Mermin, "*Solid State Physics*", (second edition, Holt Rinehart and Winston, New York, 1976).
60. R. O. Pohl, Phys. Rev. **118**, 1499 (1960).
61. B. K. Agrawal and G. S. Verma, Phys. Rev. **128**, 603 (1962).
62. M. E. Gurtin and A. C. Pipkin, Arch. Ration. Mech. Anal. **31**,113 (1968).
63. W. Hafez and M. Feng, Appl. Phys. Lett. **86**, 152101 (2005).
64. M. Feng, N. Holonyak, Jr., G. Walter and R. Chan, Appl. Phys. Lett. **87**, 131103 (2005).
65. G. A. Slack and M. A. Hussain, J. Appl. Phys. **70**, 2694 (1991).
66. http://www.ioffe.ru/SVA/NSM/Semicond/index.html for a very interesting collection of experimental data on various physical properties of the main semiconductor crystals. All information is supported by a full list of references.
67. S. Wang and N. Mingo, Appl. Phys. Lett. **94,** 203109 (2009).
68. A. Kundo, N. Mingo, D. A. Broido and D. A. Stewart, Phys. Rev. B **84**, 125426 (2011).
69. M. Zebarjadi, K. Esfarjani, A. Shakouri, J. H. Bahk, Z. Bian, G. Zeng, J. Bowers, H. Lu, J. Zide and A. Gossard, Appl. Phys. Lett. **94**, 202105 (2009).
70. S. V. Faleev and F. Leonard, Phys. Rev. B **77**, 214304 (2008).
71. W. Kim and A. Majumdar, J. Appl. Phys. **99**, 084306 (2006).

72. A. J. Minnich, J. A. Johnson, A. J. Schmidt, K. Esfarjani, M. S. Dresselhaus, K. A. Nelson and G. Chen, Phys. Rev. Lett. **107**, 095901 (2011).
73. A. J. Minnich, G. Chen, S. Mansoor and B. S. Yilbas, Phys. Rev. B **84**, 235207 (2011).
74. J. C. Ward and J. Wilks, Phil. Mag. **42**, 314, (1951).
75. J. C. Ward and J. Wilks, Phil. Mag. **43**, 48, (1952).
76. R. B. Dingle, Proc. Phys. Soc. (London) **A65**, 374, (1952).
77. M. Chester, Phys. Rev. **131**, 2013, (1963).
78. E. W. Prohofsky and J. A. Krumhansl, Phys. Rev. **133**, A1403, (1964).
79. P. C. Kwok and P. C. Martin, Phys. Rev. **142**, 495, (1966).
80. A. Griffin, Phys. Lett. **17**, 208, (1965).
81. C. P. Enz, Ann. Phys. (N.Y.) **46**, 114, (1968).
82. R. J. Hardy, Phys. Rev. B **2**, 1193, (1970).
83. J. Ranninger, J. Phys. C, **2**, 929, (1969).
84. C. C. Ackerman, B. Bertman, H. A. Fairbank and R. A. Guyer, Phys. Rev. Lett. **16**, 789, (1966).
85. V. Narayanamurti and C. M. Varma, Phys. Rev. Lett. **25**, 1105, (1970).
86. V. Narayanamurti and R. C. Dynes, Phys. Rev. Lett. **28**, 1461, (1972).
87. V. Narayanamurti and R. C. Dynes, Phys. Rev. B **12**, 1731, (1975).
88. K. Esfarjani, G. Chen and H. T. Stokes, Phys. Rev. B **84**, 085204 (2011).
89. W. Li, N. Mingo, L. Lindsay, D. A. Broido, D. A. Stewart and N. A. Katcho, Phys. Rev. B **85**, 195436 (2012).